\documentclass[aps,prd,singlecolum,nofootinbib,preprintnumbers,superscriptaddress,floatfix,nosingletitlepage]{revtex4-2}

\usepackage{epsfig}
\usepackage{wrapfig}
\usepackage{graphicx}
\usepackage{bm}
\usepackage{amssymb}
\usepackage{amsmath}
\usepackage{amsfonts}
\usepackage[dvipsnames]{xcolor}
\usepackage{subfigure}
\usepackage{dcolumn}
\usepackage[makeroom]{cancel}
\usepackage{float}

\usepackage[utf8]{inputenc}

\usepackage{amsmath}
\usepackage{amssymb}
\usepackage{amsthm}
\usepackage{amscd, amsfonts, mathrsfs}
\usepackage{cases}
\usepackage{verbatim} 
\usepackage{epsf}
\usepackage[bookmarksnumbered,pdfpagelabels=true,plainpages=false,colorlinks=true,
            linkcolor=black,citecolor=black,urlcolor=Blue]{hyperref}

\theoremstyle{plain}

\theoremstyle{definition}

\allowdisplaybreaks

\usepackage{appendix}
\newcommand{\cA}{\mathcal A}
\newcommand{\cB}{\mathcal B}

\newcommand{\cE}{\mathcal E}

\newcommand{\cL}{\mathcal L}

\newcommand{\cN}{\mathcal N}

\newcommand{\cV}{\mathcal V}

\newcommand{\ga}{\gamma}
\newcommand{\Ga}{\Gamma}
\newcommand{\de}{\delta}
\newcommand{\ep}{\varepsilon}
\newcommand{\la}{\lambda}

\newcommand{\si}{\sigma}

\newcommand{\om}{\omega}

\newcommand{\vv}{\vec{v}}

\newcommand{\pa}{\partial}

\newcommand{\lan}{\langle}
\newcommand{\ran}{\rangle}
\newcommand{\bsi}{\bar{\sigma}}
\newcommand{\bom}{\bar{\omega}}
\newcommand{\brho}{\bar{\rho}}
\newcommand{\bps}{\bar{\psi}}

\newcommand{\be}{\begin{equation}}
\newcommand{\ee}{\end{equation}}
\newcommand{\bea}{\begin{aligned}}
\newcommand{\eea}{\end{aligned}}

\newcommand{\bml}{\begin{subequations}}
\newcommand{\eml}{\end{subequations}}

\newcommand{\bbm}{\begin{bmatrix}}
\newcommand{\ebm}{\end{bmatrix}}
\newcommand{\bvm}{\begin{vmatrix}}
\newcommand{\evm}{\end{vmatrix}}

\begin{document}

\title{Far-from-equilibrium bulk-viscous transport coefficients\\ in neutron star mergers}

\date{\today}

\author{Yumu Yang}
\email{yumuy2@illinois.edu}
\affiliation{Illinois Center for Advanced Studies of the Universe \& Department of Physics, 
University of Illinois Urbana-Champaign, Urbana, IL 61801, USA}

\author{Mauricio Hippert}
\email{hippert@illinois.edu}
\affiliation{Illinois Center for Advanced Studies of the Universe \& Department of Physics, 
University of Illinois Urbana-Champaign, Urbana, IL 61801, USA}

\author{Enrico Speranza}
\email{espera@illinois.edu}
\affiliation{Illinois Center for Advanced Studies of the Universe \& Department of Physics, 
University of Illinois Urbana-Champaign, Urbana, IL 61801, USA}

\author{Jorge Noronha}
\email{jn0508@illinois.edu}
\affiliation{Illinois Center for Advanced Studies of the Universe \& Department of Physics, 
University of Illinois Urbana-Champaign, Urbana, IL 61801, USA}

\begin{abstract}
We investigate the weak-interaction-driven bulk-viscous transport properties of $npe$ matter in the neutrino transparent regime. Previous works assumed that the induced bulk viscosity correction to pressure, near beta equilibrium, is linear in deviations from the equilibrium charge fraction. We show that this is not always true for (some) realistic equations of state at densities between one and three times saturation density. This nonlinear nature of the perturbation around equilibrium motivates a far-from-beta-equilibrium description of bulk-viscous transport in neutron star mergers, which can be precisely achieved using a new Israel-Stewart formulation with resummed bulk and relaxation time transport coefficients. The computation of these transport coefficients depends on out-of-beta-equilibrium pressure corrections, which can be computed for a given equation of state. We calculate these coefficients for equations of state that satisfy the latest constraints from multi-messenger observations from LIGO/VIRGO and NICER. We show that varying the nuclear symmetry energy $J$ and its slope $L$ can significantly affect the transport coefficients and the nonlinear behavior of the out-of-equilibrium pressure corrections. Therefore, having better constraints on $J$ and $L$ will directly impact our understanding of bulk-viscous processes in neutron star mergers. 

\end{abstract}

\maketitle

\section{Introduction}

Binary neutron star collisions detected through gravitational waves \cite{LIGOScientific:2017vwq, LIGOScientific:2020aai} offer exciting prospects for constraining the dense matter equation of state (EoS) \cite{Gandolfi:2011xu, Zhang:2015ava, Kowalski:2006ju, Roca-Maza:2012uor, Zhang:2013wna, Fan:2014rha, Danielewicz:2013upa, Reinhard:2021utv, Reed:2021nqk}, both via the tidal deformability encoded in the inspiral gravitational wave signal \cite{Bauswein:2017vtn, Annala:2017llu, Most:2018hfd, LIGOScientific:2018cki, Raithel:2018ncd, De:2018uhw, Chatziioannou:2018vzf, Carson:2018xri} and through its electromagnetic counterparts \cite{Radice:2017lry, Margalit:2017dij, Rezzolla:2017aly, Ruiz:2017due, Shibata:2019ctb}. Neutron star mergers give rise to extreme densities and temperatures of 80 MeV or more \cite{Kastaun:2016yaf, Hanauske:2016gia, Perego:2019adq, Endrizzi:2019trv}, being thus of great interest for mapping out the equilibrium properties of hot and ultradense QCD matter \cite{Tan:2021ahl, Oechslin:2004yj, Most:2018eaw, Bauswein:2018bma, Most:2019onn, Weih:2019xvw, Chatziioannou:2019yko, Prakash:2021wpz, Tootle:2022pvd,MUSES:2023hyz}.

Key to determining the EoS of dense and neutron-rich matter is the nuclear symmetry energy, which characterizes the difference in energy between symmetric nuclear matter and pure neutron matter. The symmetry energy $E_{sym}$ is often modeled with a series expansion in baryon density $n_B$,
\be
E_{sym}(n_B) = J + \frac{L}{3}\left( \frac{n_B}{n_{\textrm{sat}}} - 1 \right) + \mathcal{O}\left[\left(1-\frac{n_B}{n_{\textrm{sat}}}\right)^2\right],
\ee
where $J$ is the symmetry energy at the nuclear saturation density $n_{\textrm{sat}} = 0.148\;\text{fm}^{-3}$ \cite{Piekarewicz:2008nh, Gross-Boelting:1998qhi, PREX:2021umo},  and $L$ is its slope.
Many efforts have been undertaken to constrain $J$ and $L$ \cite{Gandolfi:2011xu,Zhang:2015ava,Kowalski:2006ju,Roca-Maza:2012uor,Zhang:2013wna,Fan:2014rha,Danielewicz:2013upa,Reinhard:2021utv,Reed:2021nqk,Lim:2023dbk}, though $L$ still displays large uncertainty \cite{MUSES:2023hyz}. The slope of the symmetry energy is expected to have a large impact on the star radius and tidal deformability \cite{Lattimer:2000nx, Yagi:2013awa, Yagi:2015pkc, De:2018uhw, Raithel:2018ncd}, but previous works indicate that changing $L$ produces a small effect on the post-merger dynamics and gravitational wave emission \cite{Most:2021ktk}.

Neutron star collisions also offer new opportunities to learn about novel out-of-equilibrium properties of dense matter \cite{Alford:2017rxf}. The violent changes in temperature and density found in mergers can drive the system away from beta equilibrium \cite{Hammond:2021vtv, Most:2021ktk} as the proton fraction evolves out of phase with variations of the baryon density, which leads to dissipative work \cite{Alford:2018lhf, Alford:2021ogv, Alford:2023gxq}.
Hence, weak-interaction reaction processes that restore beta equilibrium \cite{Sawyer:1989dp, Haensel:1992zz, Alford:2018lhf, Alford:2019qtm, Alford:2019kdw, Alford:2021ogv} can in principle damp density and temperature oscillations right after the merger, which may lead to observable imprints in the gravitational waves emitted by the system \cite{Alford:2017rxf, Most:2021zvc, Most:2022yhe, Hammond:2022uua, Chabanov:2023blf}.  

It has been known for a long time that such dissipative effects associated with changing particle fractions can be modeled (at least in the context of linear response theory) as an effective bulk-viscous correction to the pressure \cite{Sawyer:1989dp, Haensel:1992zz}. Bulk viscosity has also been investigated in the context of r-modes of isolated neutron stars \cite{Madsen:1992sx, Reisenegger:2003pd, Alford:2010gw}.  
Recently, in Ref.\ \cite{Gavassino:2020kwo} (see also \cite{Celora:2022nbp, Camelio:2022ljs, Camelio:2022fds}), it was shown that sufficiently close to beta equilibrium, the correction to the beta-equilibrated pressure follows a dynamical equation of motion a la Israel-Stewart  \cite{Israel:1979wp}, with transport coefficients determined by the dense matter equation of state in beta equilibrium, as well as weak-interaction rates \cite{Gavassino:2020kwo}. Surprisingly enough, in Ref.\ \cite{Gavassino:2023xkt}, this equivalence was shown to hold also \emph{beyond} the linear response regime. As a matter of fact, \cite{Gavassino:2023xkt} demonstrated that the dynamics of a two-component, reactive relativistic fluid mixture can be rigorously
rewritten as a resummed bulk-viscous Israel-Stewart theory even far from equilibrium. However, in this case, the bulk viscosity and relaxation time transport coefficients must also depend on the out-of-beta-equilibrium pressure corrections. Therefore, simulations of neutron star mergers describing dense matter in the presence of non-conserved lepton currents due to weak decays, such as those performed in \cite{Most:2022yhe}, inherently display the physics of bulk viscosity, regardless of whether linear response arguments hold or not. In this sense, neutron star collisions are, thus, expected to be intrinsically bulk-viscous systems.

In this work, following \cite{Gavassino:2023xkt}, we determine for the first time all the weak-interaction-driven bulk-viscous transport properties of $npe$ matter in the neutrino transparent regime using realistic equations of state. Previous works assume that the induced bulk-viscous correction to the pressure, near beta equilibrium, is linear in deviations from the equilibrium charge fraction.  We show this is not the case for realistic EoSs \cite{Chen:2014sca} at densities between one and three times saturation density. In general, the nonlinear nature of the perturbation around equilibrium requires a far-from-beta-equilibrium description of bulk-viscous transport, which is investigated here using the exact Israel-Stewart formulation with resummed bulk and relaxation time transport coefficients derived in \cite{Gavassino:2023xkt}. In addition, we systematically vary $J$ and $L$ (within current constraints) to study their effects on the bulk-viscous transport coefficients. We find that the transport coefficients are quite sensitive to variations in $J$ and $L$. This implies that having better constraints on such quantities will directly benefit our understanding of bulk-viscous effects in neutron star mergers. 

This paper is organized as follows. In Section \ref{IS derivation}, we show how weak interactions give rise to a bulk-viscous description described by Israel-Stewart theory with resummed transport coefficients. In practice, given the temperatures involved in the neutrino-transparent regime, we compute the resummed transport coefficients under a zero-temperature approximation for the EoS (we leave temperature corrections to the EoS for future work). In Section \ref{EOS}, we show how to obtain realistic EoSs in the context of Walecka-like \cite{norman1997compact} relativistic mean-field theory models that satisfy the latest constraints from LIGO/VIRGO~\cite{LIGOScientific:2017vwq, LIGOScientific:2018hze, LIGOScientific:2018cki} and NICER~\cite{Miller:2019cac, Riley:2019yda, Miller:2021qha, Riley:2021pdl}, with varying $J$ and $L$. In Section \ref{Results}, we first present results for the bulk viscosity transport coefficient near beta equilibrium, which are compatible with that found in previous works \cite{Alford:2018lhf, Alford:2021ogv, Alford:2023gxq}. We then compute the resummed transport coefficients and discuss how nonlinearities in the total out-of-equilibrium pressure can impact the transport coefficients in the far-from-equilibrium regime. In particular, such nonlinearities can be so significant that they may affect the interpretation of the system as a bulk-viscous fluid. In addition, we show that by varying $J$ and $L$, one can change the equilibrium charge fraction's value, drastically affecting the transport coefficients. 

Our conclusions and outlook are presented in Section \ref{sec:conclusions}. For the sake of completeness, we present the Urca rates in Appendix \ref{A: rate}, while Appendices \ref{A: linear}, \ref{A: AC IS}, and \ref{A: equivalence} discuss further details concerning the Israel-Stewart formulation of bulk viscosity. These appendices are included to facilitate the comparison between this work and previous papers where the bulk viscosity coefficient is estimated from linear response theory applied to periodic density oscillations. Finally, in Appendix \ref{A: RMF EoS}, we briefly review the thermal field theory calculations needed to determine the equation of state.      

\emph{Notation}: We use natural units, $\hbar=c=k_B=1$, and $g_{\mu\nu}$ is the Minkowski metric with a mostly minus signature.

\section{Resummed Israel-Stewart theory from chemical imbalance}
\label{IS derivation}

We consider a system composed of protons $p$, neutrons $n$, and electrons $e$. This provides a model description of neutron star matter at temperatures that are low enough that the neutrino mean free path is larger than the radius of the star \cite{Alford:2018lhf}. In this system, flavor equilibration occurs via the direct and modified Urca processes,
\begin{gather}
    n \to p + e^- + \Bar{\nu}_e ,\\
    p + e^- \to n + \nu_e ,\\
    n + X \to p + e^- + \Bar{\nu}_e + X ,\\
    p + e^- + X \to n + \nu_e + X,
\end{gather}
where $X$ is a spectator nucleon, a neutron or a proton. 

In the case that neutrinos are not trapped, the criterion for beta equilibrium is given by the following detailed balance condition:
\be
\label{eq:beta_eq}
\mu_n = \mu_p + \mu_e ,
\ee
where $\mu_n$, $\mu_p$, and $\mu_e$ are the chemical potentials for neutrons, protons, and electrons, respectively. We can quantify violations of Eq.~\eqref{eq:beta_eq} 
with the difference $\delta \mu = \mu_n - \mu_p - \mu_e$, so that $\de \mu = 0$ characterizes beta equilibrium\footnote{The $\delta\mu=0$ case we use here is a crude approximation valid only at very low temperatures ($T \ll $ 1 MeV). At sufficiently large temperatures, beta equilibrium does not require $\delta \mu=0$, see \cite{Alford:2018lhf,Alford:2021ogv}.}. 

We now discuss how the above chemical reactions can produce bulk viscosity \cite{Gavassino:2020kwo}. 
We consider a thermally isolated system. There are two timescales associated with it: $\tau_{hydro}$, defined as the timescale of perturbations from the hydrodynamic processes such as baryon density oscillations, and $\tau_{reac}$, defined as the timescale for the system to equilibrate via chemical reactions. If $\tau_{hydro} \gg \tau_{reac}$, the reaction happens so rapidly that the system is instantaneously equilibrated. The system is in equilibrium at each time step, so there is no dissipation. If $\tau_{hydro} \ll \tau_{reac}$, the system has no time to respond to the perturbation. The system's composition is fixed, and the process is reversible, so there is no dissipation. Only when $\tau_{hydro} \approx \tau_{reac}$ the chemical reactions can change the system's composition, but they do not have enough time to keep the system in equilibrium. Consequently, the process is irreversible, the system's entropy grows, and dissipation occurs. In binary neutron star mergers, density oscillations can happen at a timescale of milliseconds, similar to those required by weak interactions to restore beta equilibrium \cite{Alford:2018lhf, Alford:2021ogv}, which corresponds to the dissipative case where $\tau_{hydro} \approx \tau_{reac}$.

Because dissipation is induced by flavor-changing processes, one can measure the effect of the reaction-induced bulk viscosity by keeping track of the change in the chemical composition of the system. We can then write the first law of thermodynamics as 
\begin{align}
\label{firstlaw}
\bea
d\ep &= T ds + \mu_n dn_n + \mu_p dn_p  + \mu_e dn_e \\
&= T ds + \mu_n dn_B - \delta \mu\, dn_e ,
\eea
\end{align}
where $\ep$ is the energy density, $s$ is the entropy density, $T$ is the temperature, $n_n$, $n_p$ and $n_e$ are the neutron, proton, and electron densities, respectively, and $n_B = n_p + n_n$ is the baryon density. In the second line of Eq.~\eqref{firstlaw} we used charge neutrality $n_p=n_e$, which implies that we are left with three degrees of freedom to describe the local thermodynamic state. In this work, we choose the energy density $\ep$, the baryon density $n_B$, and the electron fraction $Y_e=n_e/n_B$ as the three original independent thermodynamic variables.

Let us now define the equations of motion of the hydrodynamic system. In this work, we follow the derivation in Ref.~\cite{Gavassino:2023xkt} and assume the energy-momentum tensor to (formally) have the form of an ideal fluid,
\be
T^{\mu\nu} = (\ep + P)u^\mu u^\nu - P g^{\mu\nu}, 
\ee
where $P$ is the total pressure, which is not only a function of the equilibrium variables but also contains nonequilibrium contributions.

Because neutrinos get out of the system, energy and momentum are not exactly conserved, and thus, 
\be
\nabla_\mu T^{\mu \nu} = Q^\nu,
\label{eqconsEM}
\ee
where $Q^\nu$ represents the loss due to neutrinos. The equation for the energy density, obtained by contracting \eqref{eqconsEM} with $u_\nu$, is given by
\begin{equation}
\label{energy conserv}
    \begin{aligned}
    u^\mu \nabla_\mu \ep + (\ep + P) \theta = u^\mu Q_\mu,
    \end{aligned}
\end{equation}
where $\theta = \nabla_\mu u^\mu$ is the expansion rate. The baryon current is given by $J_\mu = n_B u_\mu$, and its conservation leads to
\begin{equation}
\label{baryon}
    \nabla^\mu J_\mu  = n_B \theta + u^\mu \nabla_\mu n_B = 0.
\end{equation}
Due to the weak-interaction processes mentioned above, the electron current, $J_e^\mu = n_e u^\mu$, is not conserved, namely
\begin{equation}
\label{e density}
    \begin{aligned}
    \nabla_\mu J^\mu_e = u^\mu \nabla_\mu n_e + n_e \theta = \Gamma_e (\varepsilon, n_B, Y_e) ,
    \end{aligned}
\end{equation}
where the reaction rate $\Gamma_e$ is outlined in Appendix \ref{A: rate}. Substituting Eq.\ \eqref{baryon} and the charge/electron fraction $Y_e = n_e/n_B$ into Eq.\ \eqref{e density}, we find
\begin{equation}
\label{gamma}
    \begin{aligned}
    u^\mu \nabla_\mu Y_e = \frac{\Gamma_e}{n_B}.
    \end{aligned}
\end{equation}

We are now ready to show how this system can be exactly described by a bulk-viscous Israel-Stewart \cite{Israel:1979wp} theory with new resummed transport coefficients. The derivation follows the reasoning used in the recent paper \cite{Gavassino:2023xkt}, and it works even for a system far from beta equilibrium. We define the bulk scalar $\Pi$ to be the off-beta equilibrium correction to the pressure, 
\be
P = P_{eq} + \Pi.
\ee
With the three degrees of freedom in our system, we choose $\ep$ and $n_B$ to be the variables that describe the equilibrium state and $Y_e$ to be the variable that describes the out-of-equilibrium state. We can then consider the evolution of $\Pi$, 
\be
\bea
u^\mu\nabla_\mu \Pi = \left.\frac{\pa \Pi}{\pa \ep}\right|_{n_B,Y_e} u^\mu\nabla_\mu \ep + \left.\frac{\pa \Pi}{\pa n_B}\right|_{\ep,Y_e} u^\mu\nabla_\mu n_B + \left.\frac{\pa \Pi}{\pa Y_e}\right|_{\ep,n_B} u^\mu\nabla_\mu Y_e .
\label{derivationPieq}
\eea
\ee
Substituting Eq.~\eqref{energy conserv} and Eq.~\eqref{baryon} into the equation above, one finds
\be
\bea
u^\mu\nabla_\mu \Pi &= - \left.\frac{\pa \Pi}{\pa \ep}\right|_{n_B,Y_e} (\ep + P)\theta + \left.\frac{\pa \Pi}{\pa \ep}\right|_{n_B,Y_e} u^\mu Q_\mu - \left.\frac{\pa \Pi}{\pa n_B}\right|_{\ep,Y_e} n_B \theta + \left.\frac{\pa \Pi}{\pa Y_e}\right|_{\ep,n_B} u^\mu\nabla_\mu Y_e\\
&= - \left[ \left.\frac{\pa \Pi}{\pa \ep}\right|_{n_B,Y_e} (\ep + P) + \left.\frac{\pa \Pi}{\pa n_B}\right|_{\ep,Y_e} n_B \right]\theta + \left.\frac{\pa \Pi}{\pa \ep}\right|_{n_B,Y_e} u^\mu Q_\mu + \left.\frac{\pa \Pi}{\pa Y_e}\right|_{\ep,n_B} \frac{\Ga_e}{n_B}.
\label{eqgood}
\eea
\ee

Let us now define
\be
    F = - \left.\frac{\pa \Pi}{\pa Y_e}\right|_{\ep,n_B} \frac{\Ga_e}{n_B} .
\ee
At $\beta$-equilibrium, where $\Pi = 0$, $F$ should vanish. Therefore, it must be possible to write it as $F =\Pi \,F_{\Pi}$, where the quantity $F_{\Pi}$ remains finite when $\Pi\to 0$. In other words, $F_\Pi$ can depend on higher powers of $\Pi$, but $F$ must vanish when $\Pi$ vanishes, which implies that $F_\Pi$ must remain finite (i.e., non-divergent) in this limit. One can now rewrite \eqref{eqgood} in Israel-Stewart-like form
\be
\tau_\Pi u^\mu\nabla_\mu \Pi + \Pi - \tau_\Pi \left.\frac{\pa \Pi}{\pa \ep}\right|_{n_B,Y_e} u^\mu Q_\mu = - \zeta \theta ,
\label{approxIS}
\ee
where the relaxation time, $\tau_\Pi$, and bulk viscosity coefficient, $\zeta$, are given by
\begin{gather}
    \tau_\Pi = \frac{1}{F_\Pi}, \\
    \zeta = \frac{1}{F_\Pi} \left( \left.\frac{\pa \Pi}{\pa \ep}\right|_{n_B,Y_e} (\ep + P) + \left.\frac{\pa \Pi}{\pa n_B}\right|_{\ep,Y_e} n_B \right) . 
    \label{definezetaandtaupi}
\end{gather}
We note that even though \eqref{approxIS} is a relaxation-type equation similar to that found in Israel-Stewart theory, there are new elements in \eqref{approxIS}. First, when $Q_\mu \neq 0$, a new term is added to the equation; see the third term on the left-hand side of \eqref{approxIS}. More importantly, we note that $\zeta$ and $\tau_\Pi$ in \eqref{definezetaandtaupi} depend not only on the standard thermodynamic variables, as it occurs in Israel-Stewart theory, but they also depend on the deviation from equilibrium\footnote{The attentive reader will notice that the derivation of the Israel-Stewart equation done here is equivalent but not identical to the original result in Ref.\ \cite{Gavassino:2023xkt}. Here we started by deriving an equation for $\Pi = \Pi(\varepsilon,n_B,Y_e)$, see \eqref{derivationPieq}. On the other hand, \cite{Gavassino:2023xkt} started from the equation for $Y_e=Y_e(\varepsilon,n_B,\Pi)$ in \eqref{gamma}, and then found the equation for $\Pi$ (using $Y_e^{\textrm{eq}} = Y_e(\varepsilon,n_B,0)$). These derivations are equivalent if the mapping between $\Pi$ and $Y_e$ is invertible. We have chosen the alternative derivation presented here because it is easier to numerically determine the resummed transport coefficients this way using our equations of state.} $\Pi$. Therefore, this provides an example of ``resummation" in hydrodynamics because the hydrodynamic theory maintains the same form even arbitrarily far from equilibrium. This resummation is encoded in the explicit $\Pi$ dependence of the transport coefficients $\zeta$ and $\tau_\Pi$, determined by the properties of the original out-of-beta equilibrium equation of state, $P(\varepsilon,n_B,Y_e \neq Y_e^{\textrm{eq}})$. Of course, sufficiently near equilibrium, the $\Pi$ dependence of these coefficients can be dropped, and their expressions reduce to those found in approaches based on linear response; see Appendix \ref{A: linear} and \ref{A: equivalence} for details.

For the sake of simplicity, we will only consider here the case where $Q^\mu = 0$ (this is also used in \cite{Most:2022yhe}), leaving a more detailed discussion of the $Q^\mu \neq 0$ situation for future work. In this case, Eq.\ \eqref{approxIS} reduces to
\be
\label{IS}
\tau_\Pi u^\mu\nabla_\mu \Pi + \Pi = - \zeta \theta ,
\ee
where again we emphasize that, above, $\zeta=\zeta(\varepsilon,n_B,\Pi)$ and $\tau_\Pi = \tau_\Pi(\varepsilon,n_B,\Pi)$. Summarizing, the exact equivalence uncovered in \cite{Gavassino:2023xkt} establishes that a system described by 
\begin{gather}
\nabla_\mu\left((\varepsilon+P) u^\mu u^\nu - P g^{\mu\nu}\right)= 0,\\
\nabla_\mu (n_B u^\mu) = 0, \\
u^\mu \nabla_\mu Y_e = \frac{\Gamma_e}{n_B},
\end{gather}
is exactly equivalent to the bulk-viscous system described by
\begin{gather}
\nabla_\mu\left[(\varepsilon+P) u^\mu u^\nu - (P_{eq}+\Pi) g^{\mu\nu}\right]= 0,\\
\nabla_\mu (n_B u^\mu) = 0, \\
\tau_\Pi u^\mu\nabla_\mu \Pi + \Pi = - \zeta \theta,
\end{gather}
where $\zeta$ and $\tau_\Pi$ are given by \eqref{definezetaandtaupi}. Therefore, any reactive mixture of this kind\footnote{This can be generalized to include the effects of other non-conserved currents, such as muons, see \cite{Gavassino:2023eoz}.} is inherently bulk viscous (no matter how far from equilibrium the system is). Thus, simulations that use reactive mixtures, such as \cite{Most:2022yhe}, must display bulk-viscous features. In this specific sense,  neutron star mergers are bulk-viscous systems.

It is possible to establish conditions for the far-from-equilibrium bulk viscosity theory to be causal (and strong hyperbolic) using the results of \cite{Bemfica:2019cop}. In fact, the only condition that needs to be fulfilled is
\begin{equation}
\label{causality}
    \left[ \frac{\zeta}{\tau_\Pi} + n_B \frac{\partial P}{\partial n_B} \bigg|_{\varepsilon, Y_e = Y_e^{\textrm{eq}}} \right] \frac{1}{\varepsilon+P} \leq 1 -  \frac{\partial P}{\partial \varepsilon} \bigg|_{n_B, Y_e = Y_e^{\textrm{eq}}}.
\end{equation}
When this condition holds, the far-from-equilibrium bulk-viscous theory considered here can be coupled to Einstein's equations and the initial-value problem of the full system of equations of motion in general relativity is locally well-posed \cite{Bemfica:2019cop}, which is crucial for numerical simulations. Additionally, we note that under the equivalence, \eqref{causality} corresponds to the natural statement that the speed of sound of the original reactive mixture should be non-negative and smaller than unity \cite{Gavassino:2023xkt}. 

Finally, we remark that the second law of thermodynamics is naturally implemented in this setup. Using the Gibbs relation and charge neutrality, we find
\begin{equation}
\label{Gibbs}
\begin{aligned}
    P+\varepsilon & = Ts + n_p \mu_p + n_n \mu_n + n_e \mu_e \\
    & = n_B\mu_n -n_e\delta\mu + Ts .
\end{aligned}
\end{equation}
We can write the above equation in differential form, 
\begin{equation}
    \begin{aligned}
    T u^\mu \nabla_\mu s = u^\mu \nabla_\mu \varepsilon - \mu_n u^\mu \nabla_\mu n_B + \delta \mu\, u^\mu \nabla_\mu n_e .
    \end{aligned}
\end{equation}
Using the entropy current 
\begin{equation}
    s^\mu = s u^\mu ,
\end{equation}
one can see that the entropy production rate is then
\begin{equation}
    \begin{aligned}
    \nabla_\mu s^\mu & = u^\mu \nabla_\mu s + s \theta \\
    & = \frac{1}{T} \left( u^\mu \nabla_\mu \varepsilon - \mu_n u^\mu \nabla_\mu n_B + \delta \mu\, u^\mu \nabla_\mu n_e \right) + \frac{1}{T} \left( P + \varepsilon - n_B\mu_n + \delta \mu\, n_e \right) \theta .
    \end{aligned}
\end{equation}
Employing Eqs.~\eqref{energy conserv}, \eqref{baryon}, and \eqref{e density}, the equation above can be simplified to
\begin{equation}
    \nabla_\mu s^\mu = \frac{\delta \mu \,\Gamma_e}{T} ,
\end{equation}
which should always be non-negative (see \cite{Gavassino:2023xkt} for an explicit example where this can be shown analytically).

\subsection{Cold dense matter equation of state}
In the rest of the paper, for simplicity, we will use the zero-temperature approximation for the EoS. This implies that we now only have two independent variables that we choose to be $\ep$ and $Y_e$. Consequently, the baryon density is a function of the independent variables, $n_B = n_B(\varepsilon,Y_e)$. To study dissipative effects, we define the bulk scalar to be the difference between the total pressure and the equilibrium pressure,
\be
\label{app:pi}
\Pi = P(\ep,Y_e) - P(\ep,Y_e^{\textrm{eq}}(\ep)),
\ee
where $Y_e^{\textrm{eq}}(\ep)$ is the equilibrium charge fraction. Since the equilibrium pressure is a function of $\ep$ only, the equilibrium charge fraction will also be a function of $\ep$. After the approximation \eqref{app:pi}, we obtain
\be
\bea
u^\mu\nabla_\mu \Pi &= \left.\frac{\pa P}{\pa \ep}\right|_{Y_e} u^\mu\nabla_\mu \ep + \left.\frac{\pa P}{\pa Y_e}\right|_{\ep} u^\mu\nabla_\mu Y_e - \left.\frac{\pa P}{\pa \ep}\right|_{Y_e^{\textrm{eq}}} u^\mu\nabla_\mu \ep - \left.\frac{\pa P}{\pa Y_e^{\textrm{eq}}}\right|_{\ep} \frac{\pa Y_e^{\textrm{eq}}}{\pa \ep} u^\mu\nabla_\mu \ep \\
&= - \left( \left.\frac{\pa P}{\pa \ep}\right|_{Y_e} - \left.\frac{\pa P}{\pa \ep}\right|_{Y_e^{\textrm{eq}}} - \left.\frac{\pa P}{\pa Y_e^{\textrm{eq}}}\right|_{\ep} \frac{\pa Y_e^{\textrm{eq}}}{\pa \ep} \right) (\ep + P(\ep,Y_e))\theta + \left.\frac{\pa P}{\pa Y_e}\right|_{\ep} \frac{\Ga_e}{n_B}.
\eea
\ee
If we set 
\be
F = F_\Pi \Pi = - \left.\frac{\pa P}{\pa Y_e}\right|_{\ep} \frac{\Ga_e}{n_B} ,
\ee
we obtain once more
\be
\tau_\Pi u^\mu\nabla_\mu \Pi + \Pi = - \zeta \theta ,
\ee
where the resummed transport coefficients are given by
\begin{gather}
\label{taupifinal}
\tau_\Pi = \frac{1}{F_\Pi} , \\
\label{zetafinal}
\zeta = \frac{1}{F_\Pi} \left( \left.\frac{\pa P}{\pa \ep}\right|_{Y_e} - \left.\frac{\pa P}{\pa \ep}\right|_{Y_e^{\textrm{eq}}} - \left.\frac{\pa P}{\pa Y_e^{\textrm{eq}}}\right|_{\ep} \frac{\pa Y_e^{\textrm{eq}}}{\pa \ep} \right) (\ep + P(\ep,Y_e)).
\end{gather}
The expressions above are used in this work to determine the transport coefficients, starting from a given equation of state. Note that although no explicit temperature effects exist in the EoSs used in this work, the transport coefficients will vary with the temperature. This happens because $\Gamma_e$ depends on $T$; see Appendix \ref{A: rate}. This procedure can only provide a rough estimate of the actual temperature dependence of the resummed transport coefficients, which requires the inclusion of $T$ effects in the EoS. Furthermore, we note that a more complete calculation of the rates, going beyond the Fermi surface approximation \cite{Alford:2018lhf}, would be needed to make the results more realistic. This is left for future work.

\section{Equilibrium properties of neutron star matter}
\label{EOS}

\subsection{Relativistic mean field model}

The EoS of strongly interacting matter at very high densities is not amenable to first principle QCD calculations \cite{Philipsen:2012nu}. Thus, in practice, the dense matter equation of state is determined using effective models \cite{MUSES:2023hyz}. In this paper, we employ a relativistic mean field (RMF) model \cite{norman1997compact, Chen:2014sca} with nucleons,  $\rho$, $\omega$, and $\sigma$ mesons, and electrons. This provides a simple yet convenient tool to describe the properties of $npe$ matter relevant to this work (note that we only consider the EoS for cold nuclear matter, $T\to 0$, in this work). 

In our model, the Lagrangian is of the form
\be
\cL = \cL_N + \cL_M + \cL_l.
\ee
The nucleon contribution to the Lagrangian is
\be
\cL_N = \bps[i\ga_\mu\pa^\mu - m_B + \ga^0 \mu_B + \ga_0 \frac{\tau_3}{2}\mu_I - g_\om \om^\mu \ga_\mu - g_\rho \ga^\mu \vec{\rho}_\mu\cdot\frac{\vec{\tau}}{2} + g_\si \si]\psi,
\ee
while the meson contribution is
\be
\begin{aligned}
    \cL_M &= \frac{1}{2}\pa_\mu\si \pa^\mu \si - \frac{1}{2}m_\si^2\si^2 -\frac{b}{3}m_B(g_\si\si)^3 - \frac{c}{4}(g_\si \si)^4 \\
&\;\;\;\; + \frac{1}{2}m_\om^2\om^\mu\om_\mu - \frac{1}{4}\om^{\mu\nu}\om_{\mu\nu} + \frac{1}{2}m_\rho^2\vec{\rho}^\mu\vec{\rho}_\mu - \frac{1}{4}\vec{\rho}_{\mu\nu}\vec{\rho}^{\mu\nu} + G_{\om\rho} \om^\mu\om_\mu \vec{\rho}^\mu\vec{\rho}_\mu , 
\end{aligned}
\ee
where
\begin{gather}
\om_{\mu\nu} = \pa_\mu\om_\nu - \pa_\nu\om_\mu, \\
\vec{\rho}_{\mu\nu} = \pa_\mu\vec{\rho}_\nu - \pa_\nu\vec{\rho}_\mu - g_\rho(\vec{\rho}_\mu \times \vec{\rho}_\nu).
\end{gather}
We only have electrons as leptons in the system, which are introduced using the Dirac Lagrangian 
\be
\cL_l = \bps_e(i\ga^\mu\pa_\mu - m_e)\psi_e .
\ee
 For details on the calculation of the corresponding EoS, see Appendix~\ref{A: RMF EoS}.
 
With this general form of the Lagrangian, we can find the coupling constants by comparing the resulting EoS to different observational constraints. To obtain the EoSs used in this paper, we vary $J$ and $L$ while holding all the other nuclear properties constant. 
We denote by J1 to J6 the EoSs with varying symmetry energy. Their nuclear properties are listed in Table~\ref{J EoS nuclear}, and the corresponding values of the parameters are listed in Table~\ref{J EoS}. Similarly, we denote by L1 to L5 the EoSs with different symmetry slopes. Their nuclear properties are listed in Table~\ref{L EoS nuclear}, and the corresponding values of the parameters are listed in Table~\ref{L EoS}. 

\begin{table}[h]
\begin{center}
\begin{tabular}{c c c c c c c} 
 \hline
 Quantity & J1 & J2 & J3 & J4 & J5 & J6 \\ 
 \hline
 $E_b$ [MeV] & -15.677 & -15.677 & -15.677 & -15.677 & -15.677 & -15.677 \\ 
 $K$ [MeV] & 275 & 275 & 275 & 275 & 275 & 275 \\
 $J$ [MeV] & 29 & 30 & 31 & 32 & 33 & 34 \\
 $L$ [MeV] & 51.03 & 51.03 & 51.03 & 51.03 & 51.03 & 51.03 \\
 \hline
\end{tabular}
\caption{\label{J EoS nuclear} Nuclear properties of our EoSs J1 to J6. $E_b$ is the binding energy. $K$ is the nuclear compressibility. $J$ is the nuclear symmetry energy. $L$ is the slope of the nuclear symmetry energy. See \cite{MUSES:2023hyz} for details on constraints on nuclear-matter properties.}
\end{center}
\end{table}

\begin{table}[h]
\begin{center}
\begin{tabular}{c c c c c c c} 
 \hline
 Quantity & J1 & J2 & J3 & J4 & J5 & J6 \\ 
 \hline
 $g_\si$ & 7.55 & 7.55 & 7.55 & 7.55 & 7.55 & 7.55 \\ 
 $g_\om$ & 8.36 & 8.36 & 8.36 & 8.36 & 8.36 & 8.36 \\
 $g_\rho$ & 9.58 & 9.98 & 10.37 & 10.74 & 11.11 & 11.46 \\
 $b$ & 0.0048 & 0.0048 & 0.0048 & 0.0048 & 0.0048 & 0.0048 \\
 $c$ & 0.0178 & 0.0178 & 0.0178 & 0.0178 & 0.0178 & 0.0178 \\
 $m_\si$ [MeV] & 500 & 500 & 500 & 500 & 500 & 500 \\
 $m_\om$ [MeV] & 782.65 & 782.65 & 782.65 & 782.65 & 782.65 & 782.65 \\
 $m_\rho$ [MeV] & 775.26 & 775.26 & 775.26 & 775.26 & 775.26 & 775.26 \\
 $m_B$ [MeV] & 938.9 & 938.9 & 938.9 & 938.9 & 938.9 & 938.9 \\
 $m_e$ [MeV] & 0.511 & 0.511 & 0.511 & 0.511 & 0.511 & 0.511 \\
 $G_{\om\rho}$ & 534.62 & 566.68 & 595.99 & 622.88 & 647.65 & 670.53 \\
 \hline
\end{tabular}
\caption{\label{J EoS}Parameter values of our EoSs J1 to J6. We approximate the vacuum mass of the proton and the neutron to be the same.}
\end{center}
\end{table}

\begin{table}[h]
\begin{center}
\begin{tabular}{c c c c c c} 
 \hline
 Quantity & L1 & L2 & L3 & L4 & L5 \\ 
 \hline
 $E_b$ [MeV] & -15.677 & -15.677 & -15.677 & -15.677 & -15.677 \\ 
 $K$ [MeV] & 275 & 275 & 275 & 275 & 275 \\
 $J$ [MeV] & 30.7 & 30.7 & 30.7 & 30.7 & 30.7 \\
 $L$ [MeV] & 35 & 45 & 55 & 65 & 75 \\
 \hline
\end{tabular}
\caption{\label{L EoS nuclear} Nuclear properties of our EoSs L1 to L5. $E_b$ is the binding energy. $K$ is the nuclear compressibility. $J$ is the nuclear symmetry energy. $L$ is the slope of the nuclear symmetry energy.  See \cite{MUSES:2023hyz} for details on constraints on nuclear-matter properties. }
\end{center}
\end{table}

\begin{table}[h]
\begin{center}
\begin{tabular}{c c c c c c} 
 \hline
 Quantity & L1 & L2 & L3 & L4 & L5 \\ 
 \hline
 $g_\si$ & 7.55 & 7.55 & 7.55 & 7.55 & 7.55 \\ 
 $g_\om$ & 8.36 & 8.36 & 8.36 & 8.36 & 8.36 \\
 $g_\rho$ & 11.70 & 10.73 & 9.97 & 9.35 & 8.83 \\
 $b$ & 0.0048 & 0.0048 & 0.0048 & 0.0048 & 0.0048 \\
 $c$ & 0.0178 & 0.0178 & 0.0178 & 0.0178 & 0.01788 \\
 $m_\si$ [MeV] & 500 & 500 & 500 & 500 & 500 \\
 $m_\om$ [MeV] & 782.65 & 782.65 & 782.65 & 782.65 & 782.655 \\
 $m_\rho$ [MeV] & 775.26 & 775.26 & 775.26 & 775.26 & 775.26 \\
 $m_B$ [MeV] & 938.9 & 938.9 & 938.9 & 938.9 & 938.9 \\
 $m_e$ [MeV] & 0.511 & 0.511 & 0.511 & 0.511 & 0.511 \\
 $G_{\om\rho}$ & 1143.21 & 763.30 & 487.50 & 278.16 & 113.86 \\
 \hline
\end{tabular}
\caption{\label{L EoS}Parameter values of our EoSs L1 to L5. We approximate the vacuum mass of the proton and the neutron to be the same.}
\end{center}
\end{table}

\begin{figure}[ht!]
    \centering
    \includegraphics[width=8cm]{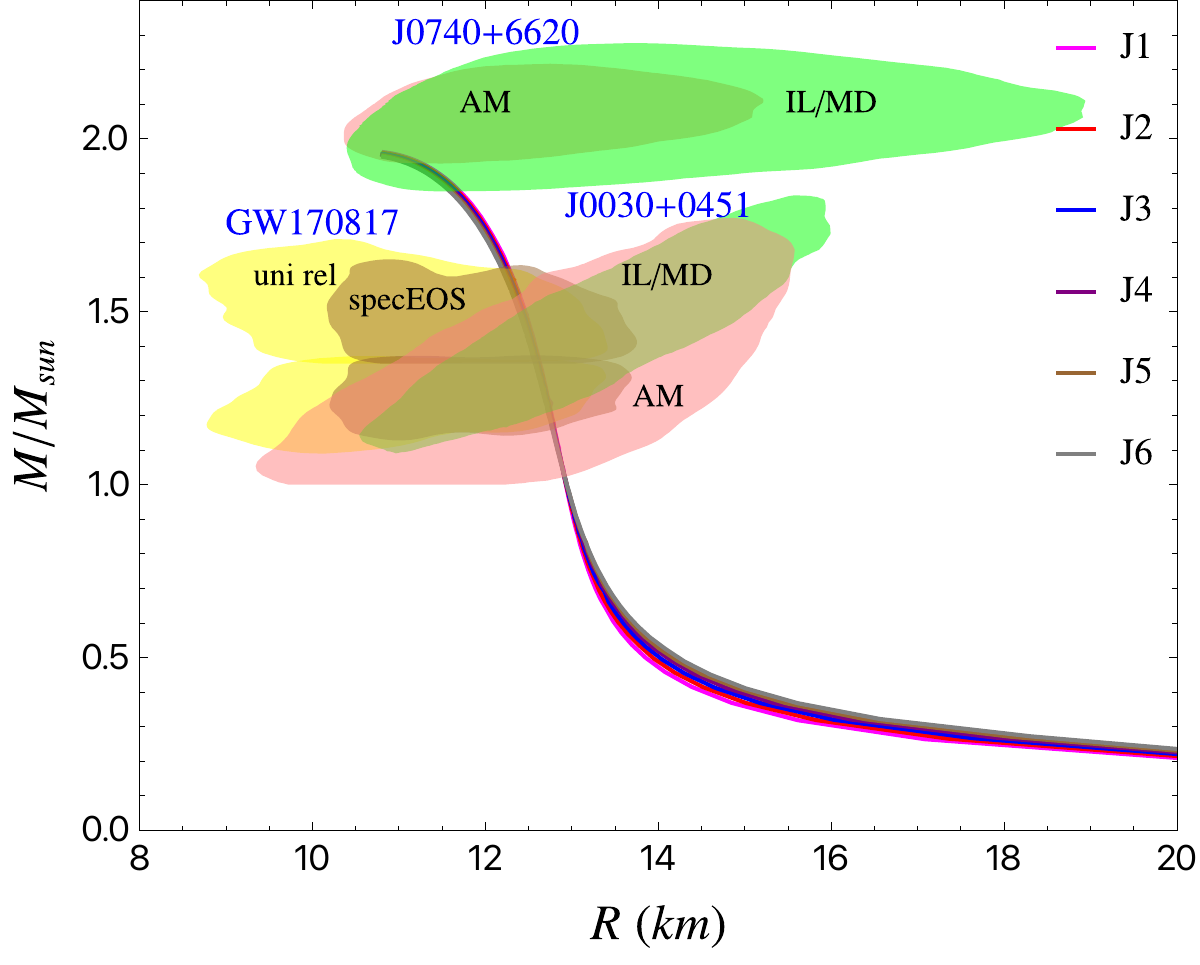} \hfill
    \includegraphics[width=8cm]{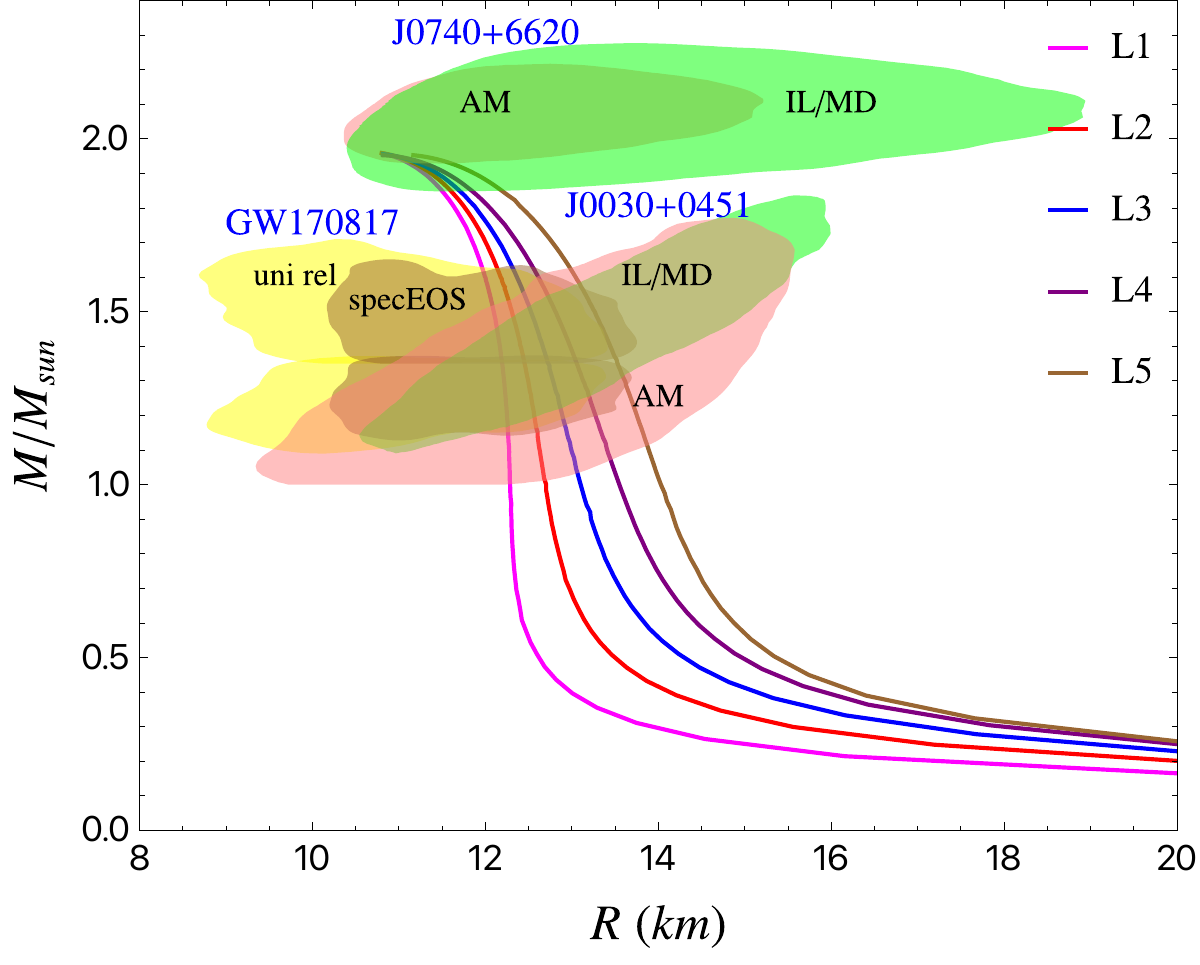} 
    \caption{Mass-radius relation of our EoS. Left: EoS with varying symmetry energy. Right: EoS with varying symmetry slope. Observational constraints on the neutron star mass-radius plane from LIGO/Virgo~\cite{LIGOScientific:2017vwq, LIGOScientific:2018hze, LIGOScientific:2018cki} and NICER~\cite{Miller:2019cac, Riley:2019yda, Miller:2021qha, Riley:2021pdl} results. The yellow and brown regions correspond to 90$\%$ confidence regions obtained from the GW170817 event by universal relations and spectral EoS approaches, respectively. The green and red regions correspond to 90$\%$ confidence regions obtained from NICER data on PSR J0030+0451 and J0740+6620. The green regions are obtained using the Illinois-Maryland analysis, and the red regions are obtained using the Amsterdam analysis.}.
    \label{TOV}
\end{figure}

\subsection{Effects of the symmetry energy and slope on mass-radius constraints}

For all of the equations of state, we calculated the mass-radius relation using the Tolman–Oppenheimer–Volkoff equation \cite{norman1997compact}. Our EoS only describes matter above one nuclear saturation density $n_{\textrm{sat}}$, so we need to connect our high-density EoS to a crust EoS. We take the sly EoS \cite{Douchin:2001sv} for matter below the neutron drip line. For the region between the neutron drip line and $n_{\textrm{sat}}$, we created a monotonically increasing polynomial for $p(\ep)$ such that the polynomial is differentiable at the intersection points with sly and our EoS. 

We plot the mass-radius relation along with all the observational constraints ~\cite{Miller:2019cac, Riley:2019yda, Miller:2021qha, Riley:2021pdl, LIGOScientific:2017vwq, LIGOScientific:2018hze, LIGOScientific:2018cki, Tan:2021ahl}. The clouds are the observational constraints, with different colors representing different analysis methods used to obtain the constraints. All of our EoSs pass these observational constraints. The left-hand side of Fig.~\ref{TOV} shows the results for our EoSs with varying $J$. One can see that $J$ does not significantly affect the mass-radius relation. The right-hand side of Fig.~\ref{TOV} shows the results for our EoSs with varying $L$. In general, larger values of $L$ make the star's radius larger.

\section{Results for Israel-Stewart transport coefficients}
\label{Results}

We now present results for the Israel-Stewart bulk-viscous transport coefficients for different equations of state. As discussed in Sec.~\ref{IS derivation}, we can obtain not only these coefficients in beta equilibrium, that is, for $Y_e-Y_e^{\textrm{eq}} = 0$, but also out of equilibrium, in which case these coefficients depend on the value of $Y_e-Y_e^{\textrm{eq}}$. 
We start by presenting our results for the transport coefficients at $Y_e=Y_e^{\textrm{eq}}$ (i.e.,  $\Pi=0$) in Sec.~\ref{sec:results_linear} before showing our full out-of-equilibrium results in Sec.~\ref{sec:outofeq}. 

\subsection{Bulk viscosity near beta equilibrium}
\label{sec:results_linear}

\begin{figure}[ht!]
    \centering
    \includegraphics[width=8cm]{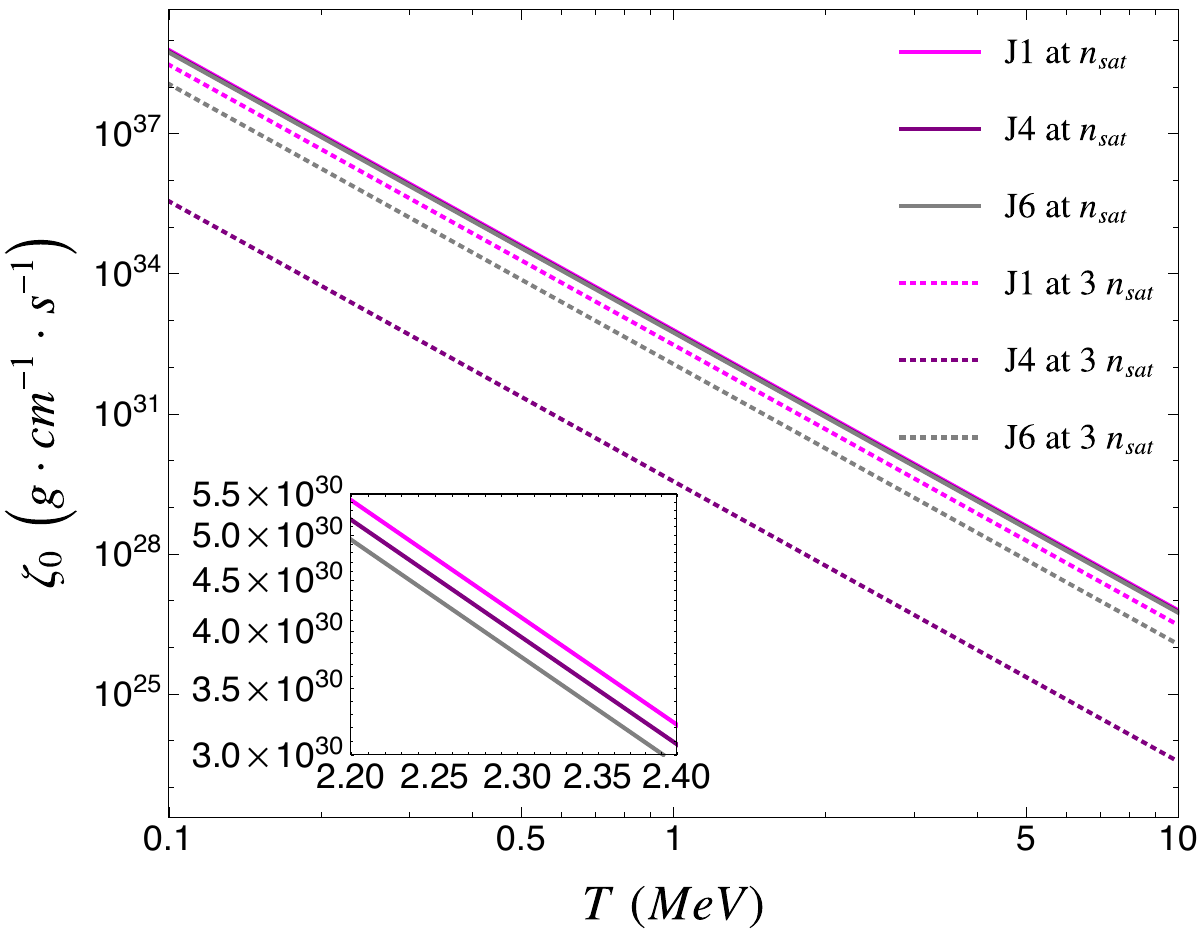} \hfill
    \includegraphics[width=8cm]{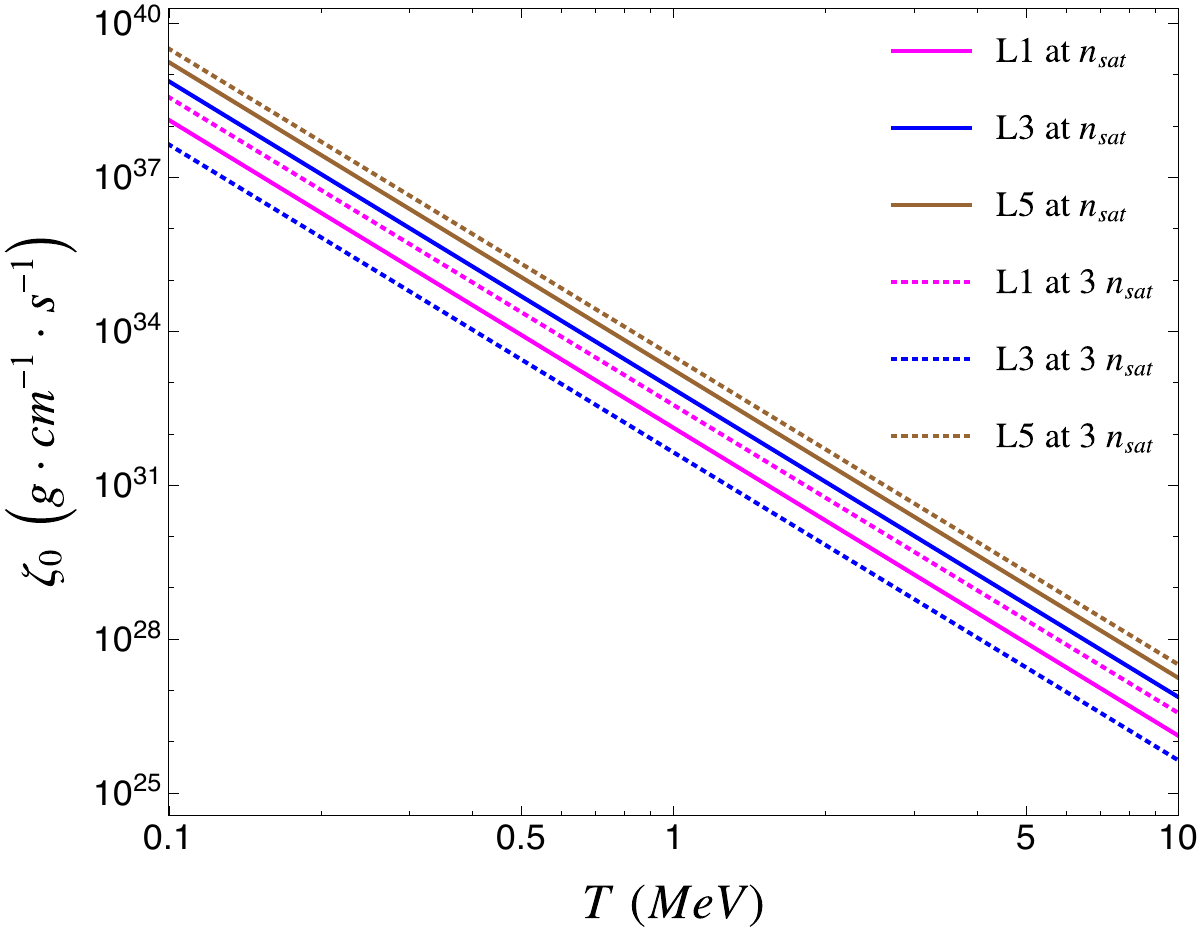} 
    \caption{Temperature dependence of the bulk viscosity $\zeta_0$ determined using parameters in beta equilibrium. Left: Results for different EoSs corresponding to several values of the symmetry energy, with the inset showing the difference between J1 and J6. Right: Similar plot for EoSs with different symmetry slopes. 
    }
    \label{bulk0}
\end{figure}

\begin{figure}[ht!]
    \centering
    \includegraphics[width=8cm]{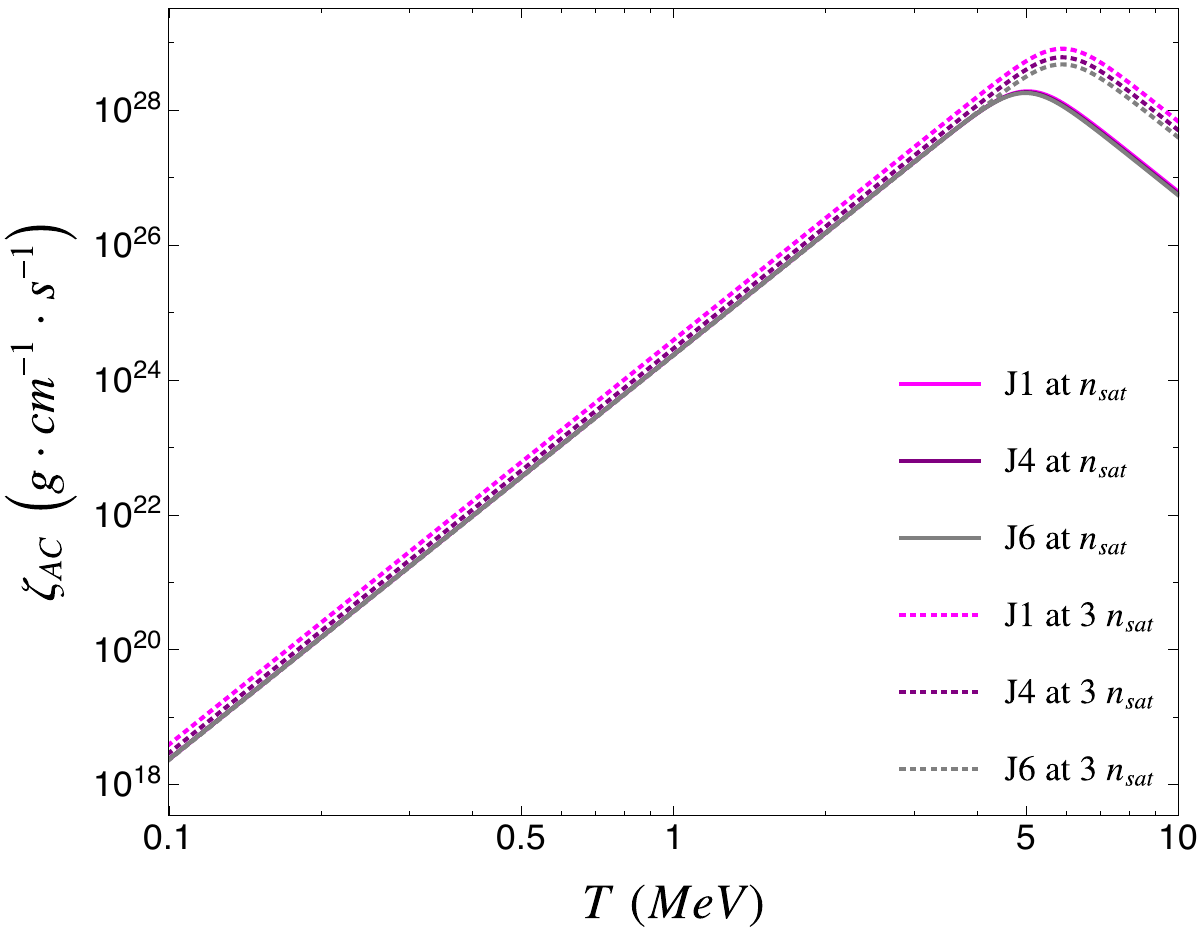} \hfill
    \includegraphics[width=8cm]{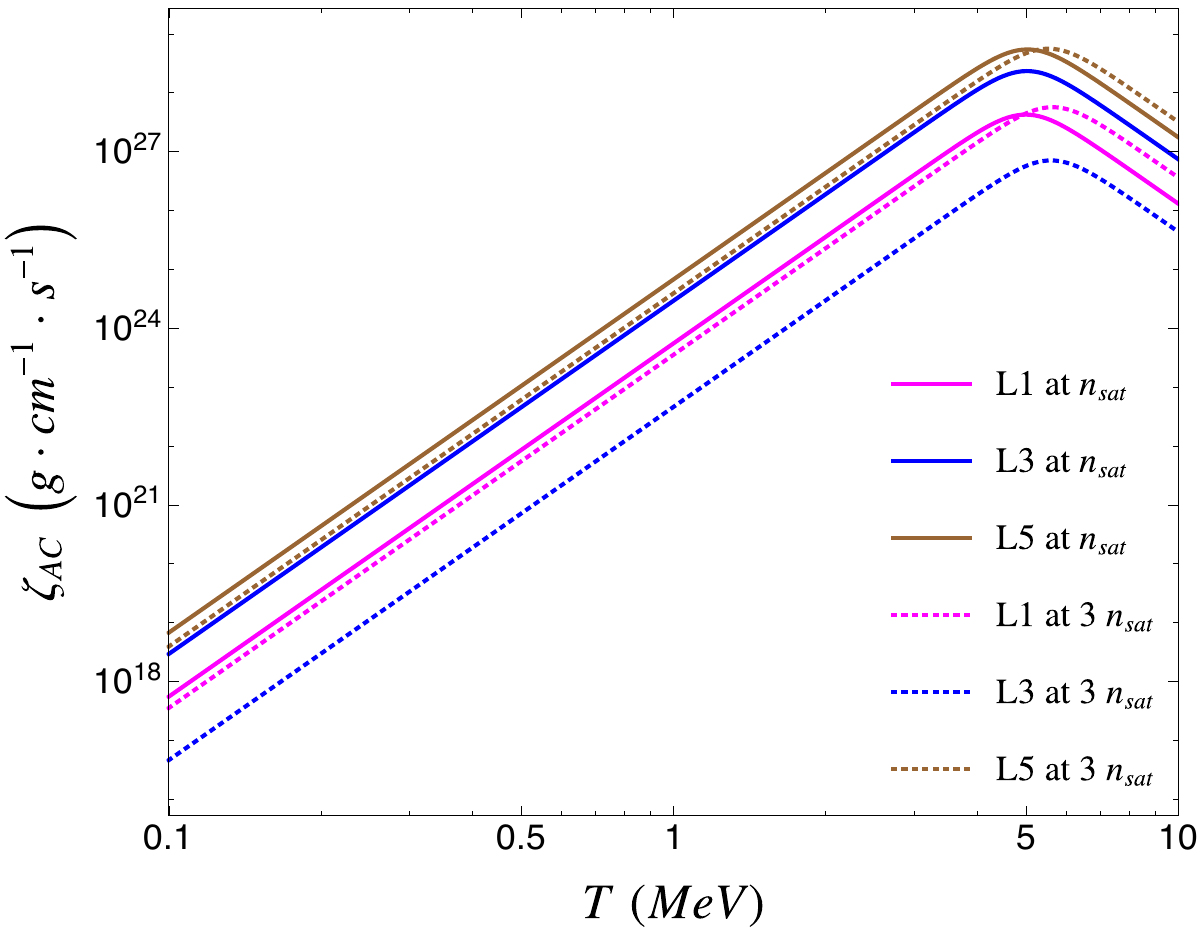} 
    \caption{Temperature dependence of the AC bulk viscosity determined using parameters in beta equilibrium. Left: Results for different EoSs obtained by varying the symmetry energy. Right: Results for different EoSs obtained by varying the symmetry slope. The oscillation frequency is fixed at $1\textrm{ kHz}$.}
    \label{AC bulk}
\end{figure}

In this section, we present the bulk viscosity results under various conditions. In this work, $\zeta$ is the most general (out of equilibrium) bulk viscosity coefficient that depends on all chosen variables, i.e., $\zeta = \zeta(\ep, n_B, \Pi)$. At zero temperature $\zeta = \zeta(\ep, \Pi)$. Additionally,  $\zeta_0$ denotes the bulk viscosity transport coefficient computed with parameters determined in beta equilibrium --- i.e., at $\Pi = 0$. Equivalently, we may also write $\zeta_0 = \zeta(\ep,\Pi = 0)$. In the context of linear response, $\zeta_0$ enters as a parameter in the frequency-dependent AC bulk viscosity $\zeta_{AC} = \zeta_0 (1+\omega^2 \tau_{\Pi,0}^2)^{-1}$ (obtained from the analysis of periodic density oscillations \cite{Sawyer:1989dp, Alford:2018lhf}), where $\om$ is the frequency and $\tau_{\Pi,0}=\tau_\Pi(\ep,\Pi=0)$ is the relaxation time computed at beta equilibrium, see the derivation in Appendix \ref{A: AC IS}.

Figure~\ref{bulk0} shows $\zeta_0$. Since we consider cold EoS,  temperature effects only enter through the calculation of the Urca rates, which is shown in Appendix \ref{A: rate}. 
Without any loss of generality, we have plotted three EoSs in each figure by varying $J$ and $L$ separately. The left-hand side of Fig.~\ref{bulk0} shows $\zeta_0$ for our EoSs with varying $J$. The labels $J1$ to $J6$ denote EoSs with increasing $J$, $J1$ being the EoS with the lowest $J$, and $J6$ with the highest. For the EoS at $n_{\textrm{sat}}$, the bulk viscosity $\zeta_0$ decreases with increasing $J$ across all temperatures. The relation between $J$ and $\zeta_0$, however, becomes non-monotonic as the density increases. In fact, as  $J$ increases, $\zeta_0$ reaches a minimum and then increases again. The right-hand side of Fig.~\ref{bulk0} shows $\zeta_0$ for our EoS with varying $L$, L1 being the EoS with the lowest value of $L$, and $L5$ with the highest value of $L$. For the EoS at $n_{\textrm{sat}}$, $\zeta_0$ increases with $L$ across all temperatures. However, at higher densities such as  $3n_{\textrm{sat}}$, as $L$ increases, $\zeta_0$ decreases and then increases again. Overall, we see that varying $J$ and $L$ strongly affect $\zeta_0$, but the effects depend on the density.

The bulk viscosity $\zeta_0$ for our EoS is compatible with those calculated from other EoSs, as shown in Sec.~\ref{other EoS}. As a consistency check of our analysis, we verified that the AC bulk viscosity $\zeta_{AC}$ is compatible with the one found in previous works \cite{Alford:2020lla}, as shown in Fig.~\ref{AC bulk}. However, we remark that it is not guaranteed that a linear approximation in deviations from beta equilibrium is applicable to all equations of state. In fact, we show later in this section that there is a regime where deviations from equilibrium do not behave linearly for realistic equations of state. This feature cannot be described using previous linear response approaches, but it can be properly captured using the resummed transport coefficients investigated in this work.

\begin{figure}[ht!]
    \centering
    \includegraphics[width=8cm]{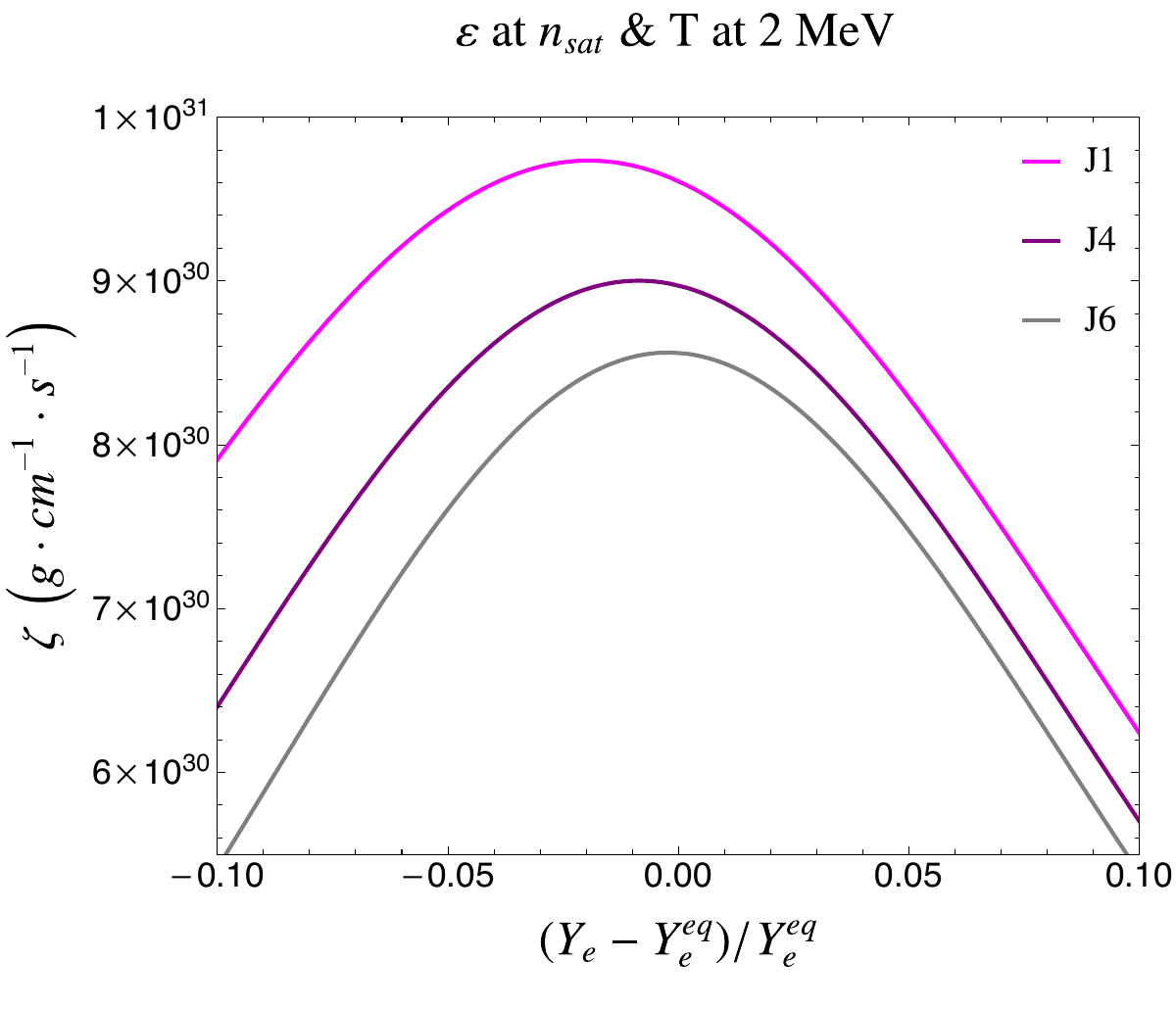} \hfill
    \includegraphics[width=8cm]{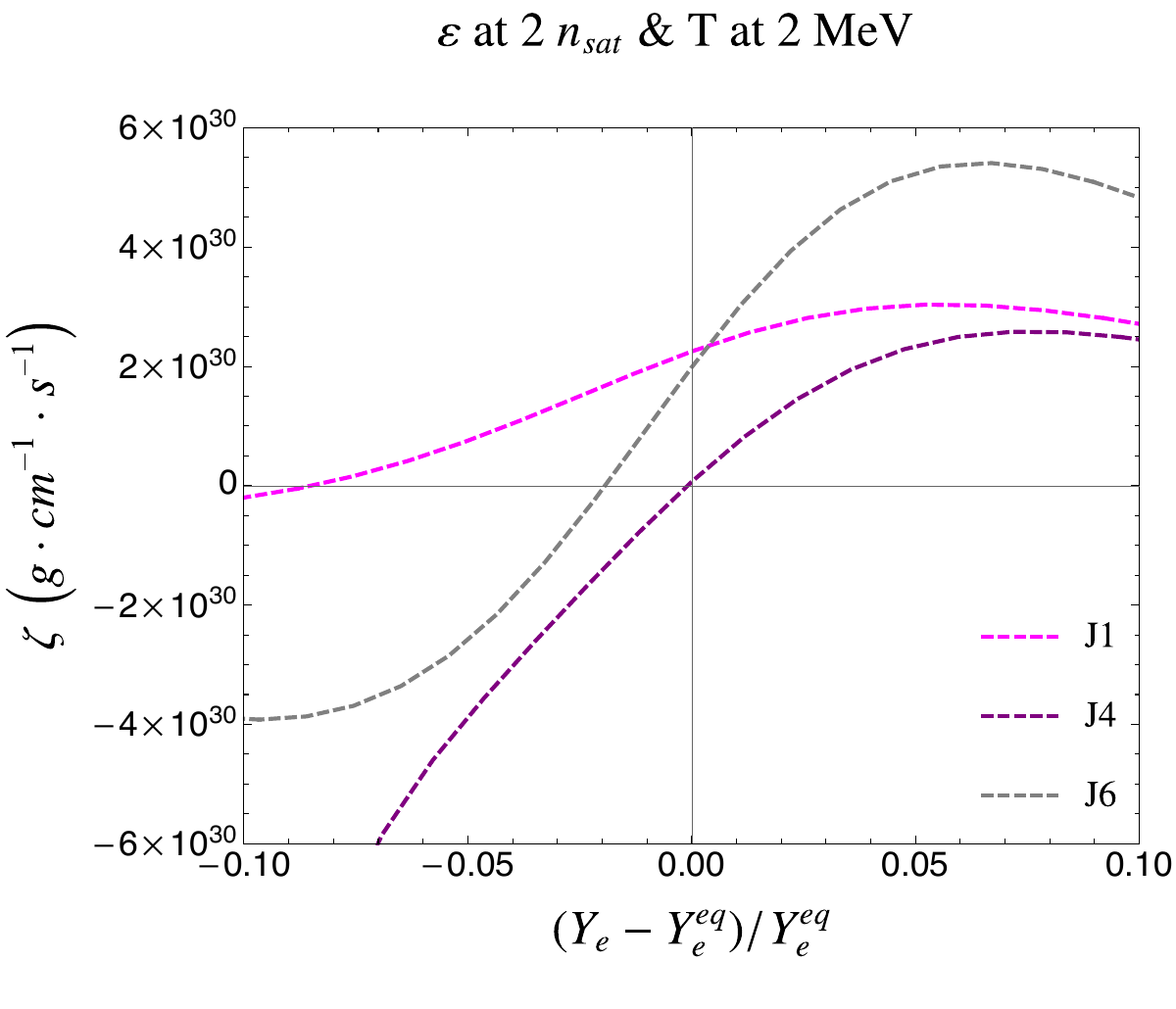} 
    \caption{Resummed $\zeta$ as a function of the deviation from the equilibrium charge fraction, at temperature $T=2$ MeV, for several symmetry energies. Left: Symmetry energy dependence of $\zeta$ evaluated at the energy density corresponding to nuclear saturation density. Right: Same quantity, computed at twice the nuclear saturation density. 
    }
    \label{J bulk}
\end{figure}

\begin{figure}[ht!]
    \centering
    \includegraphics[width=8cm]{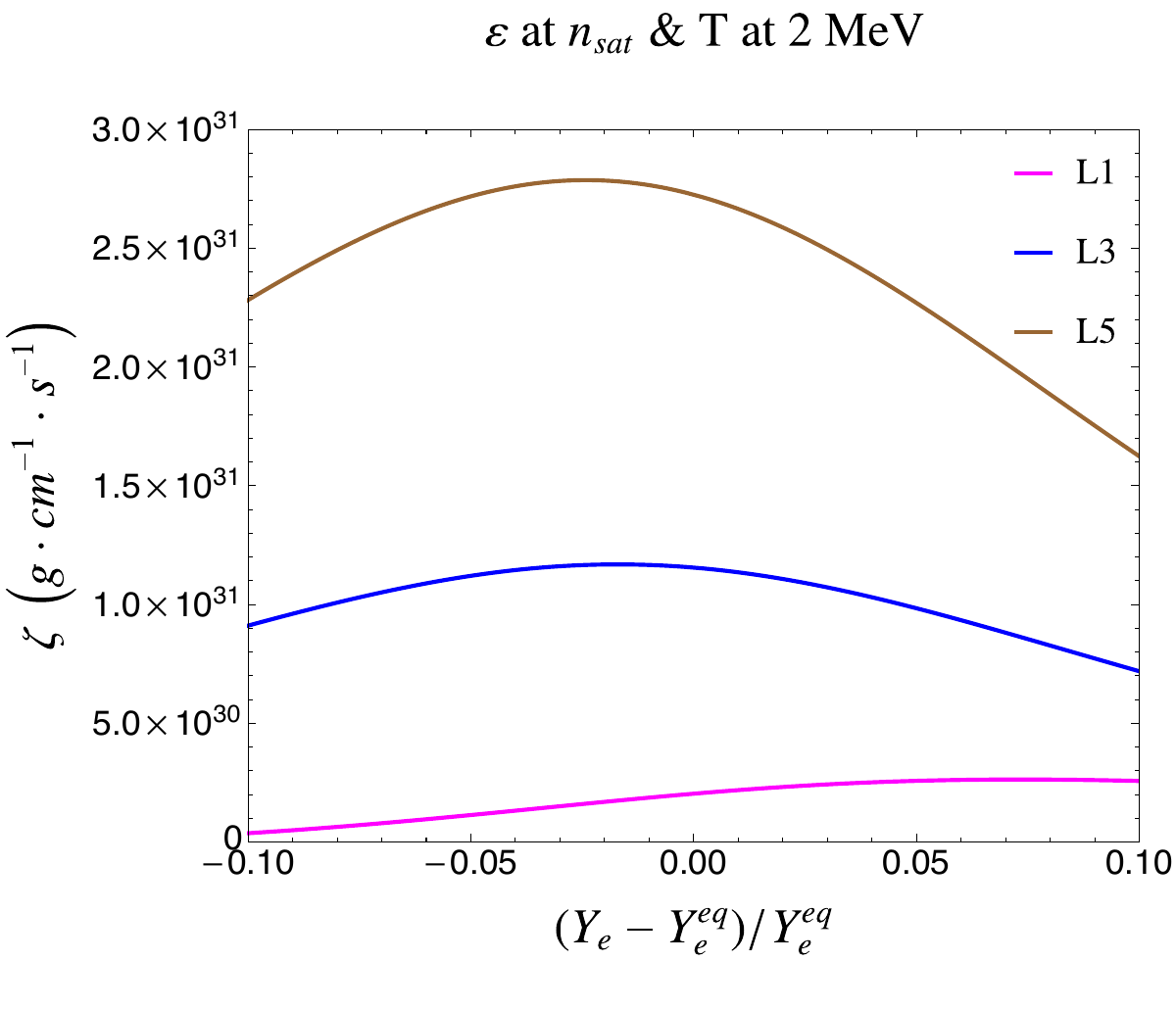} \hfill
    \includegraphics[width=8cm]{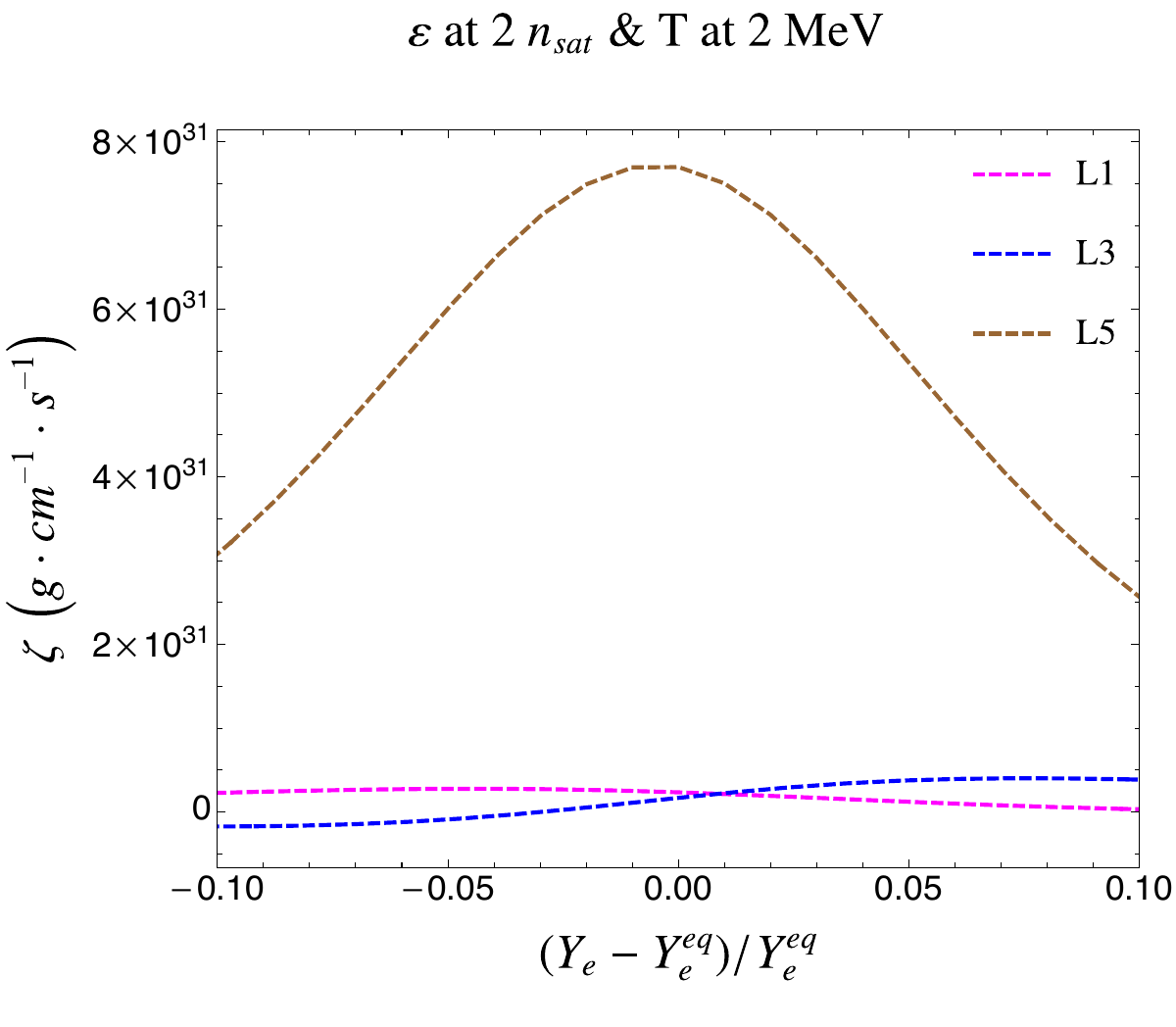} 
    \caption{Resummed $\zeta$ as a function of the deviation from the equilibrium charge fraction, for several symmetry slopes at temperature $T=2$ MeV. Left: Symmetry slope dependence of the $\zeta$ evaluated at the
energy density corresponding to nuclear saturation density. Right: Same quantity, now computed at twice the nuclear saturation density.}
    \label{L bulk}
\end{figure}

\subsection{Far-from-equilibrium transport coefficients}
\label{sec:outofeq}
\subsubsection{Energy-density dependence of transport coefficients}

\begin{figure}
    \centering
    \includegraphics[width=8cm]{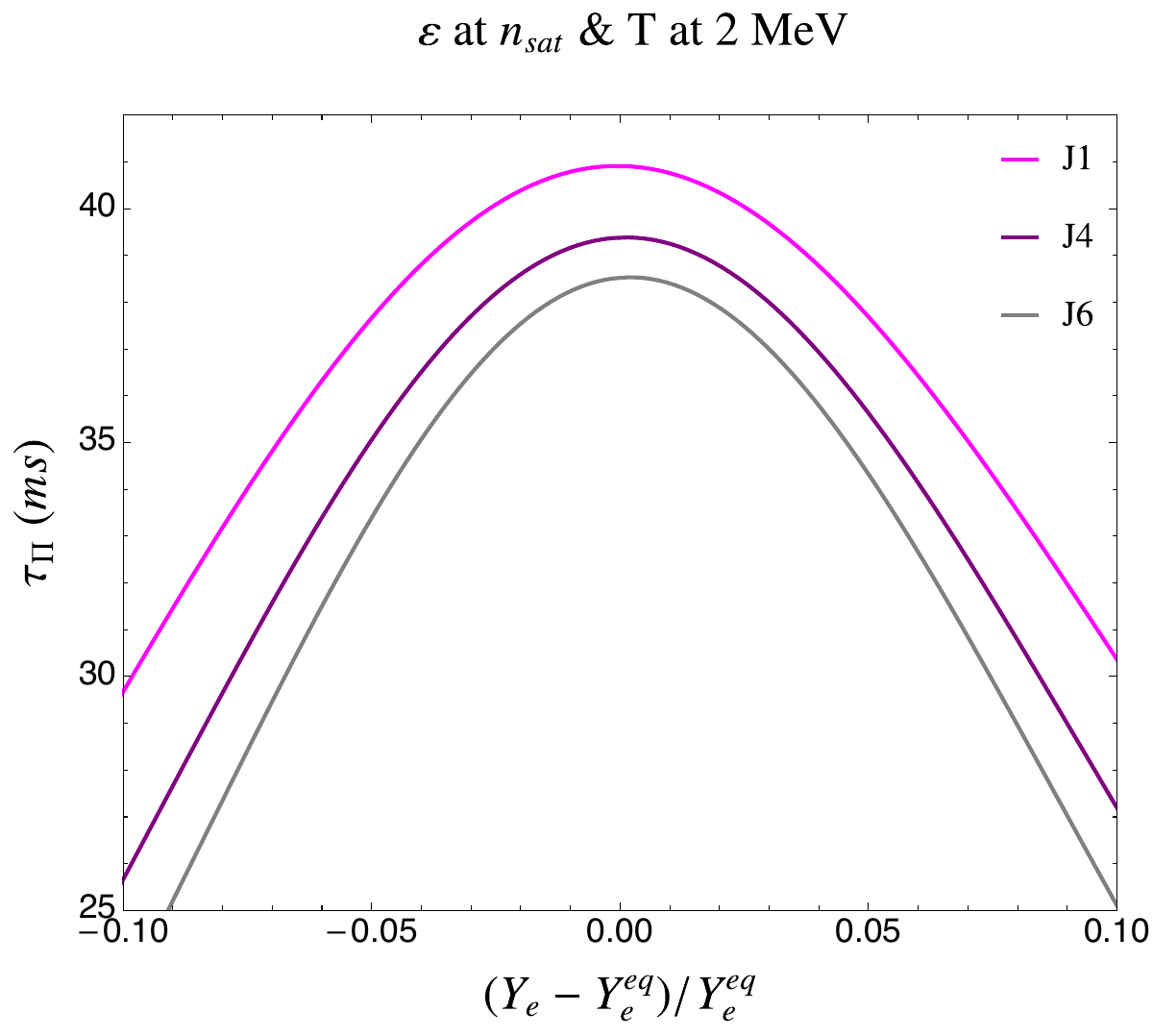} \hfill
    \includegraphics[width=8cm]{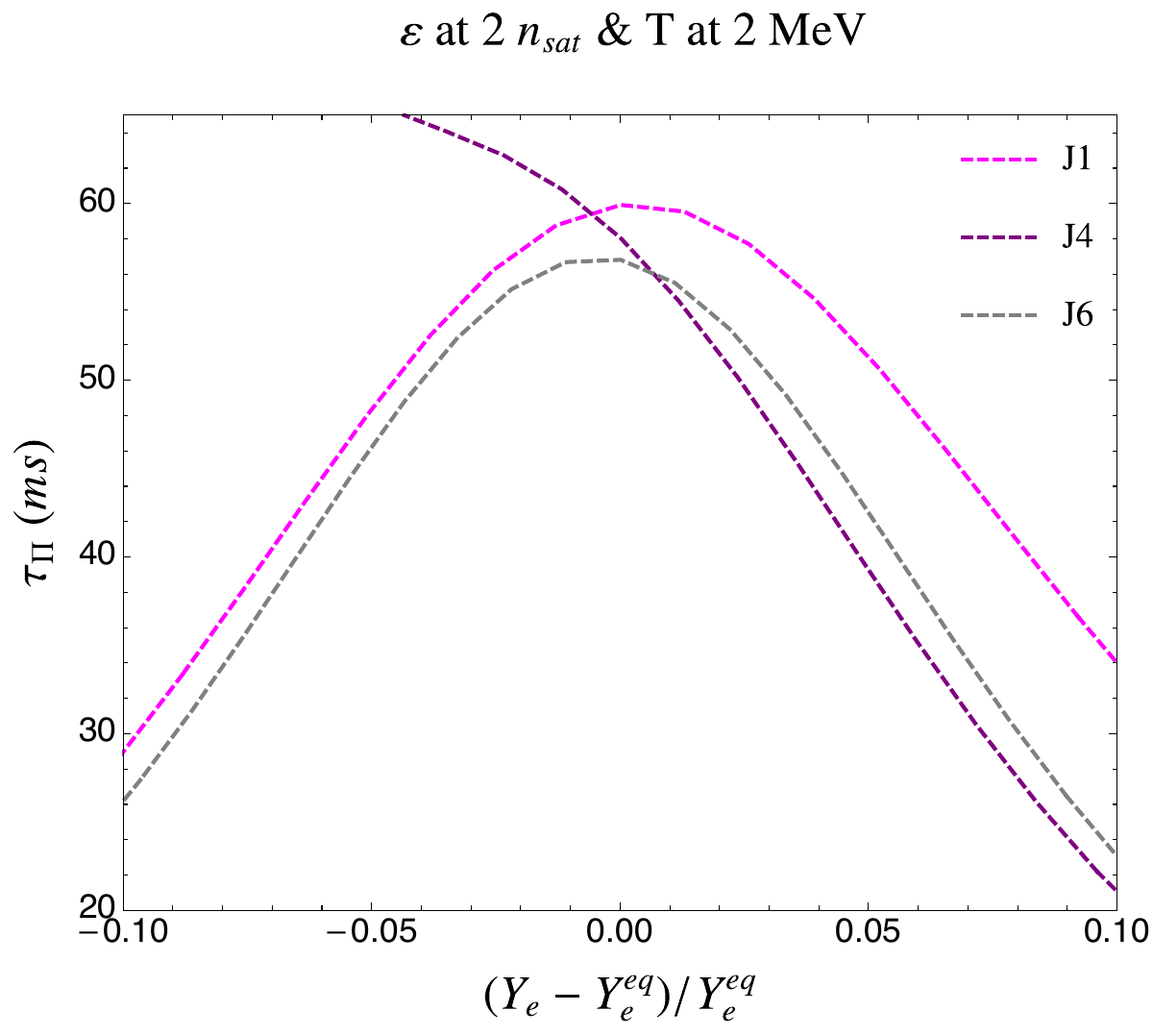} 
    \caption{Resummed $\tau_\Pi$ as a function of the deviation from the equilibrium charge fraction, for several symmetry energies, at temperature $T=2$ MeV. Left: Symmetry energy dependence of the resummed $\tau_\Pi$ for the energy density corresponding to nuclear saturation density. Right: The same quantity is evaluated at twice the nuclear saturation density.}
    \label{J tau}
\end{figure}

Variations in the equation of state become even more relevant when determining the transport coefficients away from beta equilibrium. For the sake of definiteness, we set the temperature $T = 2$ MeV (small variations around this temperature give similar results). Figures~\ref{J bulk} and \ref{L bulk} show the full out-of-equilibrium resummed $\zeta$, computed using \eqref{zetafinal}, against the deviation from the equilibrium charge fraction $Y_e^{\textrm{eq}}$, for two different energy densities. As shown in Fig.~\ref{J bulk}, the resummed $\zeta$ still decreases with increasing $J$ at saturation density (mirroring the behavior found in $\zeta_0$ before). However, one can see that at twice saturation density $\zeta$ displays non-monotonic behavior with respect to variations in $J$. Furthermore, one can see that even negative values of $\zeta$ are found (even though $\zeta_0$ is positive). This should not be a cause for concern, as this does not imply that the system has an instability or displays acausal behavior. In fact, the negative values only appear away from beta equilibrium, where the interpretation of $\zeta$ in the standard Navier-Stokes \cite{Rezzolla_Zanotti_book} sense is not valid. It is $\zeta_0$ that matters for the stability of the equilibrium state, and this quantity is always non-negative in our calculations. Finally, as discussed later in this section, causality is never violated for all EoSs we consider. Fig.~\ref{L bulk} shows that $\zeta$ increases with $L$, though one can see that the effect depends on the energy density. 

\begin{figure}[h]
    \centering
    \includegraphics[width=8cm]{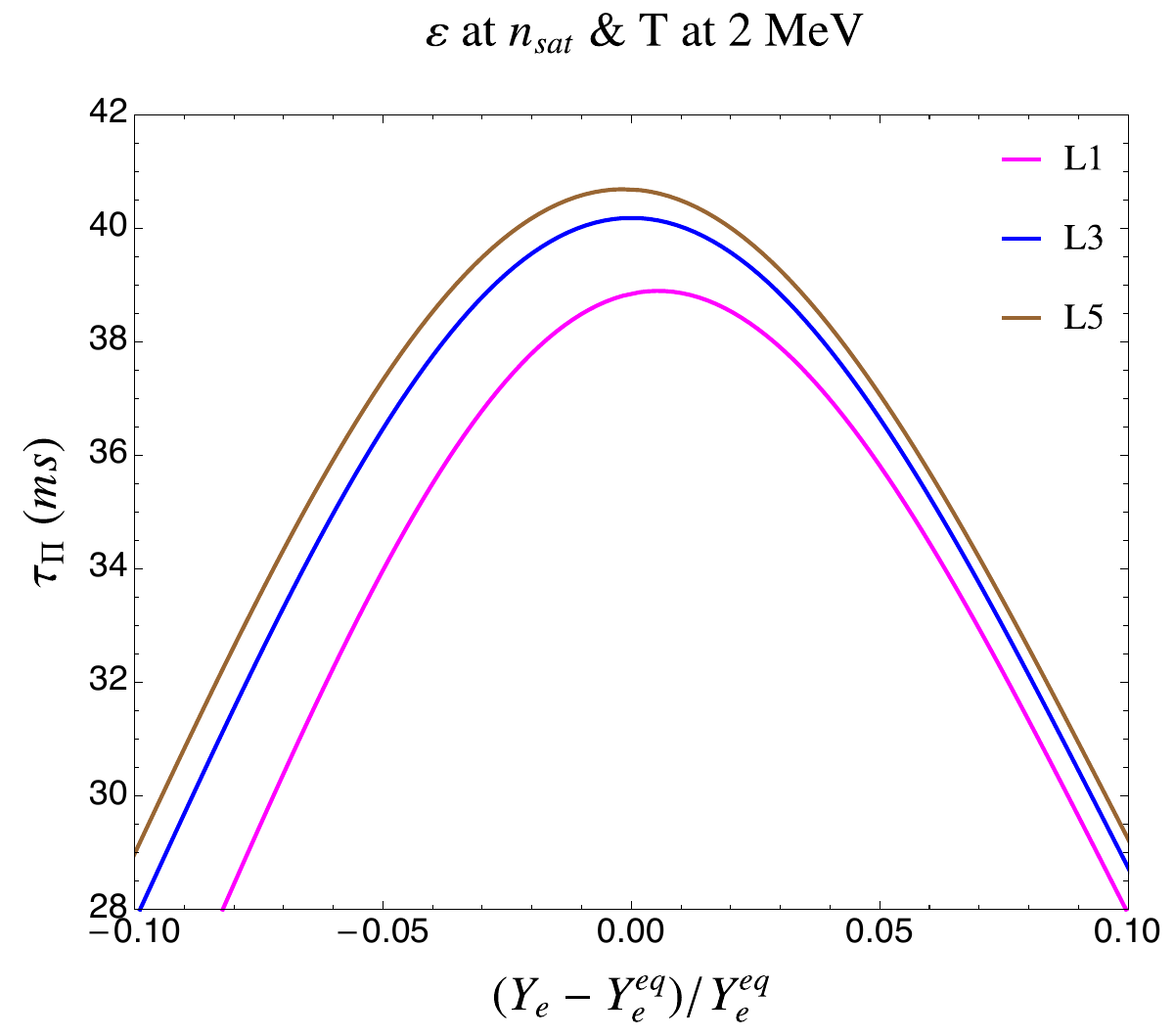} \hfill
    \includegraphics[width=8cm]{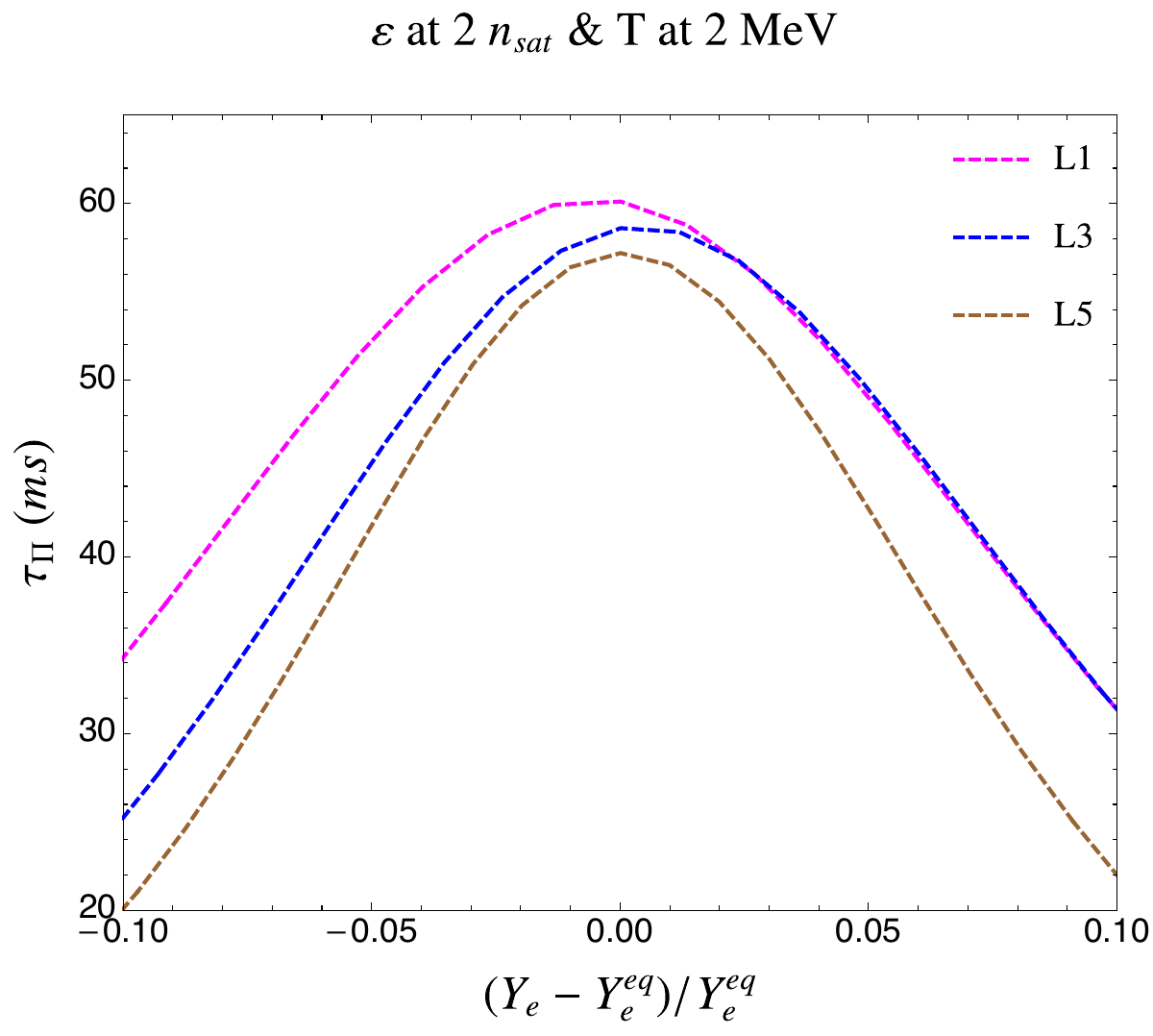} 
    \caption{Resummed $\tau_\Pi$ as a function of the deviation from the equilibrium charge fraction, for several symmetry slopes, at temperature $T=2$ Meb. Left: Resummed $\tau_\Pi$ for the energy density corresponding to nuclear saturation density. Right: The same quantity is evaluated at twice the nuclear saturation density.}
    \label{L tau}
\end{figure}

Figures~\ref{J tau} and \ref{L tau} show the resummed relaxation time $\tau_\Pi$, computed using \eqref{taupifinal}, for the same EoSs. The resummed relaxation time increases with density and also exhibits a nonlinear trend with respect to deviations from the equilibrium charge fraction. We note that the relaxation time is on the order of tens of milliseconds, so its effects will only become relevant later in the merger, potentially affecting the real-time dynamics of gravitational collapse (if that occurs).

\begin{figure}[h]
    \centering
    \includegraphics[width=8cm]{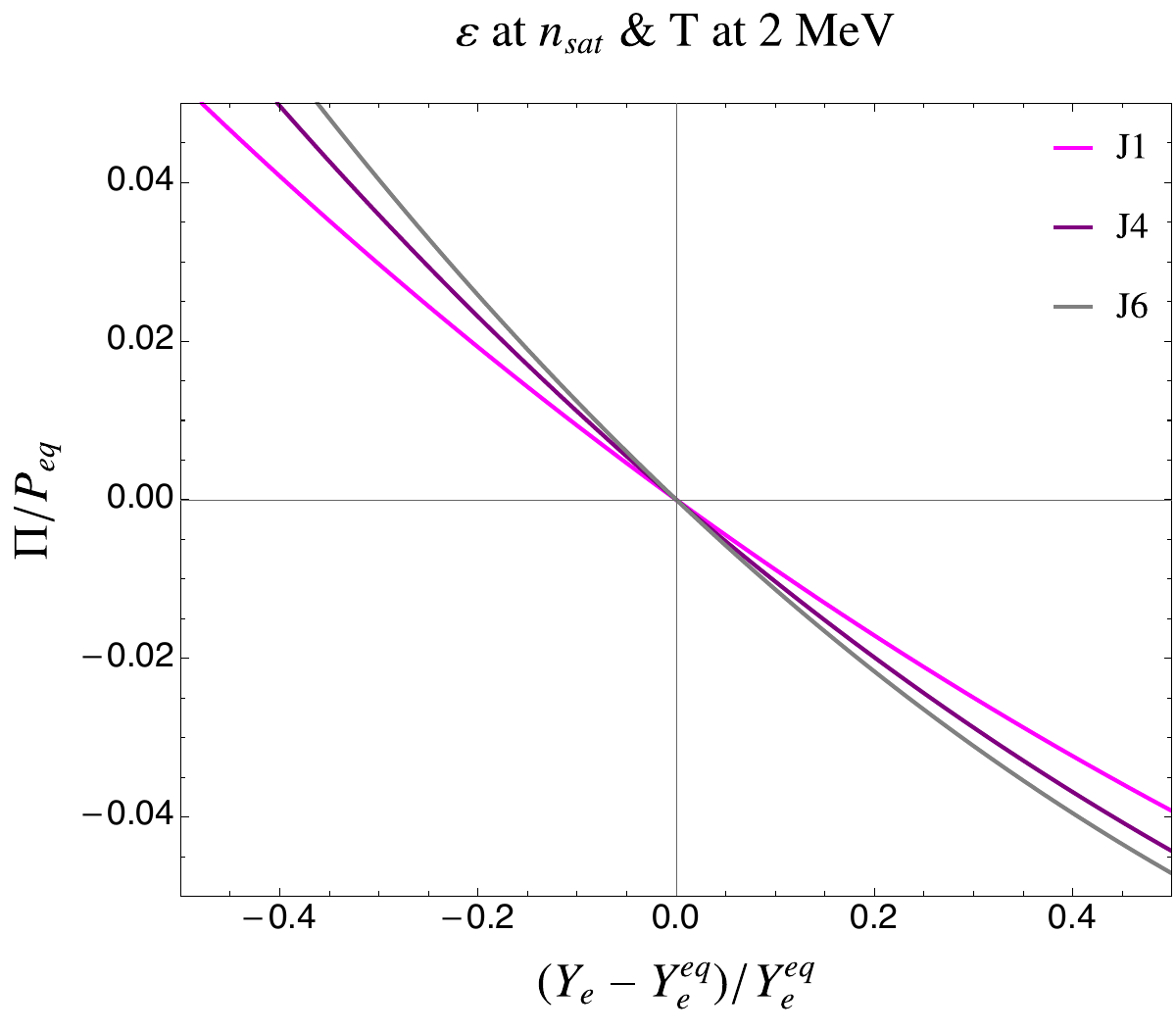} \hfill
    \includegraphics[width=8cm]{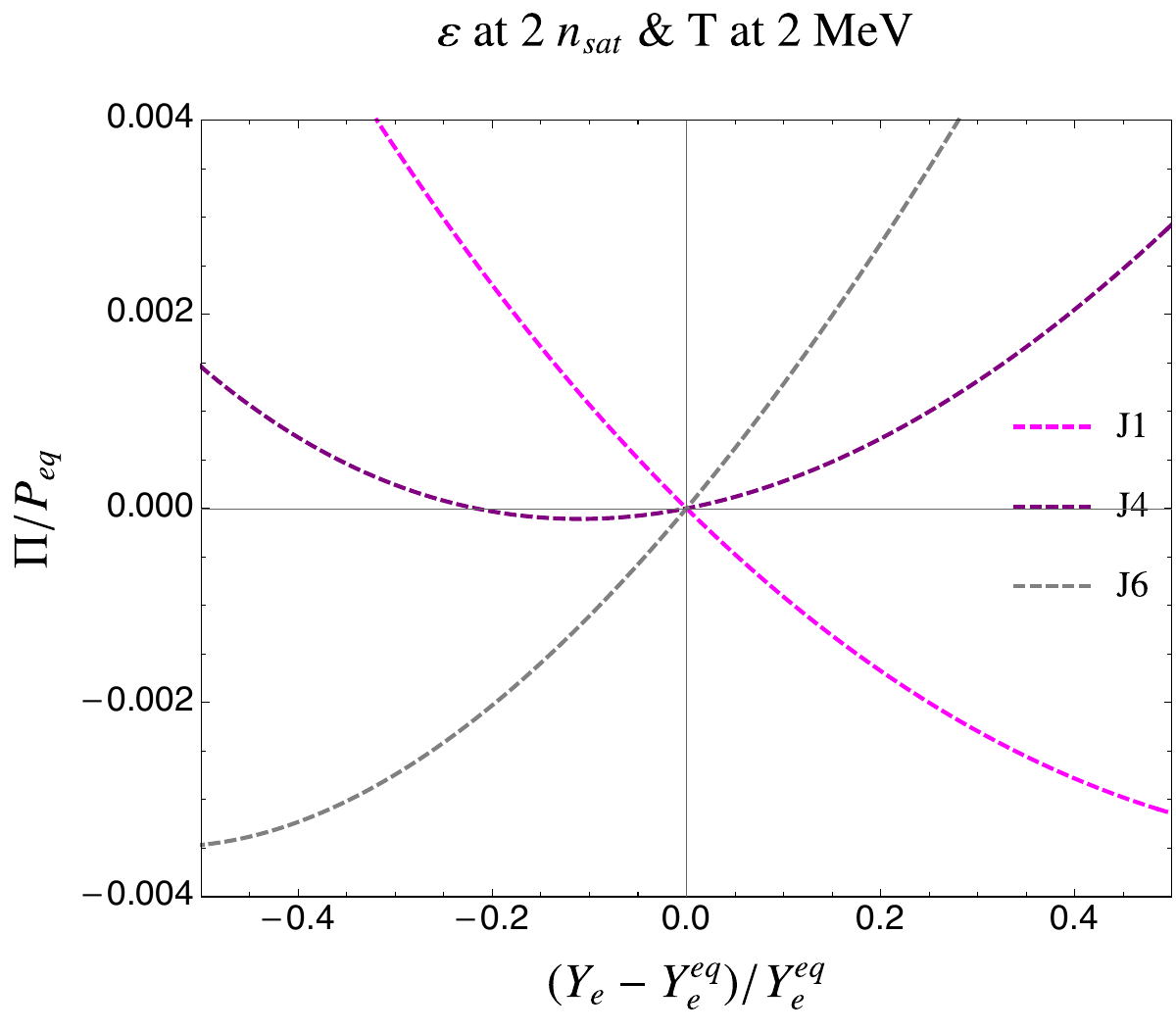} 
    \caption{Bulk scalar $\Pi$ (normalized by the equilibrium pressure) as a function of the deviation from the equilibrium charge fraction, for several symmetry energies, at temperature $T=2$ MeV. Left: $\Pi$ for the energy density corresponding to nuclear saturation density. Right: The same quantity is evaluated at twice the nuclear saturation density.}
    \label{J bulkv}
\end{figure}

\begin{figure}[h]
    \centering
    \includegraphics[width=8cm]{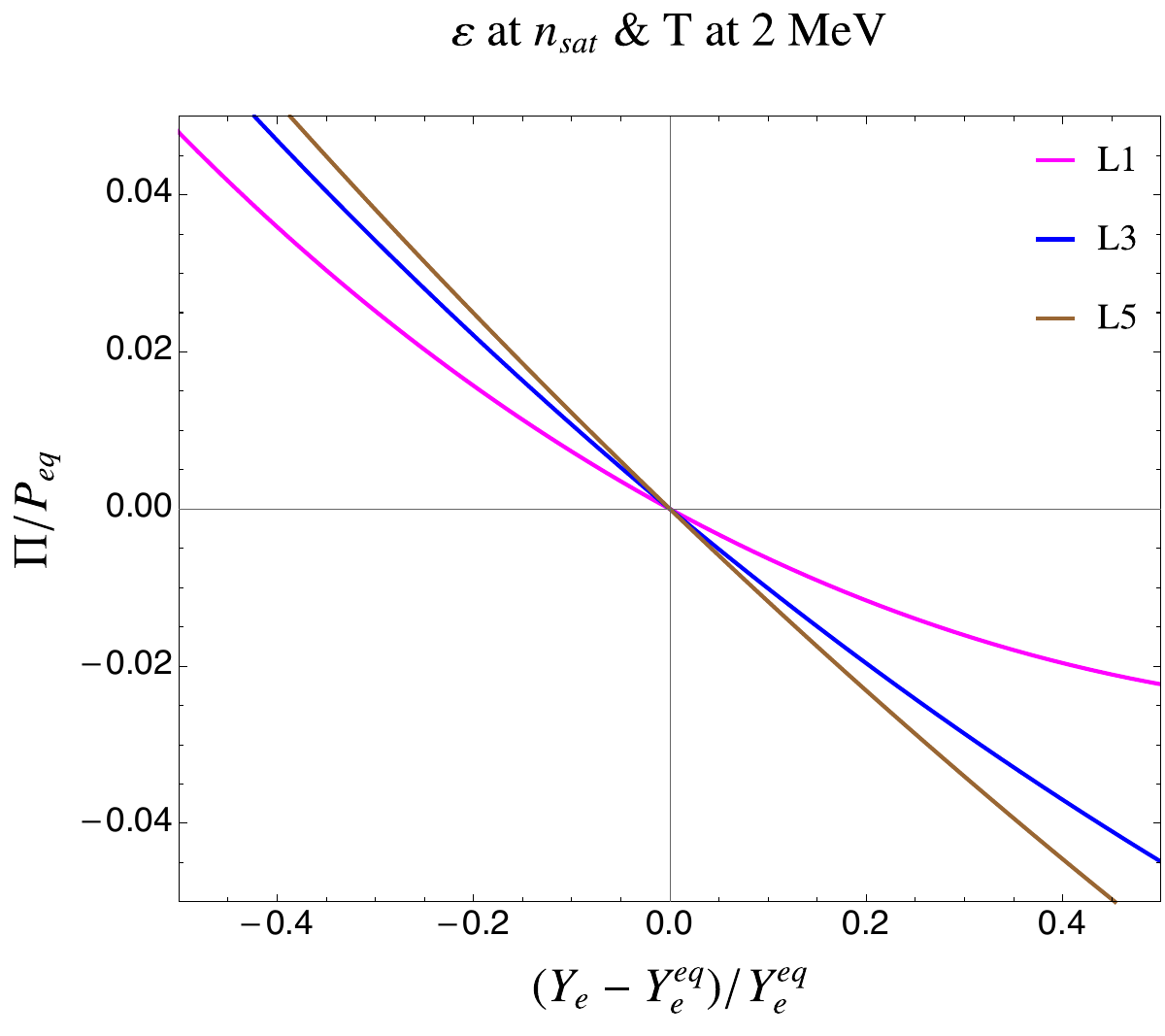} \hfill
    \includegraphics[width=8cm]{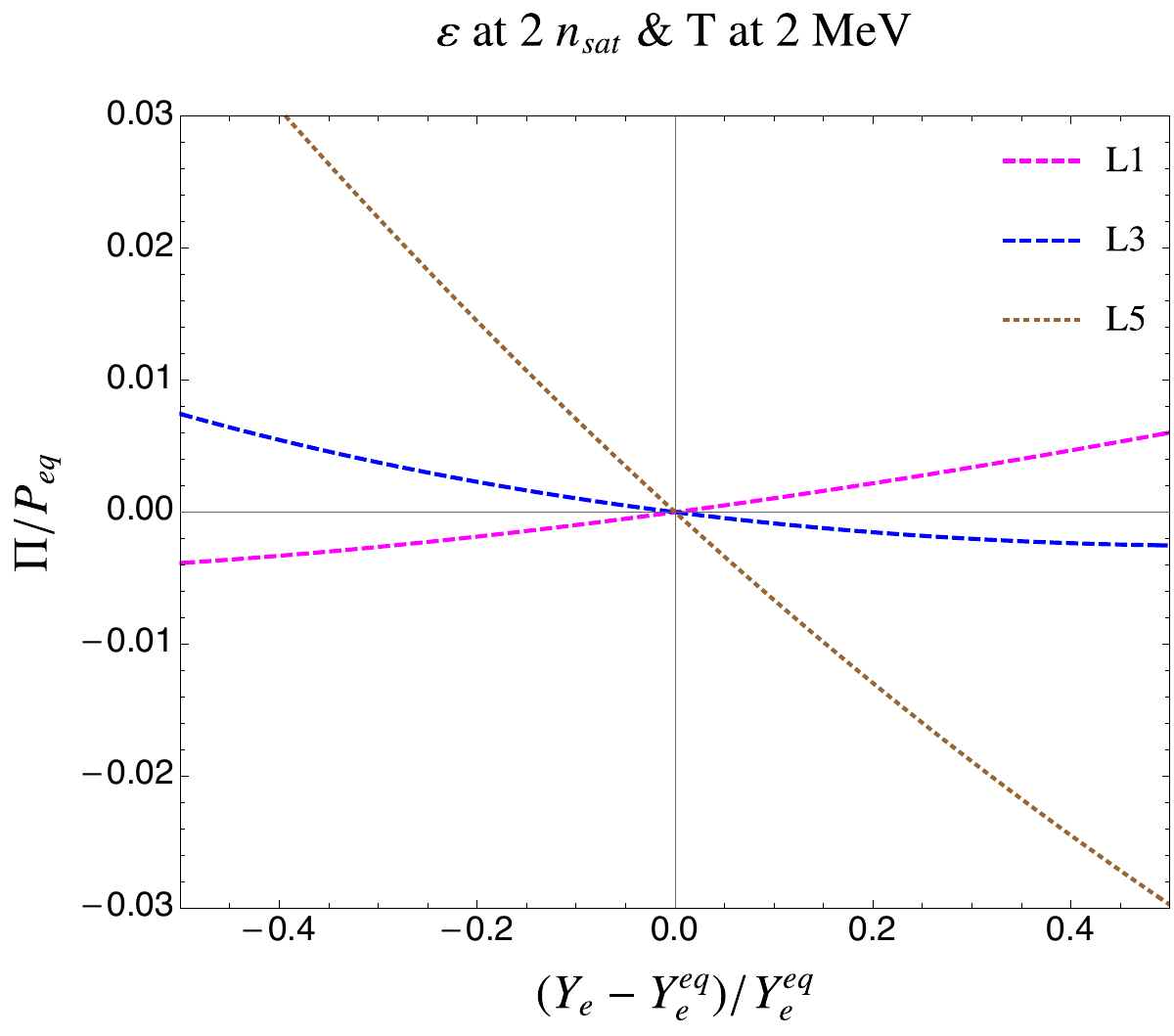} 
    \caption{Bulk scalar $\Pi$ (normalized by the equilibrium pressure) as a function of the deviation from the equilibrium charge fraction for several symmetry slopes at temperature $T=2$ MeV. Left: $\Pi$ for the energy density corresponding to nuclear saturation density. Right: The same quantity is evaluated at twice the nuclear saturation density. }
    \label{L bulkv}
\end{figure}

In order to better understand the subtleties concerning the near-equilibrium behavior of such systems, we now focus on the properties of the bulk scalar $\Pi$. In general, as $\ep$ increases, $\Pi(\varepsilon,Y_e)$ ceases to be linear in deviations from $Y_e^{\textrm{eq}}$. Figures~\ref{J bulkv} and \ref{L bulkv} show that this is the case for several values of $J$ and $L$. This indicates that this is a robust property of such systems.  

\begin{figure}[ht!]
    \centering
    \includegraphics[width=8cm]{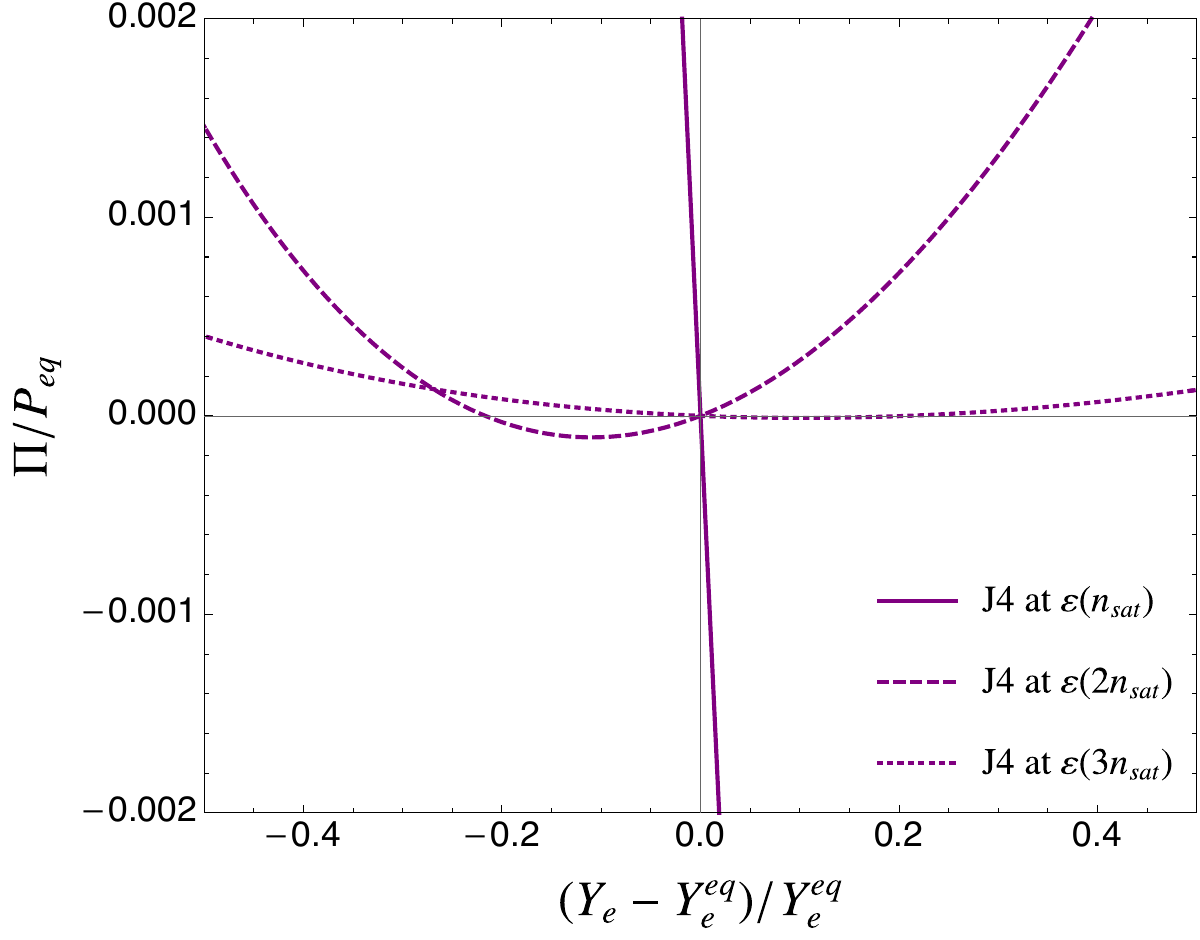}
    \caption{Bulk scalar $\Pi$ (normalized by the equilibrium pressure) of the J4 EoS as a function of the deviation from the equilibrium charge fraction, at temperature $T=2$ MeV. The different curves represent energy densities corresponding to baryon densities ranging from one to three times saturation density. Note that $\Pi$ displays a clear minimum for the J4 EoS at twice the saturation density.}
    \label{nonlinearity}
\end{figure}

Further insight into this nonlinear trend may be obtained by focusing on the results coming from a specific EoS. Figure \ref{nonlinearity} shows the bulk scalar $\Pi$ associated with the J4 EoS across three different energy densities. The behavior of $\Pi$ as a function of the deviation from the equilibrium charge fraction is very sensitive to the energy density value.  In fact, as one increases $\ep$, the linear behavior turns nonlinear, and $\Pi$ displays a minimum as a function of the deviations from the equilibrium charge density for the J4 EoS at $\ep(2n_{\textrm{sat}})$ and  $\ep(3n_{\textrm{sat}})$. 

When $\Pi$ has a minimum with respect to $Y_e$ at a fixed energy density, both the relaxation time $\tau_\Pi$ and bulk viscosity $\zeta$ formally diverge, see Eqs.\ \eqref{taupifinal} and \eqref{zetafinal}, but their ratio can remain finite. In this limit, which was discussed in \cite{Gavassino:2023xkt}, the system is probably better understood as an inviscid elastic system with bulk modulus equals $\zeta/\tau_\Pi$ rather than a bulk-viscous fluid.   

The existence of a minimum for $\Pi$, and the nearly quadratic behavior displayed by $\Pi$ versus $(Y_e-Y_e^{\textrm{eq}})/Y_e^{\textrm{eq}}$, implies that at that density the mapping between $\Pi$ and $Y_e$ is not one to one. We will take this here as an indication that the description of the system as a fluid with bulk viscosity is not warranted at that point.  

\begin{figure}[ht!]
    \centering
    \includegraphics[width=8cm]{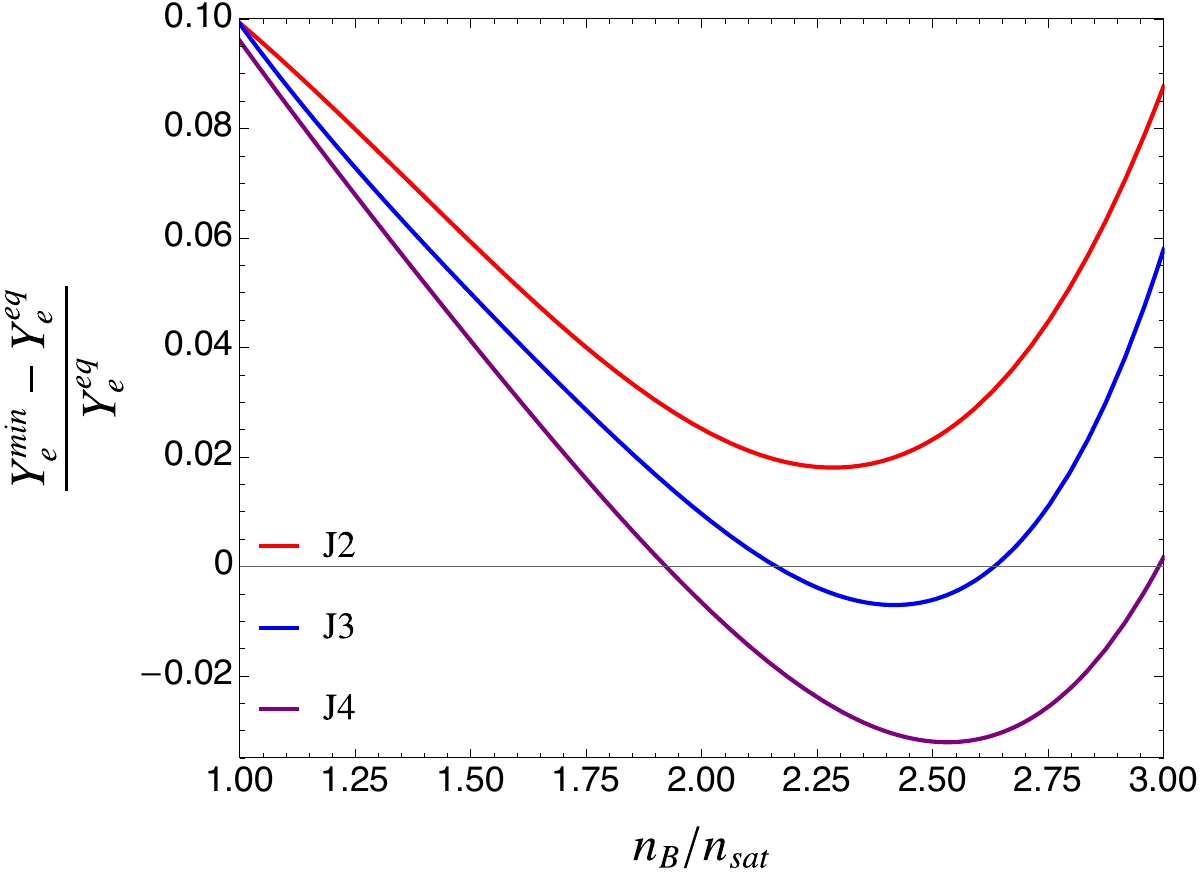} 
    \caption{The relative difference between the charge fraction at the point of minimum $\Pi$ as a function of the density for different values of the symmetry energy.}
    \label{RoV}
\end{figure}

To further study this problem, we investigate additional properties of the system near the minimum. We denote the charge fraction where $\Pi$ is at its minimum as $Y_e^{min}$. Fig.~\ref{RoV} shows the relative change in $Y_e^{min}$, with respect to $Y_e^{\textrm{eq}}$, for different symmetry energies, as a function of the density. One can see that the curve crosses zero where $Y_e^{min}=Y_e^{\textrm{eq}}$. In that case, the standard bulk-viscous description has a vanishingly small region of applicability.

In light of the equivalence put forward in \cite{Gavassino:2023xkt}, it is conceivable that this marks the transition from a bulk-viscous fluid regime to an elastic regime. In that case, one may interpret that as the signature of a far-from-equilibrium phase transition in the system.
Further work is needed to better understand the consequences of the presence of a minimum in the bulk scalar and how that affects the properties of the system.

\begin{figure}[ht!]
    \centering
    \includegraphics[width=8cm]{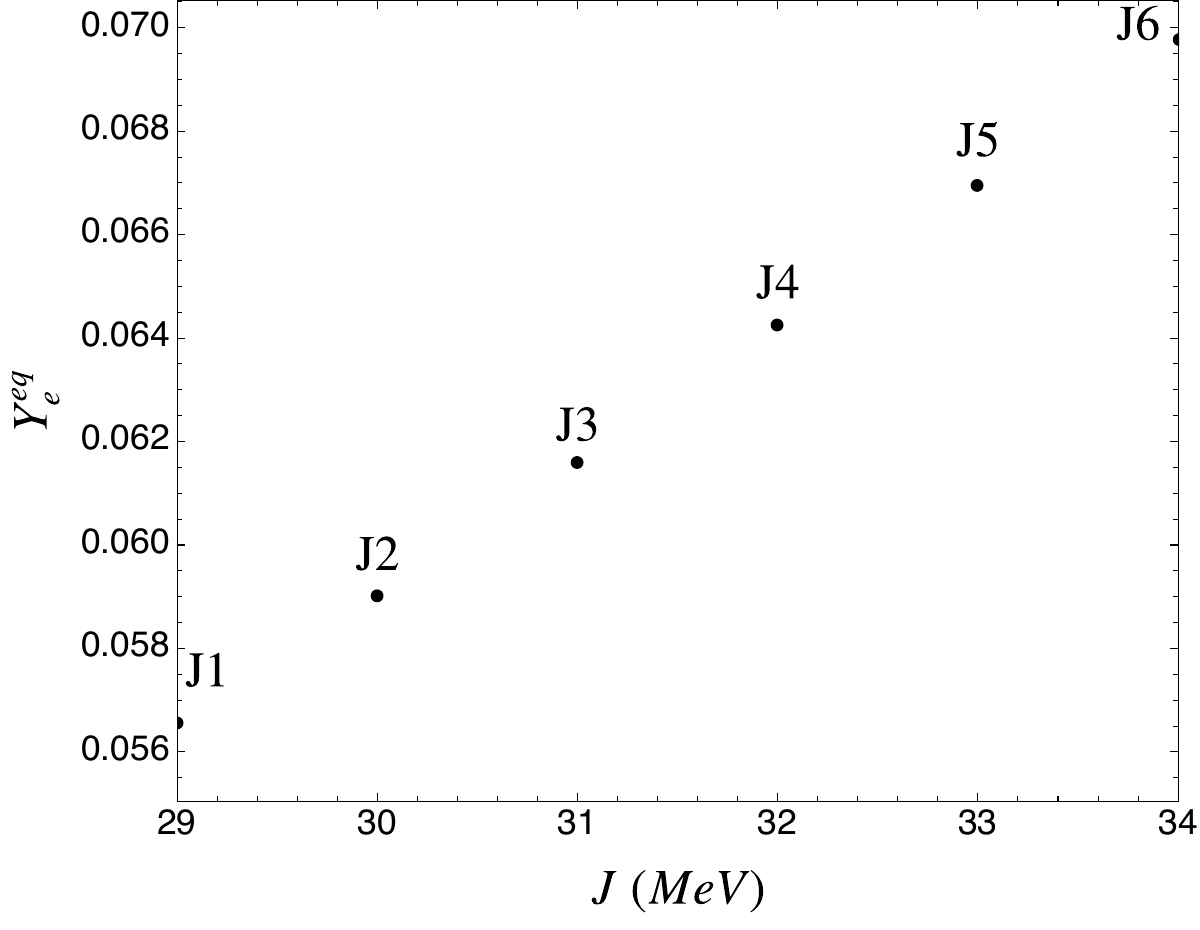} \hfill
    \includegraphics[width=8cm]{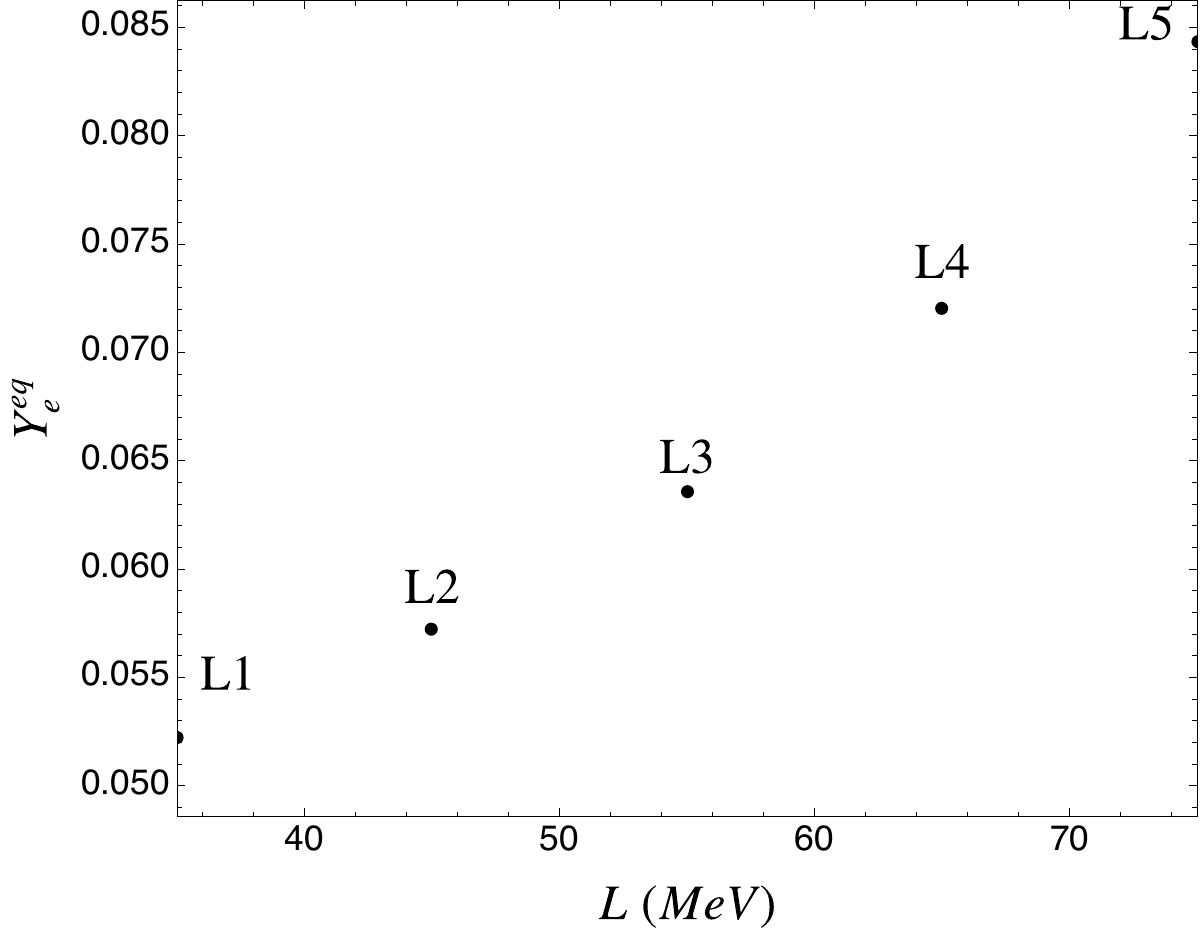} 
    \caption{Left: Equilibrium electron fraction $Y_e^{\textrm{eq}}$ as a function of the symmetry energy for $T=2$ MeV. Right: $Y_e^{\textrm{eq}}$ as a function of the symmetry slope. Each point corresponds to an equation of state.}
    \label{Ye}
\end{figure}

\subsubsection{Additional effects from the variations of the symmetry energy and slope}

We note that when $J$ and $L$ increase, $Y_e^{\textrm{eq}}$ increases, as shown in Fig.~\ref{Ye}. This is intuitively reasonable, as a larger proton fraction is required to achieve equilibrium for larger $J$ or $L$.

\begin{figure}[h]
    \centering
    \includegraphics[width=8cm]{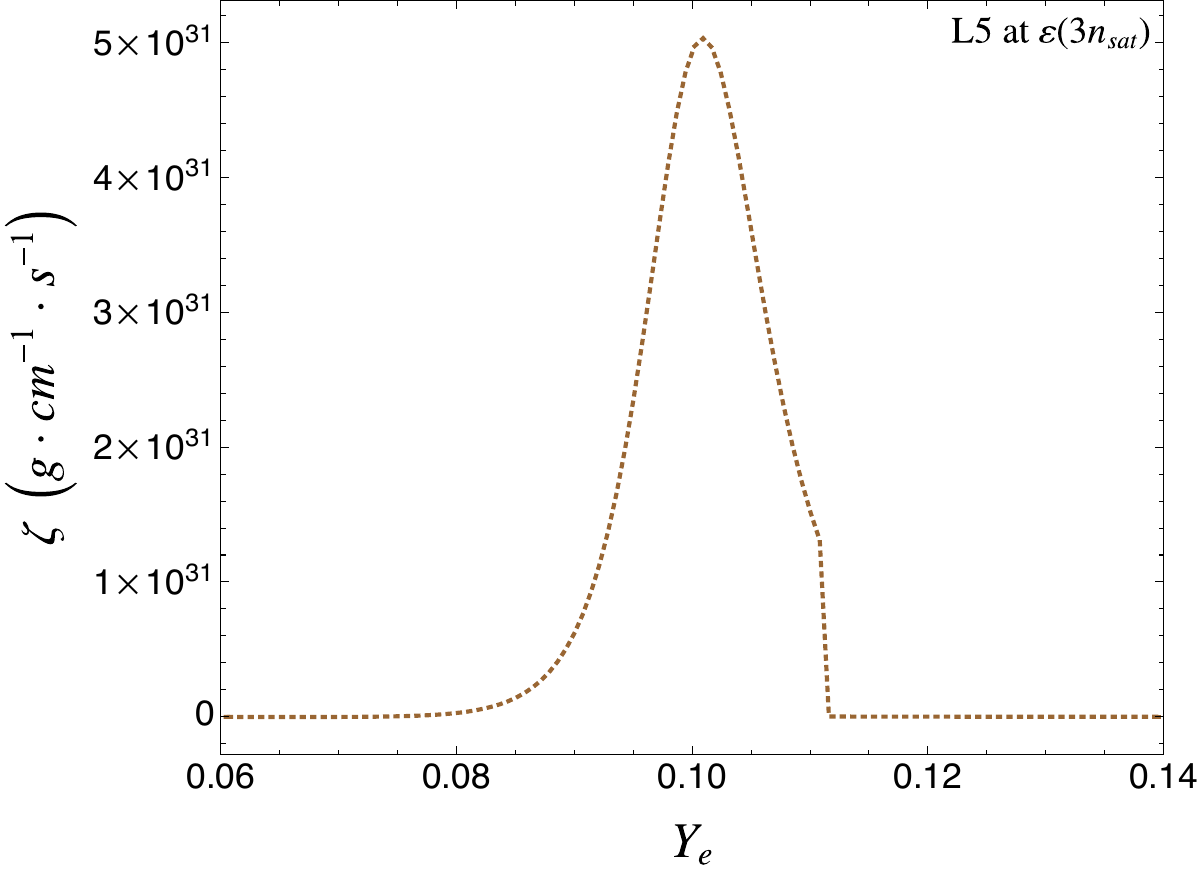} \hfill
    \includegraphics[width=8cm]{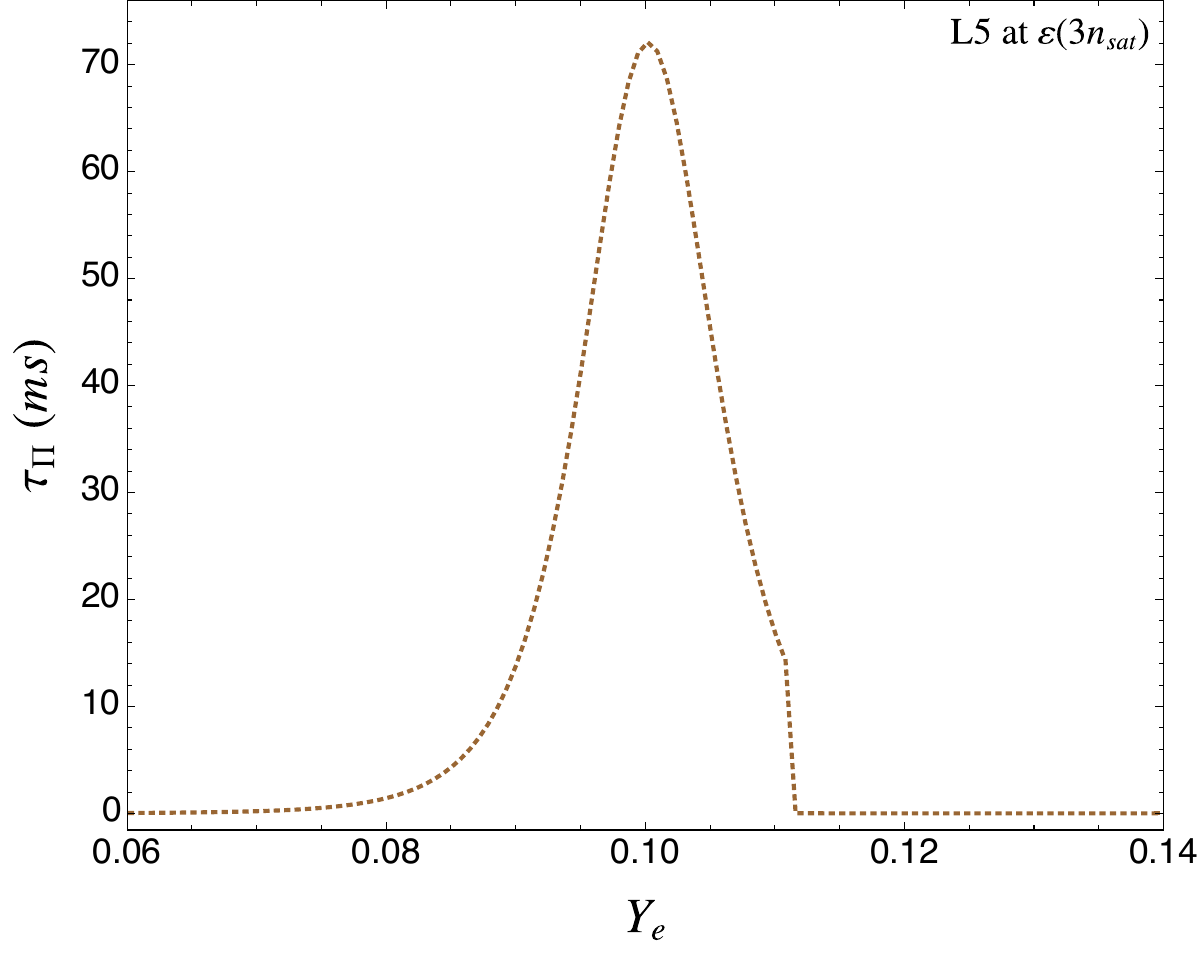} 
    \caption{Transport coefficients for the L5 EoS against the deviation from the equilibrium charge fraction, at the energy density corresponding to three times the saturation density, and temperature $T=2$ MeV. Left: Resummed $\zeta$ against the electron fraction $Y_e$. Right: Resummed $\tau_\Pi$ against $Y_e$.}
    \label{L5 direct urca}
\end{figure}

As $Y_e^{\textrm{eq}}$ increases, it becomes closer to the threshold for the direct Urca rate to be enabled, $Y_e = 1/9$. Fig.~\ref{L5 direct urca} shows that both $\tau_\Pi$ and $\zeta$ are greatly suppressed by the onset of the direct Urca rate. Once this threshold is surpassed, the direct Urca rate has a much larger magnitude than the modified Urca rate, and the chemical reactions happen significantly faster. As equilibration becomes more efficient, the relaxation time and bulk viscosity are significantly reduced.

Finally, we note that for all cases considered here, the entropy production is always non-negative, and the system should be stable under perturbations around equilibrium. In addition, near equilibrium, the transport coefficients for all of our EoS are always positive, so their physical interpretation in the resummed Israel-Stewart is well-defined. Furthermore, we remark that we have checked that all of our EoSs satisfy the causality constraint given by Eq.~\eqref{causality}. In fact, for our zero temperature approximation of the EoS, Eq.~\eqref{causality} reduces to
\be
\label{eq:causalitycond}
\frac{\zeta}{\tau_\Pi(\ep + P)} + \left.\frac{\pa P}{\pa \ep}\right|_{Y_e = Y_e^{\textrm{eq}}} \leq  1,
\ee
which can be easily evaluated for our equations of state.

\subsection{Comparison to other equations of state}
\label{other EoS}

\begin{figure}[ht!]
    \centering
    \includegraphics[width=8cm]{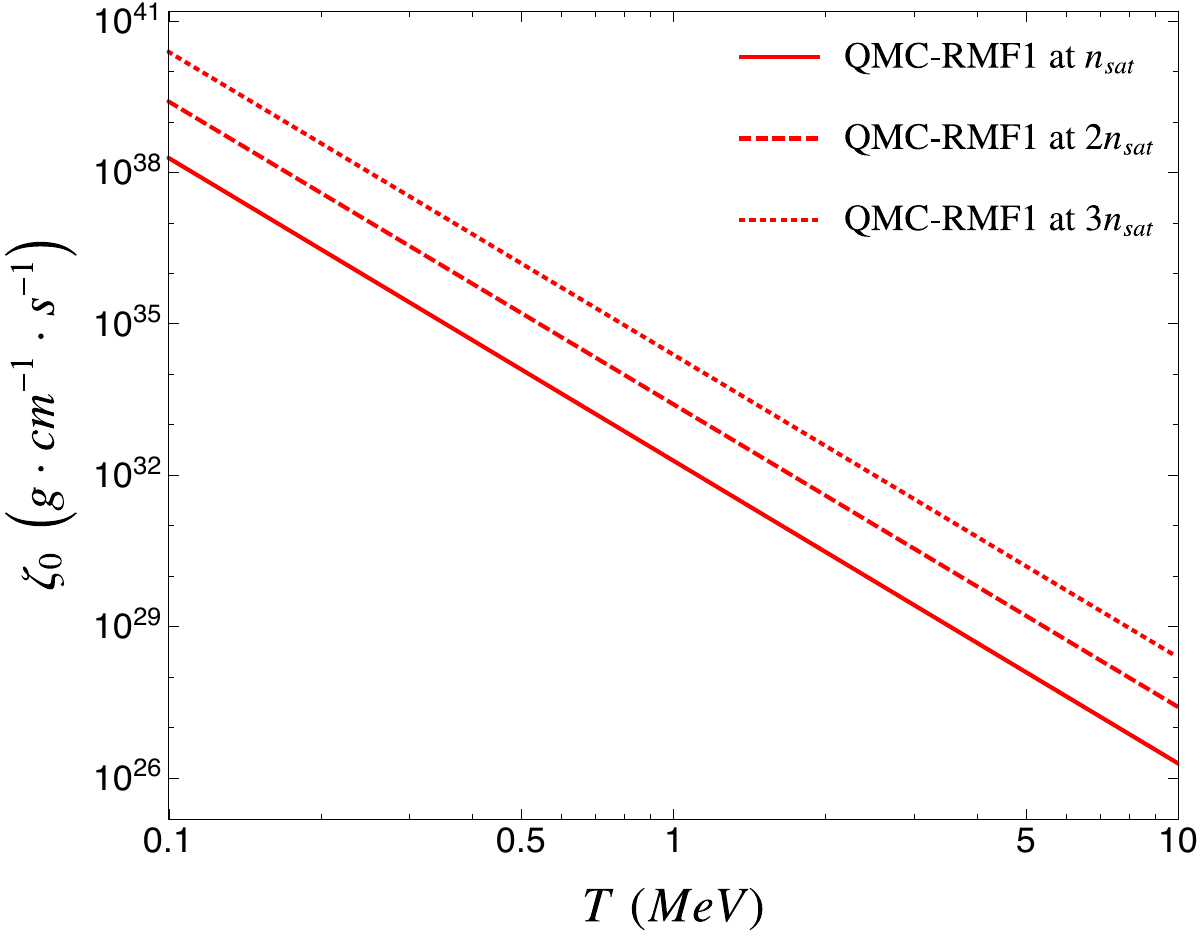} 
    \caption{Bulk viscosity $\zeta_0$ for the QMC-RMF1 EoS as a function of temperature.}
    \label{QMC1 zeta0}
\end{figure}

To ensure the robustness of our findings, we have also investigated other EoSs, where we have also observed the features described above.

\begin{figure}[ht!]
    \centering
    \includegraphics[width=8cm]{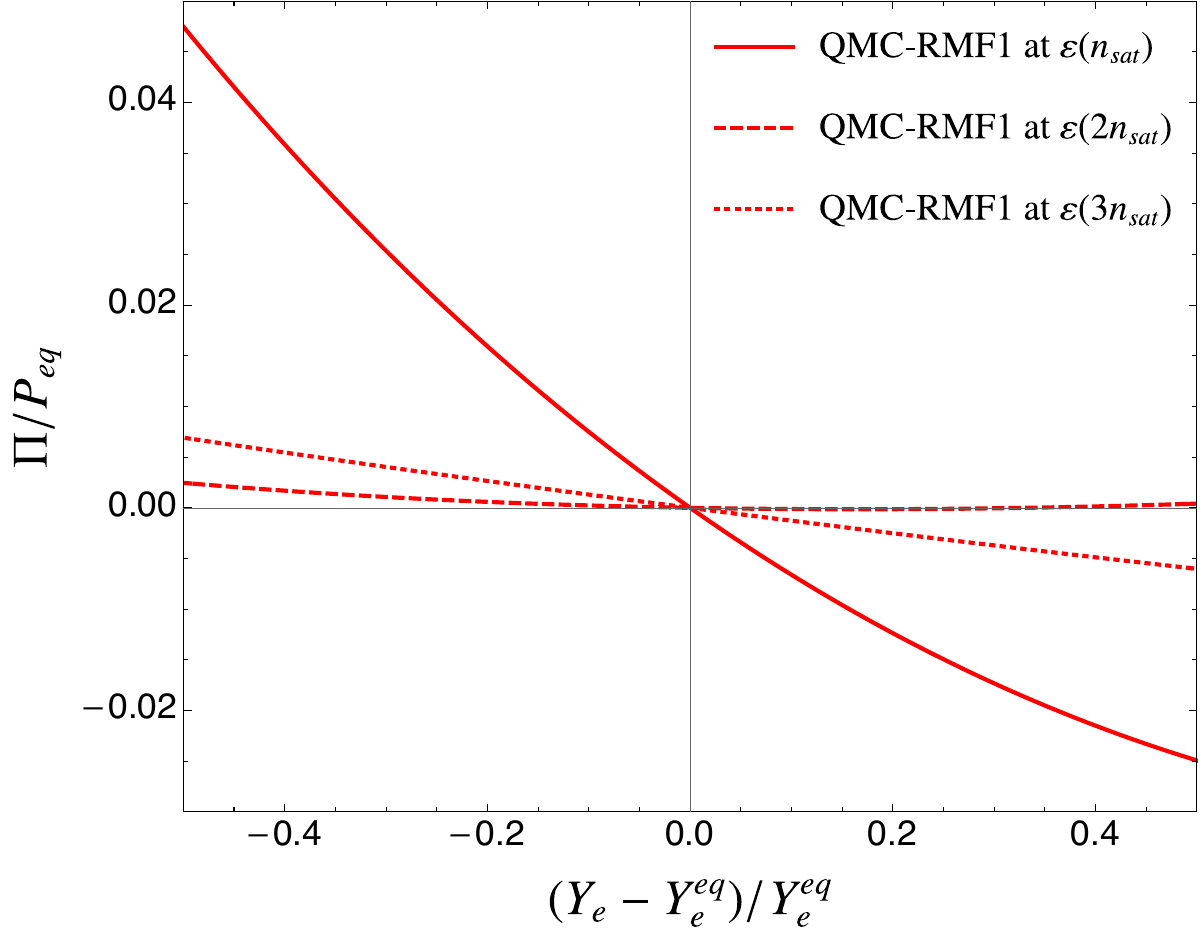} 
    \caption{Bulk scalar $\Pi$ of the QMC-RMF1 EoS (normalized by the equilibrium pressure), as a function of the deviation from the equilibrium charge fraction, at  $T\;=\;2\;$ MeV. We show the results evaluated at energy densities corresponding to number densities one to three times the nuclear saturation density.}
    \label{QMC1 pi}
\end{figure}

\begin{figure}[!ht]
    \centering
    \includegraphics[width=8cm]{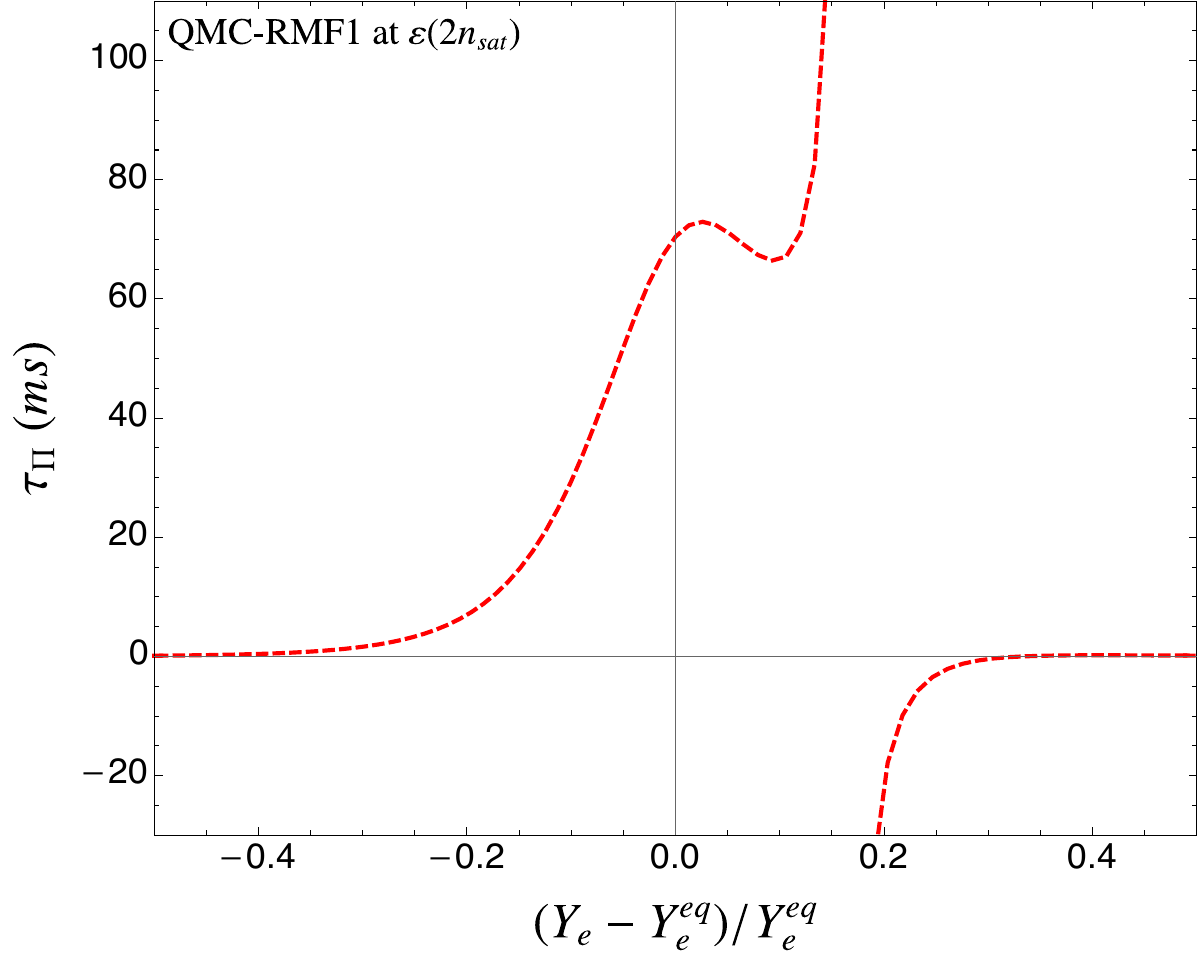}
    \caption{Resummed $\tau_\Pi$ of the QMC-RMF1 EoS, as a function of the deviation from the equilibrium charge fraction, evaluated at the energy density corresponding to a density twice the saturation density. The  temperature is $T\;=\;2\;$ MeV.}
    \label{QMC1 elastic zeta}
\end{figure}

\begin{figure}[!ht]
    \centering
    \includegraphics[width=8cm]{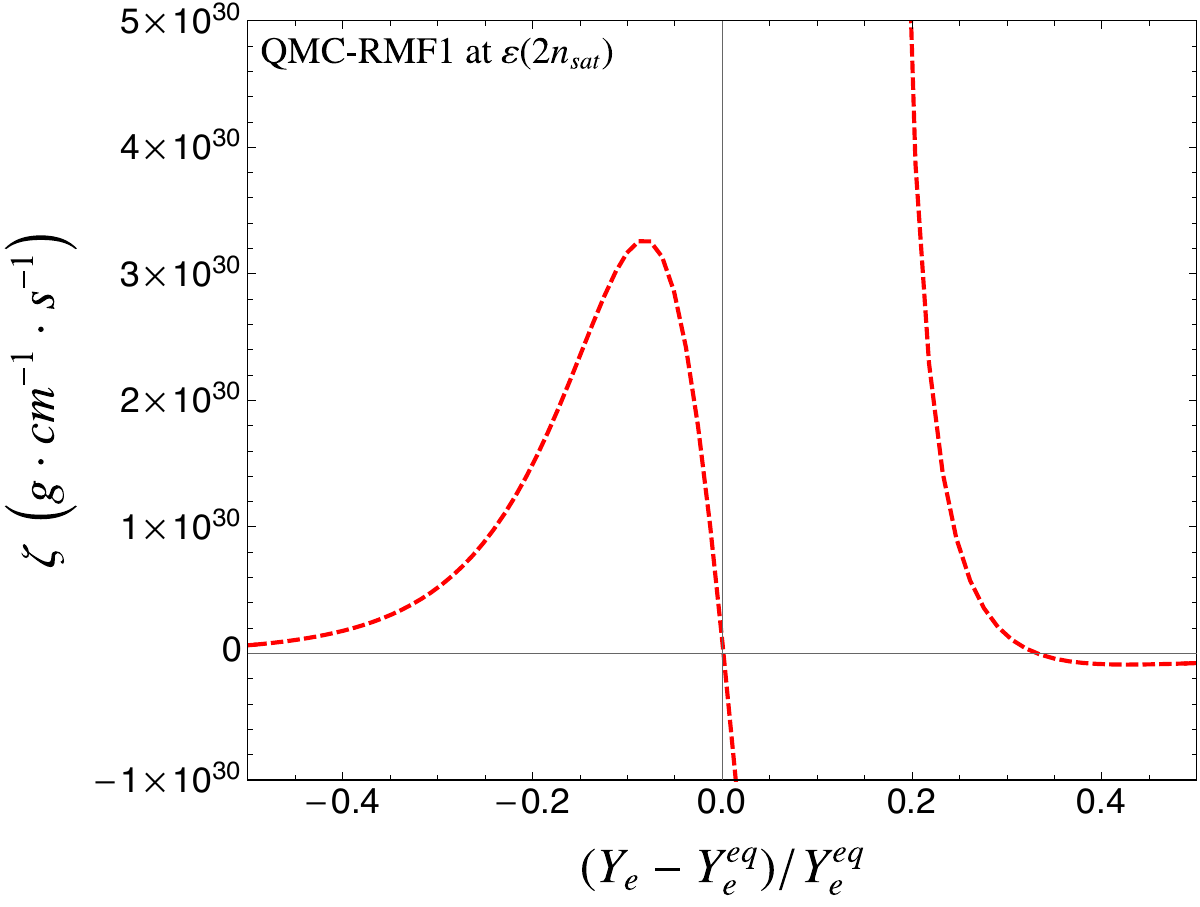}
    \caption{Resummed $\zeta$ of the QMC-RMF1 EoS, as a function of the deviation from the equilibrium charge fraction, evaluated at the energy density corresponding to a density twice the saturation density. The  temperature is $T\;=\;2\;$ MeV.}
    \label{QMC1 elastic tau}
\end{figure}

As an example, we take the EoS QMC-RMF1 \cite{Alford:2022bpp}, which can be obtained from the CompOSE website \url{compose.obspm.fr/eos/275} \cite{Typel:2013rza,Oertel:2016bki,CompOSECoreTeam:2022ddl}.\footnote{CompOSE, an acronym for CompStar Online Supernovae Equations of State, is an online repository that provides thermodynamic and microphysical properties for different EoSs.} Figure~\ref{QMC1 zeta0} shows that QMC-RMF1 has an equilibrium bulk viscosity $\zeta_0$ comparable to that found in our previous equations of state. However, while well defined, the quantity $\zeta_0$ cannot capture the bulk-viscous physics correctly when the linear response approximation is invalid. This issue happens for this equation of state. Figure~\ref{QMC1 pi} shows that QMC-RMF1 displays a nonlinear trend as the energy density increases, and at $n_B = 2\,n_{\textrm{sat}}$ the bulk scalar displays a minimum. Correspondingly, Fig.~\ref{QMC1 elastic zeta} and \ref{QMC1 elastic tau} show the resummed  $\zeta$ and $\tau_\Pi$ of QMC-RMF1 at the energy density corresponding $n_B \approx 2\,n_{\textrm{sat}}$. As expected, because of this nonlinear behavior, these quantities diverge (and turn negative) at the point where $\Pi$ displays a minimum, which is at $\delta Y_e/Y_e^{eq}\approx 0.15$ for that density. This shows that the resummed $\zeta$, which was obtained from the exact equivalence (or duality) between the reactive system and the bulk-viscous system, can capture nuances present in the EoS that cannot be seen in the linear response quantity $\zeta_0$.

\clearpage

\section{Conclusions }
\label{sec:conclusions}

In this work, we investigated the weak-interaction-driven transport properties of $npe$ matter in the neutrino transparent regime. 
Results for the bulk viscosity and bulk relaxation time were computed using realistic equations of state that satisfy the latest constraints from multi-messenger observations (see Fig.~\ref{TOV}). In our calculations, we used the resummed Israel-Stewart formulation of Ref.\ \cite{Gavassino:2023xkt}, which allows one to rewrite the reactive mixture commonly solved in numerical simulations as an exact bulk-viscous system. 
Different versions of our EoS were presented, each corresponding to different values of the nuclear symmetry energy and its slope at nuclear saturation density. 
In Figs.~\ref{bulk0} and \ref{AC bulk}, we have shown the effect of these two properties of nuclear matter on the bulk viscosity $\zeta_0$, computed using parameters in beta equilibrium, and the AC bulk viscosity $\zeta_{AC}$, which is relevant for the physics of small density oscillations. Both quantities characterize linear response behavior, and we find that varying $J$ and $L$ significantly affect these coefficients. We computed the full resummed transport coefficients for each EoS, checking again how changes in $J$ and $L$ affect the results. Overall, we find that these basic properties of the EoS greatly affect the resummed transport coefficients. Therefore, our results indicate that having better constraints on $J$ and $L$ is crucial not only for determining the nuclear equation of state but also for unveiling novel bulk-viscous phenomena in neutron star mergers.

Previous works in the literature have assumed that the induced bulk-viscous correction to the pressure, near beta equilibrium, is linear in deviations from the equilibrium charge fraction\footnote{We note that an amplitude-dependent bulk viscosity in dense matter was considered in \cite{Alford:2010gw}.}. However, in Fig.~\ref{J bulkv}, we showed that for realistic EoSs, at densities between one and three times saturation density, the out-of-equilibrium correction to the pressure becomes nonlinear in $Y_e-Y_e^{\mathrm{eq}}$. This result limits the applicability of a linear approximation in deviations from equilibrium and requires a bulk-viscous description that can be meaningful even far from equilibrium, such as the one derived in \cite{Gavassino:2023xkt}. 
Figures~\ref{J bulk} through \ref{L tau} show, for the first time, how the bulk-viscous transport coefficients change out of equilibrium for realistic equations of state.  

Furthermore, we showed that there are regions in density where the description of the reactive system as a standard bulk-viscous fluid ceases to hold. This occurs when $\Pi$ displays a minimum as a function of $Y_e-Y_e^{\mathrm{eq}}$, which was shown to happen in realistic equations of state. In that case, following \cite{Gavassino:2023xkt}, the system may be closer to an elastic regime where $\zeta/\tau_\Pi$ gives the corresponding bulk modulus. This shows how rich the dynamics encoded in the simple $npe$ reactive mixture can be, opening the door for investigating new transport phenomena in neutron star mergers. 

The far-from-equilibrium bulk-viscous effects investigated here stem from particle-number changing processes, which depend on the microscopic excitations of the system. Therefore, the far-from-equilibrium effects considered here should be sensitive to the presence of deconfined degrees of freedom in the dense matter equation of state. Finally, one may also investigate the interplay between far-from-equilibrium bulk viscosity and superfluid/superconducting pairing patterns in ultradense matter \cite{Alford:2011df}.

\section*{Acknowledgements}
We thank J.~Noronha-Hostler, L.~Gavassino, and M.~Alford for enlightening discussions, L.~Brodie and A.~Haber for providing assistance with the QMC-RMF1 EoS, and C.~Conde for providing us with the QLIMR module developed within the MUSES collaboration, which we have used to solve the TOV equation. 
JN is partly supported by the U.S. Department of Energy, Office of Science, Office for Nuclear Physics
under Award No. DE-SC0021301 and DE-SC002386. MH and JN were partly supported by the National Science Foundation (NSF) within the framework of the MUSES collaboration, under grant number OAC-2103680.  JN, MH, and ES thank KITP Santa Barbara for its hospitality during “The Many Faces
of Relativistic Fluid Dynamics” Program, where this work’s last stages were completed. This
research was partly supported by the National Science Foundation under Grant No. NSF
PHY-1748958.

\newpage

\appendix

\section{Urca rates}
\label{A: rate}
We use the Fermi surface approximation to calculate the Urca rates; see \cite{Alford:2018lhf, Alford:2021ogv, Harris:2020rus}. The total rate is the sum of the individual rates,
\be
\Ga_e = \Ga_{dU} + \Ga_{mU,n} + \Ga_{mU,p},
\ee
which are then found to be,
\begin{gather}
\Ga_{dU} = \frac{G^2(1+3g_A^2)}{240\pi^5}E^*_{Fn}E^*_{Fp}p_{Fe}\vartheta_{dU}\de\mu(17
\pi^4T^4 + 10\pi^2\de\mu^2T^2 + \de\mu^4) , \\
\Ga_{mU,n} = \frac{1}{5760\pi^9}G^2 g_A^2 f^4 \frac{(E^*_{Fn})^3 E^*_{Fp}}{m_\pi^4} \frac{p_{Fn}^4 p_{Fp}}{(p_{Fn}^2+m_\pi^2)^2}\vartheta_n \de\mu( 1835 \pi^6 T^6 + 945\pi^4 \de\mu^2 T^4 + 105\pi^2\de\mu^4 T^2 + 3\de\mu^6 ) ,\\
\Ga_{mU,p} = \frac{1}{40320\pi^9}G^2 g_A^2 f^4 \frac{E^*_{Fn} (E^*_{Fp})^3}{m_\pi^4} \frac{p_{Fn}(p_{Fn}-p_{Fp})^4}{((p_{Fn}-p_{Fp})^2+m_\pi^2)^2}\vartheta_p \de\mu ( 1835\pi^6 T^6 + 945\pi^4 \de\mu^2 T^4 + 105\pi^2\de\mu^4T^2 + 3\de\mu^6 ) ,
\end{gather}
where the pion-nucleon couling constant $f\approx1$, $G^2 = G_F^2 cos^2\theta_c = 1.1\times 10^{-22} MeV^{-4}$ with $G_F$ being the Fermi coupling constant and $\theta_c$ the Cabibbo angle, the axial vector coupling constant is $g_A = 1.26$, $p_{FN}$ is the nucleon Fermi momentum, and $E^*_{FN} = \sqrt{p_{FN}^2 + m_N^{*2}}$ is the nucleon energy. 

In the Fermi surface approximation, the direct Urca rate only operates above a threshold density, where
\be
  \vartheta_{dU} =
  \begin{cases}
  1 & \text{if $p_{Fn}<p_{Fp} + p_{Fe}$} \\
  0 & \text{if $p_{Fn}>p_{Fp} + p_{Fe}$} ,
  \end{cases}
\ee
The modified Urca rate is also affected by the density, for which we have introduced
\be
  \vartheta_n =
  \begin{cases}
  1 & \text{if $p_{Fn}>p_{Fp} + p_{Fe}$} \\
  1 - \frac{3}{8} \frac{(p_{Fp} + p_{Fe} - p_{Fn})^2}{p_{Fp}p_{Fe}} & \text{if $p_{Fn}<p_{Fp} + p_{Fe}$} ,
  \end{cases}
\ee
\be
  \vartheta_p =
  \begin{cases}
  0 & \text{if $p_{Fn}>3p_{Fp} + p_{Fe}$} \\
  \frac{(3p_{Fp} + p_{Fe} - p_{Fn})^2}{p_{Fn}p_{Fe}} & \text{if} 
  \begin{cases}
  p_{Fn}>3p_{Fp} - p_{Fe} \\
  p_{Fn}<3p_{Fp} + p_{Fe}
  \end{cases} \\
  4 \frac{3p_{Fp} - p_{Fn}}{p_{Fn}} & \text{if}
  \begin{cases}
  3p_{Fp} - p_{Fe} > p_{Fn} \\
  p_{Fn} > p_{Fp} + p_{Fe}
  \end{cases} \\
  2 + 3\frac{2p_{Fp} - p_{Fn}}{p_{Fe}} - 3\frac{(p_{Fp} - p_{Fe})^2}{p_{Fn}p_{Fe}} & \text{if $p_{Fn}<p_{Fp} + p_{Fe}$} .
  \end{cases}
\ee

\newpage
\section{Israel-Stewart formalism in the linear $\de \mu$ regime}
\label{A: linear}

Assuming a linear perturbation around the chemical equilibrium, one can also derive an Israel-Stewart equation for the bulk scalar, see \cite{Gavassino:2020kwo}. For the sake of completeness, here we briefly go through the derivation of this result. 

For a system that is close to equilibrium in terms of $\de\mu$, one can expand the pressure with respect to $\de\mu$ around beta equilibrium,
\be
P(\varepsilon, n_B, \delta \mu) = P|_{\delta \mu = 0} + P_1 \delta \mu ,
\ee
where $P_1 = \left.\frac{\partial P}{\partial \delta \mu}\right|_{\varepsilon, n_B, \delta \mu = 0}$. If we assume isotropy of the fluid and ignore second-order terms in $\de\mu$, the stress-energy tensor takes the form of a perfect fluid, 
\begin{equation}
\bea
    T^{\mu \nu} &= \varepsilon u^\mu u^\nu + P(g^{\mu \nu}+u^\mu u^\nu) \\
    &\approx \ep u^\mu u^\nu + (P|_{\de\mu=0} + \Pi)(g^{\mu \nu}+u^\mu u^\nu) ,
\eea
\end{equation}
where $\Pi = P_1 \delta \mu$ and is the bulk scalar induced by the chemical reactions.

We can define partial equilibrium states with all non-conserved variables. Writing $\delta \mu = \delta \mu(\varepsilon, n_B, Y_e)$ for one such state, one finds
\begin{equation}
\label{beta chain}
    u^\mu \nabla_\mu \delta \mu = \frac{\partial \delta \mu}{\partial \varepsilon}\bigg|_{n_B, Y_e} u^\mu \nabla_\mu \varepsilon + \frac{\partial \delta \mu}{\partial n_B}\bigg|_{\varepsilon, Y_e} u^\mu \nabla_\mu n_B + \frac{\partial \delta \mu}{\partial Y_e}\bigg|_{\epsilon, n_B} u^\mu \nabla_\mu Y_e.
\end{equation}
Substituting Eq.~\eqref{energy conserv}, Eq.~\eqref{baryon}, and Eq.~\eqref{gamma} into Eq.~\eqref{beta chain}, we find 
\begin{equation}
\label{beta approx}
    \begin{aligned}
    u^\mu \nabla_\mu \delta \mu = \frac{\partial \delta \mu}{\partial Y_e}\bigg|_{\epsilon, n_B} \frac{\Gamma_e}{n_B} - n_B \theta\, \left( \left.\frac{\partial \delta \mu}{\partial n_B}\right|_{\varepsilon, Y_e} + \left.\frac{\varepsilon + P}{n_B}  \frac{\partial \delta \mu}{\partial \varepsilon}\right|_{n_B, Y_e} \right) .
    \end{aligned}
\end{equation}
Furthermore, approximating the rate of weak processes to first order in $\delta \mu$, one can write
\begin{equation}
\label{gamma approx}
    \Gamma_e = \la \delta \mu.
\end{equation}
Here, one can see that both the out-of-equilibrium correction to the pressure and the change in charge fraction are approximated to be linear in $\de\mu$. 

Using Eq.~\eqref{gamma approx}, we find that Eq.~\eqref{beta approx} becomes
\begin{equation}
\label{beta approx 2}
    \begin{aligned}
    u^\mu \nabla_\mu \delta \mu = - \mathcal{A} \delta \mu - n_B \theta \mathcal{B} ,
    \end{aligned}
\end{equation}
where $\mathcal{A} = - \frac{\la}{n_B} \left. \frac{\partial \delta \mu}{\partial Y_e}\right|_{\varepsilon, n_B}$ and $\mathcal{B} = \left.\frac{\partial \delta \mu}{\partial n_B}\right|_{\varepsilon, Y_e} + \left.\frac{\varepsilon + P}{n_B}  \frac{\partial \delta \mu}{\partial \varepsilon}\right|_{n_B, Y_e}$. 

Because we want to derive an equation of motion for $\Pi$, we also need to consider the expansion of $P_1$. In this case, since $P_1$ is determined at $\delta \mu = 0$, we can write $P_1 = \frac{\partial P}{\partial \delta \mu}\bigg|_{\varepsilon, n_B, \delta \mu = 0}$ as
\begin{equation}
    P_1 = P_1(\varepsilon, n_B).
\end{equation}
Then
\begin{equation}
    u^\mu \nabla_\mu P_1 = \frac{\partial P_1}{\partial \varepsilon} \bigg|_{n_B} u^\mu \nabla_\mu \varepsilon + \frac{\partial P_1}{\partial n_B}\bigg|_\varepsilon u^\mu \nabla_\mu n_B.
\end{equation}
Substituting Eq.~\eqref{energy conserv} and Eq.~\eqref{baryon} into the equation above one finds
\begin{equation}
\label{P1 expansion}
    u^\mu \nabla_\mu P_1 = -\frac{\partial P_1}{\partial \varepsilon} \bigg|_{n_B} \left[(\varepsilon+P|_{\delta \mu = 0})\theta\right] - \frac{\partial P_1}{\partial n_B}\bigg|_\varepsilon n_B \theta.
\end{equation}
Now, we multiply Eq.~\eqref{beta approx 2} by $\frac{P_1}{\mathcal{A}}$  and add $\frac{\delta \mu}{\mathcal{A}}u^\mu \nabla_\mu P_1$ to both sides, which gives
\begin{equation}
\label{IS deriv}
    \begin{aligned}
    \frac{1}{\mathcal{A}} u^\mu \nabla_\mu \Pi + \Pi = - \frac{P_1}{\mathcal{A}}n_B\mathcal{B}\theta + \frac{\delta \mu}{\mathcal{A}}u^\mu \nabla_\mu P_1,
    \end{aligned}
\end{equation}
where we wrote $\Pi = \delta \mu P_1$.

Finally, one can substitute Eq.~\eqref{P1 expansion} into Eq.~\eqref{IS deriv} to find
\begin{equation}
\label{IS deriv 2}
    \begin{aligned}
    \frac{1}{\mathcal{A}} u^\mu \nabla_\mu \Pi + \Pi +  \theta \frac{\delta \mu}{\mathcal{A}}\left[ \frac{\partial P_1}{\partial \varepsilon}\bigg|_{n_B} (\varepsilon+P|_{\delta \mu=0}) + \frac{\partial P_1}{\partial n_B}\bigg|_\varepsilon n_B \right] = - \theta \frac{P_1}{\mathcal{A}}n_B\mathcal{B}
    \end{aligned}.
\end{equation}
Defining
\begin{gather}
    \tau_{\Pi,0} = \frac{1}{\mathcal{A}} \label{eq:tauPi} ,\\
    \zeta_0 = \frac{P_1}{\mathcal{A}}n_B\mathcal{B} \label{eq:zeta} , \\
    \delta_{\Pi\Pi}=\frac{\tau_{\Pi,0}} {P_1}\left[ \frac{\partial P_1}{\partial \varepsilon}\bigg|_{n_B}(\varepsilon+P|_{\delta \mu=0}) + \left.\frac{\partial P_1}{\partial n_B}\right|_\varepsilon n_B \right] ,
\end{gather}
one can see that Eq.~\eqref{IS deriv 2} becomes 
\begin{equation}
\label{eq:ISfinalform}
    \tau_{\Pi,0} u^\mu \nabla_\mu \Pi + \delta_{\Pi\Pi}\theta\, \Pi + \Pi = - \zeta_0 \theta .
\end{equation}
which is the Israel-Stewart equation with coefficients determined by parameters computed in beta equilibrium.

\newpage
\section{AC bulk viscosity from Israel-Stewart theory}
\label{A: AC IS}

Previous works have considered bulk viscosity to be an emerging, effective phenomenon associated with periodic density perturbations. In that context, the bulk viscosity has a frequency dependence \cite{Harris:2020rus, Sawyer:1989dp, Sad:2009hba}, which here we call $\zeta_{\mathrm{AC}}$. 

One can derive the AC bulk viscosity from Israel-Stewart theory using perturbation of the spacetime metric. In fact, one can define the corresponding Green's function from the linear correction to $T^{\mu\nu}$ in response to a metric perturbation $\delta g_{\mu\nu}\propto e^{-i\omega t}$. This allows us to extract the transport coefficients from the retarded Green's function in Fourier space, using Kubo formulas \cite{Stephanov:2017ghc}.

The metric perturbation induces the following expansion rate
\begin{equation}
   \theta = \frac{ \partial_0 \sqrt{-\det(\eta_{\mu\nu} + \delta g_{\mu\nu})}}{\sqrt{-\det(\eta_{\mu\nu})}} = \frac{\partial_0\sqrt{1 - 2\,\eta^{\alpha\beta}\delta g_{\alpha\beta}}}{1} = -i\omega\,\eta^{\alpha\beta}\delta g_{\alpha\beta}.
\end{equation}
Substituting this result into Eq.~\eqref{eq:ISfinalform}, we find
\begin{equation}
    -i\omega \,\tau_{\Pi,0}\, \Pi + \delta_{\Pi\Pi}\Pi (-i\omega\,\eta^{\alpha\beta}\delta g_{\alpha\beta}) + \Pi = i\omega\,\zeta_0\,\eta^{\alpha\beta}\delta g_{\alpha\beta}.
\end{equation}
But $\Pi$ is a correction to the pressure $\Pi=\delta p = \delta T^i_i/3$, which, to linear order in $\delta g_{\alpha\beta}$, becomes
\begin{equation}
\label{eq:linearresp}
    \frac{1}{3}\delta T^i_i = \frac{i\omega\, \zeta_0}{1-i\omega\,\tau_{\Pi,0}}\eta^{\alpha\beta}\delta g_{\alpha\beta}. 
\end{equation}

From Eq.~\eqref{eq:linearresp}, one can then extract the retarded Green's function:
\begin{equation}
    G^{\theta}_R(\omega) = \frac{i\omega}{1-i\omega \tau_{\Pi,0}}\, \zeta_0\,.
\end{equation}
Note that one can obtain $\zeta_0$ from the Kubo formula
\begin{equation}
\zeta_0 = \lim_{\omega\to 0}\frac{1}{\omega} \operatorname{Im}\,G_R^\theta(\omega).
\end{equation}
The AC bulk viscosity is defined by extending the Kubo formula to nonzero frequencies 
\begin{equation}
\label{eq:kubozeta}
    \zeta_{\textrm{AC}}(\omega) = \frac{1}{\omega} \operatorname{Im}\,G_R^\theta(\omega) = \frac{1}{1+\omega^2 \tau_{\Pi,0}^2}\,\zeta_0\,,
\end{equation}
which correctly recovers the DC bulk viscosity in the limit $\omega\to 0$. 
Replacing Eqs.~\eqref{eq:tauPi} and \eqref{eq:zeta} in Eq.~\eqref{eq:kubozeta}, one finds
\begin{equation}
\label{eq:ACzeta}
    \zeta_{\textrm{AC}}(\omega) = \frac{n_B\,P_1\, \mathcal{A} \, \mathcal{B}}{\mathcal{A}^2+\omega^2}.
\end{equation}
This expression matches the AC bulk viscosity usually employed in the analysis of small-density oscillations, as we show in the next section.
Finally, we note that in the high-frequency limit $\omega\gg \mathcal{A}$, weak interactions become irrelevant, and the electron fraction $Y_e$ becomes an independent variable. For that reason, weak processes no longer contribute to dissipation and $\zeta_{AC}\to 0$.

\newpage
\section{AC bulk viscosity from periodic density oscillations and the equivalence to Israel-Stewart bulk viscosity}
\label{A: equivalence}

It is interesting to note that another way to derive the AC bulk viscosity was used in \cite{Sawyer:1989dp, Sad:2009hba,Harris:2020rus}. We sketch the derivation here as follows. First, one considers a periodic perturbation to the baryon density $n_B(t) = n_0 + Re(\delta n_0 e^{i\omega t})$, which leads to a deviation from $\beta$ equilibrium characterized by $\delta \mu$. 

If one excludes the effects of strong interactions, i.e. the system is never out of thermal equilibrium, the quasi-equilibrium state can be characterized by $n_B$ and $Y_e$,
\be
n_B = n_p+n_n, \;\; Y_e = n_e/n_B,
\ee
Of course, we note that there is also the constraint of charge neutrality,
\be
n_p = n_e, 
\ee
and all the number densities can be found in terms of $n_B$ and $Y_e$,
\be
\label{num def}
n_e = Y_e n_B, \;\; n_p = Y_e n_B, \;\; n_n = (1-Y_e)n_B.
\ee
These number densities can be expressed in terms of the corresponding chemical potentials, $n_i = n_i(\mu_i)$. In beta equilibrium, the three chemical potentials are related as $\mu_n = \mu_p + \mu_e$, so the deviation from equilibrium can be calculated as
\be
\label{out eq}
\begin{aligned}
\delta \mu & = \de \mu_n - \de \mu_p - \de \mu_e \\
& = \frac{\pa \mu_n}{\pa n_n} \de n_n -  \frac{\pa \mu_p}{\pa n_p} \de n_p -  \frac{\pa \mu_e}{\pa n_e} \de n_e.
\end{aligned}
\ee
One can use Eq.~\eqref{num def} for Eq.~\eqref{out eq} to obtain,
\be
\delta \mu = \frac{C}{n_B} \de n_B + B \de Y_e,
\ee
where C and B are given by
\be
\begin{aligned}
C = (n_B-n_e) \frac{\pa \mu_n}{\pa n_n} - n_e \frac{\pa \mu_p}{\pa n_p} - n_e \frac{\pa \mu_e}{\pa n_e}, \\
B = - n_B \left( \frac{\pa \mu_n}{\pa n_n} + \frac{\pa \mu_p}{\pa n_p} + \frac{\pa \mu_e}{\pa n_e} \right).
\end{aligned}
\ee
When $\de \mu$ is nonzero, the two Urca processes would have slightly different rates, and this net effect can be characterized by the change of $Y_e$,
\be
\begin{aligned}
\Ga_e = \Gamma_{\overline{\nu}} - \Gamma_{\nu} = \la \de \mu, \\
n_B \frac{d (\de Y_e)}{dt} = \la \de \mu. 
\end{aligned}
\ee
Thus, we obtain the differential equation,
\be
n_B \frac{d (\de Y_e)}{dt} = \la \left(C \frac{\de n_B}{n_B} + B \de Y_e\right).
\ee
Assuming a similar periodic perturbation for $Y_e$, i.e. $Y_e = Y_{e,0} + Re(\de Y_{e,0} e^{i \om t})$, one can solve the differential equation to find
\be
\de Y_{e,0} = \frac{\de n_B}{n_B} \frac{C}{i\left( \frac{n_B \om}{\la} \right) - B}.
\ee
If we assume the Navier-Stokes equation for a periodic process, the bulk viscosity $\zeta_{AC}$ is defined as the coefficient in the expression for dissipation of Helmholtz's free energy density averaged over one period, $\tau = 2\pi/\om$, 
\be
\lan\dot{\cE}_{diss}\ran = - \frac{\zeta_{AC}}{\tau} \int^\tau_0 dt (\nabla \cdot \vv)^2,
\ee
where $\vv$ is the hydrodynamic velocity associated with density oscillations. Substituting the baryon continuity equation,
\be
\frac{\pa n_B}{\pa t} + n_B \nabla\cdot\vec{v} = 0,
\ee
into the energy-density dissipation equation, we obtain
\be
\lan\dot{\cE}_{diss}\ran = - \frac{\zeta_{AC} \om^2}{2} \left( \frac{\de n_0}{n_B} \right)^2.
\ee
To calculate the energy density dissipation, one needs to consider the $PdV$ work done on the system
\be
\lan\dot{\cE}_{diss}\ran = \frac{n_B}{\tau} \int^\tau_0 P \dot{\cV} dt,
\ee
where $\cV \equiv 1/n_B$ is the specific volume.
Notice that the variations in pressure $P$ can be written as
\be
\begin{aligned}
\de P & = \frac{\pa P}{\pa \mu_p} \de \mu_p + \frac{\pa P}{\pa \mu_n} \de \mu_n + \frac{\pa P}{\pa \mu_e} \de \mu_e \\
& = \frac{\pa P}{\pa n_B} \de n_B - n_B C \de Y_e.
\end{aligned}
\ee
The energy-density dissipation is then
\be
\begin{aligned}
\lan\dot{\cE}_{diss}\ran &= \frac{n_B}{\tau} \int^\tau_0 (P+\de P)\frac{d}{dt}(V+\de V) dt \\
&= -\frac{1}{2} \left( \frac{\de n_0}{n_B} \right)^2 \frac{\la \om^2 C^2}{\om^2 + (\la B/n_B)^2}.
\end{aligned}
\ee
We can now compare the two equations for the energy-density dissipation found above, which determines
\be
\zeta_{AC} = \frac{\la C^2}{\om^2 + (\la B/n_B)^2}.
\ee
Thus, one can see that this bulk viscosity depends on the frequency of the periodic perturbation.

By comparing this result for the bulk viscosity with the one derived from Israel-Stewart formalism, one can see that this derivation only allows two independent variables, $n_B$ and $Y_e$. With these simplifications to the Israel-Stewart formalism, one can show that
\be
\bea
\cA &= - \frac{\la}{n_B} \left.\frac{\partial \delta \mu}{\partial Y_e}\right|_{\varepsilon, n_B} \\
&= - \frac{\la}{n_B} B ,
\eea
\ee
and
\be
\bea
n_B P_1 \cA \cB &= n_B \left.\frac{\pa P}{\pa \delta \mu}\right|_{n_B,\delta \mu = 0} \left(- \frac{\la}{n_B} B\right) \left.\frac{\partial \delta \mu}{\partial n_B}\right|_{Y_e} \\
&= \la C^2 . 
\eea
\ee
Therefore, we finally arrived at
\be
\zeta_{AC} = \frac{\la C^2}{\om^2 + (\la B/n_B)^2} = \frac{n_B\,P_1\, \mathcal{A} \, \mathcal{B}}{\mathcal{A}^2+\omega^2},
\ee
which demonstrates that the two methods to derive the frequency-dependent bulk viscosity discussed in this work agree with each other.

\newpage
\section{Procedure to obtain our EoS from thermal field theory}
\label{A: RMF EoS}

In this appendix, we review the calculation of the EoS in a relativistic mean-field model.

To obtain our EoS from thermal field theory, for convenience, we momentarily consider a small temperature $T$, which will be later taken to zero.  In that case, the partition function can be rewritten as an Euclidean path integral \cite{Kapusta:2006pm}
\be
Z = 
\displaystyle\int
[d\phi]\,\exp\left[ \int^{1/T}_0 d\tau \int d^3x \left( \cL + \mu_i \,\cN_i
\right) \right],
\ee
where $\phi$ represents all the fields under consideration,\footnote{The boundary conditions in the compactified imaginary time direction $\tau$ are periodic for bosons and anti-periodic for fermions.} 
$\mu_i$ is the chemical potential for particle species $i$, and $\cN_i$ is the charge density corresponding to $\mu_i$.

For all of our RMF EoSs, the Lagrangian is of the following form,
\be
\cL = \cL_N + \cL_M + \cL_l,
\ee
where $\cL_M$, $\cL_N$, and $\cL_l$ are the Lagrangians of mesons, nucleons (including interactions), and leptons, respectively.

We are interested in uniform matter in its ground state, so we replace the meson fields by their mean values in this state such that
\begin{gather}
\si \to \bsi , \\
\om^0 \to \bom , \\
\rho^0_i \to \brho\, \delta_{i3}.
\end{gather}
We can assume isotropy of space, so the spatial part of vector mesons vanishes:
\begin{gather}
\om^i \to 0 , \\
\rho^i_j \to 0.
\end{gather}
Under these approximations, the nucleons behave as a free gas with effective mass and chemical potentials given by
\begin{gather}
    m^*_B = m_B - g_\si \Bar{\si} , \\
    \mu_B^* = \mu_B - g_\om \Bar{\om} , \\
    \mu_I^* = \mu_I - g_\rho \Bar{\rho} ,
\end{gather}
where $m_B = m_N$ is the baryon/nucleon mass, and $\mu_B$ and $\mu_I$ are chemical potentials for baryon and isospin, respectively. 

At zero temperature, or in the Fermi surface approximation, the effective chemical potentials for proton and neutron can be found from their respective densities, $n_p = n_B/2 + n_I$ and $n_n = n_B/2 - n_I$, where $n_I$ is the isospin density. 

The Fermi momenta $k^F_{p.n}$ and the effective chemical potentials $\mu_{p.n}^*$ are given by
\begin{gather}
    k^F_{p.n} = 2\pi \left( \frac{3}{8\pi} n_{p,n} \right)^{1/3} , \\
    \mu_{p.n}^* = \sqrt{k^{F\;2}_{p.n} + m^{*\;2}_B} .
\end{gather} 
One can then find the effective baryon and isospin chemical potentials, respectively, 
\begin{gather}
    \mu_B^* = \frac{\mu_p^* + \mu_n^*}{2} , \\
    \mu_I^* = \mu_p^* - \mu_n^* .
\end{gather}

In the mean-field approximation, the meson part $\mathcal{L}_M=\mathcal{L}_M(\bar\sigma,\bar\omega,\bar \rho)$ (excluding interaction with nucleons) is constant, and we can write
\be
\bea
Z &= 
 \int
 [d\phi]\exp\left[ \int^{1/T}_0 d\tau \int d^3x \left( \cL_N + \cL_l + \mu_i \,\cN_i
 \right) \right] \exp \left[ \frac{V}{T} \cL_M \right] .
\eea
\ee

The pressure
is obtained as follows
\be
\bea
P &= \frac{\pa (T \ln Z)}{\pa V} \\
&= \frac{\pa}{\pa V} \left( T \ln 
\int
[d\phi]\exp\left[ \int^{\beta}_0 d\tau \int d^3x \left( \cL_N + \cL_l + \mu_i\, \cN_i
\right) \right] \right) + \cL_M,
\eea
\ee
where $V$ is the volume of the system.
We can now take the $T\to 0$ limit, in which case we can analytically compute the fermionic contribution to the pressure. In practice, the EoS takes the form
\be
P = P_{free}(\mu_{p.n}^*, m_B^*) + P_{free}(\mu_e, m_e) + \cL_M(\bar \sigma, \bar\omega, \bar\rho) ,
\ee
where $P_{free}(\mu_{p.n}^*, m_B^*)$ is the pressure of a free gas of nucleons with the effective parameters, and $P_{free}(\mu_e, m_e)$ is the pressure of a free gas of electrons, both computed at zero temperature. The values of the condensates are found by maximizing the pressure with respect to $\bar \sigma$, $\bar\omega$,  and $\bar\rho$.

\bibliography{references,notInspire}

\begin{thebibliography}{90}%
\makeatletter
\providecommand \@ifxundefined [1]{%
 \@ifx{#1\undefined}
}%
\providecommand \@ifnum [1]{%
 \ifnum #1\expandafter \@firstoftwo
 \else \expandafter \@secondoftwo
 \fi
}%
\providecommand \@ifx [1]{%
 \ifx #1\expandafter \@firstoftwo
 \else \expandafter \@secondoftwo
 \fi
}%
\providecommand \natexlab [1]{#1}%
\providecommand \enquote  [1]{``#1''}%
\providecommand \bibnamefont  [1]{#1}%
\providecommand \bibfnamefont [1]{#1}%
\providecommand \citenamefont [1]{#1}%
\providecommand \href@noop [0]{\@secondoftwo}%
\providecommand \href [0]{\begingroup \@sanitize@url \@href}%
\providecommand \@href[1]{\@@startlink{#1}\@@href}%
\providecommand \@@href[1]{\endgroup#1\@@endlink}%
\providecommand \@sanitize@url [0]{\catcode `\\12\catcode `\$12\catcode
  `\&12\catcode `\#12\catcode `\^12\catcode `\_12\catcode `\%12\relax}%
\providecommand \@@startlink[1]{}%
\providecommand \@@endlink[0]{}%
\providecommand \url  [0]{\begingroup\@sanitize@url \@url }%
\providecommand \@url [1]{\endgroup\@href {#1}{\urlprefix }}%
\providecommand \urlprefix  [0]{URL }%
\providecommand \Eprint [0]{\href }%
\providecommand \doibase [0]{https://doi.org/}%
\providecommand \selectlanguage [0]{\@gobble}%
\providecommand \bibinfo  [0]{\@secondoftwo}%
\providecommand \bibfield  [0]{\@secondoftwo}%
\providecommand \translation [1]{[#1]}%
\providecommand \BibitemOpen [0]{}%
\providecommand \bibitemStop [0]{}%
\providecommand \bibitemNoStop [0]{.\EOS\space}%
\providecommand \EOS [0]{\spacefactor3000\relax}%
\providecommand \BibitemShut  [1]{\csname bibitem#1\endcsname}%
\let\auto@bib@innerbib\@empty
\bibitem [{\citenamefont {Abbott}\ \emph {et~al.}(2017)\citenamefont {Abbott}
  \emph {et~al.}}]{LIGOScientific:2017vwq}%
  \BibitemOpen
  \bibfield  {author} {\bibinfo {author} {\bibfnamefont {B.~P.}\ \bibnamefont
  {Abbott}} \emph {et~al.} (\bibinfo {collaboration} {LIGO Scientific,
  Virgo}),\ }\bibfield  {title} {\bibinfo {title} {{GW170817: Observation of
  Gravitational Waves from a Binary Neutron Star Inspiral}},\ }\href
  {https://doi.org/10.1103/PhysRevLett.119.161101} {\bibfield  {journal}
  {\bibinfo  {journal} {Phys. Rev. Lett.}\ }\textbf {\bibinfo {volume} {119}},\
  \bibinfo {pages} {161101} (\bibinfo {year} {2017})},\ \Eprint
  {https://arxiv.org/abs/1710.05832} {arXiv:1710.05832 [gr-qc]} \BibitemShut
  {NoStop}%
\bibitem [{\citenamefont {Abbott}\ \emph {et~al.}(2020)\citenamefont {Abbott}
  \emph {et~al.}}]{LIGOScientific:2020aai}%
  \BibitemOpen
  \bibfield  {author} {\bibinfo {author} {\bibfnamefont {B.~P.}\ \bibnamefont
  {Abbott}} \emph {et~al.} (\bibinfo {collaboration} {LIGO Scientific,
  Virgo}),\ }\bibfield  {title} {\bibinfo {title} {{GW190425: Observation of a
  Compact Binary Coalescence with Total Mass $\sim 3.4 M_{\odot}$}},\ }\href
  {https://doi.org/10.3847/2041-8213/ab75f5} {\bibfield  {journal} {\bibinfo
  {journal} {Astrophys. J. Lett.}\ }\textbf {\bibinfo {volume} {892}},\
  \bibinfo {pages} {L3} (\bibinfo {year} {2020})},\ \Eprint
  {https://arxiv.org/abs/2001.01761} {arXiv:2001.01761 [astro-ph.HE]}
  \BibitemShut {NoStop}%
\bibitem [{\citenamefont {Gandolfi}\ \emph {et~al.}(2012)\citenamefont
  {Gandolfi}, \citenamefont {Carlson},\ and\ \citenamefont
  {Reddy}}]{Gandolfi:2011xu}%
  \BibitemOpen
  \bibfield  {author} {\bibinfo {author} {\bibfnamefont {S.}~\bibnamefont
  {Gandolfi}}, \bibinfo {author} {\bibfnamefont {J.}~\bibnamefont {Carlson}},\
  and\ \bibinfo {author} {\bibfnamefont {S.}~\bibnamefont {Reddy}},\ }\bibfield
   {title} {\bibinfo {title} {{The maximum mass and radius of neutron stars and
  the nuclear symmetry energy}},\ }\href
  {https://doi.org/10.1103/PhysRevC.85.032801} {\bibfield  {journal} {\bibinfo
  {journal} {Phys. Rev. C}\ }\textbf {\bibinfo {volume} {85}},\ \bibinfo
  {pages} {032801} (\bibinfo {year} {2012})},\ \Eprint
  {https://arxiv.org/abs/1101.1921} {arXiv:1101.1921 [nucl-th]} \BibitemShut
  {NoStop}%
\bibitem [{\citenamefont {Zhang}\ and\ \citenamefont
  {Chen}(2015)}]{Zhang:2015ava}%
  \BibitemOpen
  \bibfield  {author} {\bibinfo {author} {\bibfnamefont {Z.}~\bibnamefont
  {Zhang}}\ and\ \bibinfo {author} {\bibfnamefont {L.-W.}\ \bibnamefont
  {Chen}},\ }\bibfield  {title} {\bibinfo {title} {{Electric dipole
  polarizability in $^{208}$Pb as a probe of the symmetry energy and neutron
  matter around $\rho_0/3$}},\ }\href
  {https://doi.org/10.1103/PhysRevC.92.031301} {\bibfield  {journal} {\bibinfo
  {journal} {Phys. Rev. C}\ }\textbf {\bibinfo {volume} {92}},\ \bibinfo
  {pages} {031301} (\bibinfo {year} {2015})},\ \Eprint
  {https://arxiv.org/abs/1504.01077} {arXiv:1504.01077 [nucl-th]} \BibitemShut
  {NoStop}%
\bibitem [{\citenamefont {Kowalski}\ \emph {et~al.}(2007)\citenamefont
  {Kowalski} \emph {et~al.}}]{Kowalski:2006ju}%
  \BibitemOpen
  \bibfield  {author} {\bibinfo {author} {\bibfnamefont {S.}~\bibnamefont
  {Kowalski}} \emph {et~al.},\ }\bibfield  {title} {\bibinfo {title}
  {{Experimental determination of the symmetry energy of a low density nuclear
  gas}},\ }\href {https://doi.org/10.1103/PhysRevC.75.014601} {\bibfield
  {journal} {\bibinfo  {journal} {Phys. Rev. C}\ }\textbf {\bibinfo {volume}
  {75}},\ \bibinfo {pages} {014601} (\bibinfo {year} {2007})},\ \Eprint
  {https://arxiv.org/abs/nucl-ex/0602023} {arXiv:nucl-ex/0602023} \BibitemShut
  {NoStop}%
\bibitem [{\citenamefont {Roca-Maza}\ \emph {et~al.}(2013)\citenamefont
  {Roca-Maza}, \citenamefont {Brenna}, \citenamefont {Agrawal}, \citenamefont
  {Bortignon}, \citenamefont {Col\`o}, \citenamefont {Cao}, \citenamefont
  {Paar},\ and\ \citenamefont {Vretenar}}]{Roca-Maza:2012uor}%
  \BibitemOpen
  \bibfield  {author} {\bibinfo {author} {\bibfnamefont {X.}~\bibnamefont
  {Roca-Maza}}, \bibinfo {author} {\bibfnamefont {M.}~\bibnamefont {Brenna}},
  \bibinfo {author} {\bibfnamefont {B.~K.}\ \bibnamefont {Agrawal}}, \bibinfo
  {author} {\bibfnamefont {P.~F.}\ \bibnamefont {Bortignon}}, \bibinfo {author}
  {\bibfnamefont {G.}~\bibnamefont {Col\`o}}, \bibinfo {author} {\bibfnamefont
  {L.-G.}\ \bibnamefont {Cao}}, \bibinfo {author} {\bibfnamefont
  {N.}~\bibnamefont {Paar}},\ and\ \bibinfo {author} {\bibfnamefont
  {D.}~\bibnamefont {Vretenar}},\ }\bibfield  {title} {\bibinfo {title} {{Giant
  Quadrupole Resonances in 208Pb, the nuclear symmetry energy and the neutron
  skin thickness}},\ }\href {https://doi.org/10.1103/PhysRevC.87.034301}
  {\bibfield  {journal} {\bibinfo  {journal} {Phys. Rev. C}\ }\textbf {\bibinfo
  {volume} {87}},\ \bibinfo {pages} {034301} (\bibinfo {year} {2013})},\
  \Eprint {https://arxiv.org/abs/1212.4377} {arXiv:1212.4377 [nucl-th]}
  \BibitemShut {NoStop}%
\bibitem [{\citenamefont {Zhang}\ and\ \citenamefont
  {Chen}(2013)}]{Zhang:2013wna}%
  \BibitemOpen
  \bibfield  {author} {\bibinfo {author} {\bibfnamefont {Z.}~\bibnamefont
  {Zhang}}\ and\ \bibinfo {author} {\bibfnamefont {L.-W.}\ \bibnamefont
  {Chen}},\ }\bibfield  {title} {\bibinfo {title} {{Constraining the symmetry
  energy at subsaturation densities using isotope binding energy difference and
  neutron skin thickness}},\ }\href
  {https://doi.org/10.1016/j.physletb.2013.08.002} {\bibfield  {journal}
  {\bibinfo  {journal} {Phys. Lett. B}\ }\textbf {\bibinfo {volume} {726}},\
  \bibinfo {pages} {234} (\bibinfo {year} {2013})},\ \Eprint
  {https://arxiv.org/abs/1302.5327} {arXiv:1302.5327 [nucl-th]} \BibitemShut
  {NoStop}%
\bibitem [{\citenamefont {Fan}\ \emph {et~al.}(2014)\citenamefont {Fan},
  \citenamefont {Dong},\ and\ \citenamefont {Zuo}}]{Fan:2014rha}%
  \BibitemOpen
  \bibfield  {author} {\bibinfo {author} {\bibfnamefont {X.}~\bibnamefont
  {Fan}}, \bibinfo {author} {\bibfnamefont {J.}~\bibnamefont {Dong}},\ and\
  \bibinfo {author} {\bibfnamefont {W.}~\bibnamefont {Zuo}},\ }\bibfield
  {title} {\bibinfo {title} {{Density-dependent symmetry energy at
  subsaturation densities from nuclear mass differences}},\ }\href
  {https://doi.org/10.1103/PhysRevC.89.017305} {\bibfield  {journal} {\bibinfo
  {journal} {Phys. Rev. C}\ }\textbf {\bibinfo {volume} {89}},\ \bibinfo
  {pages} {017305} (\bibinfo {year} {2014})},\ \Eprint
  {https://arxiv.org/abs/1403.2055} {arXiv:1403.2055 [nucl-th]} \BibitemShut
  {NoStop}%
\bibitem [{\citenamefont {Danielewicz}\ and\ \citenamefont
  {Lee}(2014)}]{Danielewicz:2013upa}%
  \BibitemOpen
  \bibfield  {author} {\bibinfo {author} {\bibfnamefont {P.}~\bibnamefont
  {Danielewicz}}\ and\ \bibinfo {author} {\bibfnamefont {J.}~\bibnamefont
  {Lee}},\ }\bibfield  {title} {\bibinfo {title} {{Symmetry Energy II: Isobaric
  Analog States}},\ }\href {https://doi.org/10.1016/j.nuclphysa.2013.11.005}
  {\bibfield  {journal} {\bibinfo  {journal} {Nucl. Phys. A}\ }\textbf
  {\bibinfo {volume} {922}},\ \bibinfo {pages} {1} (\bibinfo {year} {2014})},\
  \Eprint {https://arxiv.org/abs/1307.4130} {arXiv:1307.4130 [nucl-th]}
  \BibitemShut {NoStop}%
\bibitem [{\citenamefont {Reinhard}\ \emph {et~al.}(2021)\citenamefont
  {Reinhard}, \citenamefont {Roca-Maza},\ and\ \citenamefont
  {Nazarewicz}}]{Reinhard:2021utv}%
  \BibitemOpen
  \bibfield  {author} {\bibinfo {author} {\bibfnamefont {P.-G.}\ \bibnamefont
  {Reinhard}}, \bibinfo {author} {\bibfnamefont {X.}~\bibnamefont
  {Roca-Maza}},\ and\ \bibinfo {author} {\bibfnamefont {W.}~\bibnamefont
  {Nazarewicz}},\ }\bibfield  {title} {\bibinfo {title} {{Information Content
  of the Parity-Violating Asymmetry in Pb208}},\ }\href
  {https://doi.org/10.1103/PhysRevLett.127.232501} {\bibfield  {journal}
  {\bibinfo  {journal} {Phys. Rev. Lett.}\ }\textbf {\bibinfo {volume} {127}},\
  \bibinfo {pages} {232501} (\bibinfo {year} {2021})},\ \Eprint
  {https://arxiv.org/abs/2105.15050} {arXiv:2105.15050 [nucl-th]} \BibitemShut
  {NoStop}%
\bibitem [{\citenamefont {Reed}\ \emph {et~al.}(2021)\citenamefont {Reed},
  \citenamefont {Fattoyev}, \citenamefont {Horowitz},\ and\ \citenamefont
  {Piekarewicz}}]{Reed:2021nqk}%
  \BibitemOpen
  \bibfield  {author} {\bibinfo {author} {\bibfnamefont {B.~T.}\ \bibnamefont
  {Reed}}, \bibinfo {author} {\bibfnamefont {F.~J.}\ \bibnamefont {Fattoyev}},
  \bibinfo {author} {\bibfnamefont {C.~J.}\ \bibnamefont {Horowitz}},\ and\
  \bibinfo {author} {\bibfnamefont {J.}~\bibnamefont {Piekarewicz}},\
  }\bibfield  {title} {\bibinfo {title} {{Implications of PREX-2 on the
  Equation of State of Neutron-Rich Matter}},\ }\href
  {https://doi.org/10.1103/PhysRevLett.126.172503} {\bibfield  {journal}
  {\bibinfo  {journal} {Phys. Rev. Lett.}\ }\textbf {\bibinfo {volume} {126}},\
  \bibinfo {pages} {172503} (\bibinfo {year} {2021})},\ \Eprint
  {https://arxiv.org/abs/2101.03193} {arXiv:2101.03193 [nucl-th]} \BibitemShut
  {NoStop}%
\bibitem [{\citenamefont {Bauswein}\ \emph {et~al.}(2017)\citenamefont
  {Bauswein}, \citenamefont {Just}, \citenamefont {Janka},\ and\ \citenamefont
  {Stergioulas}}]{Bauswein:2017vtn}%
  \BibitemOpen
  \bibfield  {author} {\bibinfo {author} {\bibfnamefont {A.}~\bibnamefont
  {Bauswein}}, \bibinfo {author} {\bibfnamefont {O.}~\bibnamefont {Just}},
  \bibinfo {author} {\bibfnamefont {H.-T.}\ \bibnamefont {Janka}},\ and\
  \bibinfo {author} {\bibfnamefont {N.}~\bibnamefont {Stergioulas}},\
  }\bibfield  {title} {\bibinfo {title} {{Neutron-star radius constraints from
  GW170817 and future detections}},\ }\href
  {https://doi.org/10.3847/2041-8213/aa9994} {\bibfield  {journal} {\bibinfo
  {journal} {Astrophys. J. Lett.}\ }\textbf {\bibinfo {volume} {850}},\
  \bibinfo {pages} {L34} (\bibinfo {year} {2017})},\ \Eprint
  {https://arxiv.org/abs/1710.06843} {arXiv:1710.06843 [astro-ph.HE]}
  \BibitemShut {NoStop}%
\bibitem [{\citenamefont {Annala}\ \emph {et~al.}(2018)\citenamefont {Annala},
  \citenamefont {Gorda}, \citenamefont {Kurkela},\ and\ \citenamefont
  {Vuorinen}}]{Annala:2017llu}%
  \BibitemOpen
  \bibfield  {author} {\bibinfo {author} {\bibfnamefont {E.}~\bibnamefont
  {Annala}}, \bibinfo {author} {\bibfnamefont {T.}~\bibnamefont {Gorda}},
  \bibinfo {author} {\bibfnamefont {A.}~\bibnamefont {Kurkela}},\ and\ \bibinfo
  {author} {\bibfnamefont {A.}~\bibnamefont {Vuorinen}},\ }\bibfield  {title}
  {\bibinfo {title} {{Gravitational-wave constraints on the neutron-star-matter
  Equation of State}},\ }\href {https://doi.org/10.1103/PhysRevLett.120.172703}
  {\bibfield  {journal} {\bibinfo  {journal} {Phys. Rev. Lett.}\ }\textbf
  {\bibinfo {volume} {120}},\ \bibinfo {pages} {172703} (\bibinfo {year}
  {2018})},\ \Eprint {https://arxiv.org/abs/1711.02644} {arXiv:1711.02644
  [astro-ph.HE]} \BibitemShut {NoStop}%
\bibitem [{\citenamefont {Most}\ \emph {et~al.}(2018)\citenamefont {Most},
  \citenamefont {Weih}, \citenamefont {Rezzolla},\ and\ \citenamefont
  {Schaffner-Bielich}}]{Most:2018hfd}%
  \BibitemOpen
  \bibfield  {author} {\bibinfo {author} {\bibfnamefont {E.~R.}\ \bibnamefont
  {Most}}, \bibinfo {author} {\bibfnamefont {L.~R.}\ \bibnamefont {Weih}},
  \bibinfo {author} {\bibfnamefont {L.}~\bibnamefont {Rezzolla}},\ and\
  \bibinfo {author} {\bibfnamefont {J.}~\bibnamefont {Schaffner-Bielich}},\
  }\bibfield  {title} {\bibinfo {title} {{New constraints on radii and tidal
  deformabilities of neutron stars from GW170817}},\ }\href
  {https://doi.org/10.1103/PhysRevLett.120.261103} {\bibfield  {journal}
  {\bibinfo  {journal} {Phys. Rev. Lett.}\ }\textbf {\bibinfo {volume} {120}},\
  \bibinfo {pages} {261103} (\bibinfo {year} {2018})},\ \Eprint
  {https://arxiv.org/abs/1803.00549} {arXiv:1803.00549 [gr-qc]} \BibitemShut
  {NoStop}%
\bibitem [{\citenamefont {Abbott}\ \emph {et~al.}(2018)\citenamefont {Abbott}
  \emph {et~al.}}]{LIGOScientific:2018cki}%
  \BibitemOpen
  \bibfield  {author} {\bibinfo {author} {\bibfnamefont {B.~P.}\ \bibnamefont
  {Abbott}} \emph {et~al.} (\bibinfo {collaboration} {LIGO Scientific,
  Virgo}),\ }\bibfield  {title} {\bibinfo {title} {{GW170817: Measurements of
  neutron star radii and equation of state}},\ }\href
  {https://doi.org/10.1103/PhysRevLett.121.161101} {\bibfield  {journal}
  {\bibinfo  {journal} {Phys. Rev. Lett.}\ }\textbf {\bibinfo {volume} {121}},\
  \bibinfo {pages} {161101} (\bibinfo {year} {2018})},\ \Eprint
  {https://arxiv.org/abs/1805.11581} {arXiv:1805.11581 [gr-qc]} \BibitemShut
  {NoStop}%
\bibitem [{\citenamefont {Raithel}\ \emph {et~al.}(2018)\citenamefont
  {Raithel}, \citenamefont {\"Ozel},\ and\ \citenamefont
  {Psaltis}}]{Raithel:2018ncd}%
  \BibitemOpen
  \bibfield  {author} {\bibinfo {author} {\bibfnamefont {C.}~\bibnamefont
  {Raithel}}, \bibinfo {author} {\bibfnamefont {F.}~\bibnamefont {\"Ozel}},\
  and\ \bibinfo {author} {\bibfnamefont {D.}~\bibnamefont {Psaltis}},\
  }\bibfield  {title} {\bibinfo {title} {{Tidal deformability from GW170817 as
  a direct probe of the neutron star radius}},\ }\href
  {https://doi.org/10.3847/2041-8213/aabcbf} {\bibfield  {journal} {\bibinfo
  {journal} {Astrophys. J. Lett.}\ }\textbf {\bibinfo {volume} {857}},\
  \bibinfo {pages} {L23} (\bibinfo {year} {2018})},\ \Eprint
  {https://arxiv.org/abs/1803.07687} {arXiv:1803.07687 [astro-ph.HE]}
  \BibitemShut {NoStop}%
\bibitem [{\citenamefont {De}\ \emph {et~al.}(2018)\citenamefont {De},
  \citenamefont {Finstad}, \citenamefont {Lattimer}, \citenamefont {Brown},
  \citenamefont {Berger},\ and\ \citenamefont {Biwer}}]{De:2018uhw}%
  \BibitemOpen
  \bibfield  {author} {\bibinfo {author} {\bibfnamefont {S.}~\bibnamefont
  {De}}, \bibinfo {author} {\bibfnamefont {D.}~\bibnamefont {Finstad}},
  \bibinfo {author} {\bibfnamefont {J.~M.}\ \bibnamefont {Lattimer}}, \bibinfo
  {author} {\bibfnamefont {D.~A.}\ \bibnamefont {Brown}}, \bibinfo {author}
  {\bibfnamefont {E.}~\bibnamefont {Berger}},\ and\ \bibinfo {author}
  {\bibfnamefont {C.~M.}\ \bibnamefont {Biwer}},\ }\bibfield  {title} {\bibinfo
  {title} {{Tidal Deformabilities and Radii of Neutron Stars from the
  Observation of GW170817}},\ }\href
  {https://doi.org/10.1103/PhysRevLett.121.091102} {\bibfield  {journal}
  {\bibinfo  {journal} {Phys. Rev. Lett.}\ }\textbf {\bibinfo {volume} {121}},\
  \bibinfo {pages} {091102} (\bibinfo {year} {2018})},\ \bibinfo {note}
  {[Erratum: Phys.Rev.Lett. 121, 259902 (2018)]},\ \Eprint
  {https://arxiv.org/abs/1804.08583} {arXiv:1804.08583 [astro-ph.HE]}
  \BibitemShut {NoStop}%
\bibitem [{\citenamefont {Chatziioannou}\ \emph {et~al.}(2018)\citenamefont
  {Chatziioannou}, \citenamefont {Haster},\ and\ \citenamefont
  {Zimmerman}}]{Chatziioannou:2018vzf}%
  \BibitemOpen
  \bibfield  {author} {\bibinfo {author} {\bibfnamefont {K.}~\bibnamefont
  {Chatziioannou}}, \bibinfo {author} {\bibfnamefont {C.-J.}\ \bibnamefont
  {Haster}},\ and\ \bibinfo {author} {\bibfnamefont {A.}~\bibnamefont
  {Zimmerman}},\ }\bibfield  {title} {\bibinfo {title} {{Measuring the neutron
  star tidal deformability with equation-of-state-independent relations and
  gravitational waves}},\ }\href {https://doi.org/10.1103/PhysRevD.97.104036}
  {\bibfield  {journal} {\bibinfo  {journal} {Phys. Rev. D}\ }\textbf {\bibinfo
  {volume} {97}},\ \bibinfo {pages} {104036} (\bibinfo {year} {2018})},\
  \Eprint {https://arxiv.org/abs/1804.03221} {arXiv:1804.03221 [gr-qc]}
  \BibitemShut {NoStop}%
\bibitem [{\citenamefont {Carson}\ \emph {et~al.}(2019)\citenamefont {Carson},
  \citenamefont {Steiner},\ and\ \citenamefont {Yagi}}]{Carson:2018xri}%
  \BibitemOpen
  \bibfield  {author} {\bibinfo {author} {\bibfnamefont {Z.}~\bibnamefont
  {Carson}}, \bibinfo {author} {\bibfnamefont {A.~W.}\ \bibnamefont
  {Steiner}},\ and\ \bibinfo {author} {\bibfnamefont {K.}~\bibnamefont
  {Yagi}},\ }\bibfield  {title} {\bibinfo {title} {{Constraining nuclear matter
  parameters with GW170817}},\ }\href
  {https://doi.org/10.1103/PhysRevD.99.043010} {\bibfield  {journal} {\bibinfo
  {journal} {Phys. Rev. D}\ }\textbf {\bibinfo {volume} {99}},\ \bibinfo
  {pages} {043010} (\bibinfo {year} {2019})},\ \Eprint
  {https://arxiv.org/abs/1812.08910} {arXiv:1812.08910 [gr-qc]} \BibitemShut
  {NoStop}%
\bibitem [{\citenamefont {Radice}\ \emph {et~al.}(2018)\citenamefont {Radice},
  \citenamefont {Perego}, \citenamefont {Zappa},\ and\ \citenamefont
  {Bernuzzi}}]{Radice:2017lry}%
  \BibitemOpen
  \bibfield  {author} {\bibinfo {author} {\bibfnamefont {D.}~\bibnamefont
  {Radice}}, \bibinfo {author} {\bibfnamefont {A.}~\bibnamefont {Perego}},
  \bibinfo {author} {\bibfnamefont {F.}~\bibnamefont {Zappa}},\ and\ \bibinfo
  {author} {\bibfnamefont {S.}~\bibnamefont {Bernuzzi}},\ }\bibfield  {title}
  {\bibinfo {title} {{GW170817: Joint Constraint on the Neutron Star Equation
  of State from Multimessenger Observations}},\ }\href
  {https://doi.org/10.3847/2041-8213/aaa402} {\bibfield  {journal} {\bibinfo
  {journal} {Astrophys. J. Lett.}\ }\textbf {\bibinfo {volume} {852}},\
  \bibinfo {pages} {L29} (\bibinfo {year} {2018})},\ \Eprint
  {https://arxiv.org/abs/1711.03647} {arXiv:1711.03647 [astro-ph.HE]}
  \BibitemShut {NoStop}%
\bibitem [{\citenamefont {Margalit}\ and\ \citenamefont
  {Metzger}(2017)}]{Margalit:2017dij}%
  \BibitemOpen
  \bibfield  {author} {\bibinfo {author} {\bibfnamefont {B.}~\bibnamefont
  {Margalit}}\ and\ \bibinfo {author} {\bibfnamefont {B.~D.}\ \bibnamefont
  {Metzger}},\ }\bibfield  {title} {\bibinfo {title} {{Constraining the Maximum
  Mass of Neutron Stars From Multi-Messenger Observations of GW170817}},\
  }\href {https://doi.org/10.3847/2041-8213/aa991c} {\bibfield  {journal}
  {\bibinfo  {journal} {Astrophys. J. Lett.}\ }\textbf {\bibinfo {volume}
  {850}},\ \bibinfo {pages} {L19} (\bibinfo {year} {2017})},\ \Eprint
  {https://arxiv.org/abs/1710.05938} {arXiv:1710.05938 [astro-ph.HE]}
  \BibitemShut {NoStop}%
\bibitem [{\citenamefont {Rezzolla}\ \emph {et~al.}(2018)\citenamefont
  {Rezzolla}, \citenamefont {Most},\ and\ \citenamefont
  {Weih}}]{Rezzolla:2017aly}%
  \BibitemOpen
  \bibfield  {author} {\bibinfo {author} {\bibfnamefont {L.}~\bibnamefont
  {Rezzolla}}, \bibinfo {author} {\bibfnamefont {E.~R.}\ \bibnamefont {Most}},\
  and\ \bibinfo {author} {\bibfnamefont {L.~R.}\ \bibnamefont {Weih}},\
  }\bibfield  {title} {\bibinfo {title} {{Using gravitational-wave observations
  and quasi-universal relations to constrain the maximum mass of neutron
  stars}},\ }\href {https://doi.org/10.3847/2041-8213/aaa401} {\bibfield
  {journal} {\bibinfo  {journal} {Astrophys. J. Lett.}\ }\textbf {\bibinfo
  {volume} {852}},\ \bibinfo {pages} {L25} (\bibinfo {year} {2018})},\ \Eprint
  {https://arxiv.org/abs/1711.00314} {arXiv:1711.00314 [astro-ph.HE]}
  \BibitemShut {NoStop}%
\bibitem [{\citenamefont {Ruiz}\ \emph {et~al.}(2018)\citenamefont {Ruiz},
  \citenamefont {Shapiro},\ and\ \citenamefont {Tsokaros}}]{Ruiz:2017due}%
  \BibitemOpen
  \bibfield  {author} {\bibinfo {author} {\bibfnamefont {M.}~\bibnamefont
  {Ruiz}}, \bibinfo {author} {\bibfnamefont {S.~L.}\ \bibnamefont {Shapiro}},\
  and\ \bibinfo {author} {\bibfnamefont {A.}~\bibnamefont {Tsokaros}},\
  }\bibfield  {title} {\bibinfo {title} {{GW170817, General Relativistic
  Magnetohydrodynamic Simulations, and the Neutron Star Maximum Mass}},\ }\href
  {https://doi.org/10.1103/PhysRevD.97.021501} {\bibfield  {journal} {\bibinfo
  {journal} {Phys. Rev. D}\ }\textbf {\bibinfo {volume} {97}},\ \bibinfo
  {pages} {021501} (\bibinfo {year} {2018})},\ \Eprint
  {https://arxiv.org/abs/1711.00473} {arXiv:1711.00473 [astro-ph.HE]}
  \BibitemShut {NoStop}%
\bibitem [{\citenamefont {Shibata}\ \emph {et~al.}(2019)\citenamefont
  {Shibata}, \citenamefont {Zhou}, \citenamefont {Kiuchi},\ and\ \citenamefont
  {Fujibayashi}}]{Shibata:2019ctb}%
  \BibitemOpen
  \bibfield  {author} {\bibinfo {author} {\bibfnamefont {M.}~\bibnamefont
  {Shibata}}, \bibinfo {author} {\bibfnamefont {E.}~\bibnamefont {Zhou}},
  \bibinfo {author} {\bibfnamefont {K.}~\bibnamefont {Kiuchi}},\ and\ \bibinfo
  {author} {\bibfnamefont {S.}~\bibnamefont {Fujibayashi}},\ }\bibfield
  {title} {\bibinfo {title} {{Constraint on the maximum mass of neutron stars
  using GW170817 event}},\ }\href {https://doi.org/10.1103/PhysRevD.100.023015}
  {\bibfield  {journal} {\bibinfo  {journal} {Phys. Rev. D}\ }\textbf {\bibinfo
  {volume} {100}},\ \bibinfo {pages} {023015} (\bibinfo {year} {2019})},\
  \Eprint {https://arxiv.org/abs/1905.03656} {arXiv:1905.03656 [astro-ph.HE]}
  \BibitemShut {NoStop}%
\bibitem [{\citenamefont {Kastaun}\ \emph {et~al.}(2016)\citenamefont
  {Kastaun}, \citenamefont {Ciolfi},\ and\ \citenamefont
  {Giacomazzo}}]{Kastaun:2016yaf}%
  \BibitemOpen
  \bibfield  {author} {\bibinfo {author} {\bibfnamefont {W.}~\bibnamefont
  {Kastaun}}, \bibinfo {author} {\bibfnamefont {R.}~\bibnamefont {Ciolfi}},\
  and\ \bibinfo {author} {\bibfnamefont {B.}~\bibnamefont {Giacomazzo}},\
  }\bibfield  {title} {\bibinfo {title} {{Structure of Stable Binary Neutron
  Star Merger Remnants: a Case Study}},\ }\href
  {https://doi.org/10.1103/PhysRevD.94.044060} {\bibfield  {journal} {\bibinfo
  {journal} {Phys. Rev. D}\ }\textbf {\bibinfo {volume} {94}},\ \bibinfo
  {pages} {044060} (\bibinfo {year} {2016})},\ \Eprint
  {https://arxiv.org/abs/1607.02186} {arXiv:1607.02186 [astro-ph.HE]}
  \BibitemShut {NoStop}%
\bibitem [{\citenamefont {Hanauske}\ \emph {et~al.}(2017)\citenamefont
  {Hanauske}, \citenamefont {Takami}, \citenamefont {Bovard}, \citenamefont
  {Rezzolla}, \citenamefont {Font}, \citenamefont {Galeazzi},\ and\
  \citenamefont {St\"ocker}}]{Hanauske:2016gia}%
  \BibitemOpen
  \bibfield  {author} {\bibinfo {author} {\bibfnamefont {M.}~\bibnamefont
  {Hanauske}}, \bibinfo {author} {\bibfnamefont {K.}~\bibnamefont {Takami}},
  \bibinfo {author} {\bibfnamefont {L.}~\bibnamefont {Bovard}}, \bibinfo
  {author} {\bibfnamefont {L.}~\bibnamefont {Rezzolla}}, \bibinfo {author}
  {\bibfnamefont {J.~A.}\ \bibnamefont {Font}}, \bibinfo {author}
  {\bibfnamefont {F.}~\bibnamefont {Galeazzi}},\ and\ \bibinfo {author}
  {\bibfnamefont {H.}~\bibnamefont {St\"ocker}},\ }\bibfield  {title} {\bibinfo
  {title} {{Rotational properties of hypermassive neutron stars from binary
  mergers}},\ }\href {https://doi.org/10.1103/PhysRevD.96.043004} {\bibfield
  {journal} {\bibinfo  {journal} {Phys. Rev. D}\ }\textbf {\bibinfo {volume}
  {96}},\ \bibinfo {pages} {043004} (\bibinfo {year} {2017})},\ \Eprint
  {https://arxiv.org/abs/1611.07152} {arXiv:1611.07152 [gr-qc]} \BibitemShut
  {NoStop}%
\bibitem [{\citenamefont {Perego}\ \emph {et~al.}(2019)\citenamefont {Perego},
  \citenamefont {Bernuzzi},\ and\ \citenamefont {Radice}}]{Perego:2019adq}%
  \BibitemOpen
  \bibfield  {author} {\bibinfo {author} {\bibfnamefont {A.}~\bibnamefont
  {Perego}}, \bibinfo {author} {\bibfnamefont {S.}~\bibnamefont {Bernuzzi}},\
  and\ \bibinfo {author} {\bibfnamefont {D.}~\bibnamefont {Radice}},\
  }\bibfield  {title} {\bibinfo {title} {{Thermodynamics conditions of matter
  in neutron star mergers}},\ }\href
  {https://doi.org/10.1140/epja/i2019-12810-7} {\bibfield  {journal} {\bibinfo
  {journal} {Eur. Phys. J. A}\ }\textbf {\bibinfo {volume} {55}},\ \bibinfo
  {pages} {124} (\bibinfo {year} {2019})},\ \Eprint
  {https://arxiv.org/abs/1903.07898} {arXiv:1903.07898 [gr-qc]} \BibitemShut
  {NoStop}%
\bibitem [{\citenamefont {Endrizzi}\ \emph {et~al.}(2020)\citenamefont
  {Endrizzi}, \citenamefont {Perego}, \citenamefont {Fabbri}, \citenamefont
  {Branca}, \citenamefont {Radice}, \citenamefont {Bernuzzi}, \citenamefont
  {Giacomazzo}, \citenamefont {Pederiva},\ and\ \citenamefont
  {Lovato}}]{Endrizzi:2019trv}%
  \BibitemOpen
  \bibfield  {author} {\bibinfo {author} {\bibfnamefont {A.}~\bibnamefont
  {Endrizzi}}, \bibinfo {author} {\bibfnamefont {A.}~\bibnamefont {Perego}},
  \bibinfo {author} {\bibfnamefont {F.~M.}\ \bibnamefont {Fabbri}}, \bibinfo
  {author} {\bibfnamefont {L.}~\bibnamefont {Branca}}, \bibinfo {author}
  {\bibfnamefont {D.}~\bibnamefont {Radice}}, \bibinfo {author} {\bibfnamefont
  {S.}~\bibnamefont {Bernuzzi}}, \bibinfo {author} {\bibfnamefont
  {B.}~\bibnamefont {Giacomazzo}}, \bibinfo {author} {\bibfnamefont
  {F.}~\bibnamefont {Pederiva}},\ and\ \bibinfo {author} {\bibfnamefont
  {A.}~\bibnamefont {Lovato}},\ }\bibfield  {title} {\bibinfo {title}
  {{Thermodynamics conditions of matter in the neutrino decoupling region
  during neutron star mergers}},\ }\href
  {https://doi.org/10.1140/epja/s10050-019-00018-6} {\bibfield  {journal}
  {\bibinfo  {journal} {Eur. Phys. J. A}\ }\textbf {\bibinfo {volume} {56}},\
  \bibinfo {pages} {15} (\bibinfo {year} {2020})},\ \Eprint
  {https://arxiv.org/abs/1908.04952} {arXiv:1908.04952 [astro-ph.HE]}
  \BibitemShut {NoStop}%
\bibitem [{\citenamefont {Tan}\ \emph {et~al.}(2022)\citenamefont {Tan},
  \citenamefont {Dore}, \citenamefont {Dexheimer}, \citenamefont
  {Noronha-Hostler},\ and\ \citenamefont {Yunes}}]{Tan:2021ahl}%
  \BibitemOpen
  \bibfield  {author} {\bibinfo {author} {\bibfnamefont {H.}~\bibnamefont
  {Tan}}, \bibinfo {author} {\bibfnamefont {T.}~\bibnamefont {Dore}}, \bibinfo
  {author} {\bibfnamefont {V.}~\bibnamefont {Dexheimer}}, \bibinfo {author}
  {\bibfnamefont {J.}~\bibnamefont {Noronha-Hostler}},\ and\ \bibinfo {author}
  {\bibfnamefont {N.}~\bibnamefont {Yunes}},\ }\bibfield  {title} {\bibinfo
  {title} {{Extreme matter meets extreme gravity: Ultraheavy neutron stars with
  phase transitions}},\ }\href {https://doi.org/10.1103/PhysRevD.105.023018}
  {\bibfield  {journal} {\bibinfo  {journal} {Phys. Rev. D}\ }\textbf {\bibinfo
  {volume} {105}},\ \bibinfo {pages} {023018} (\bibinfo {year} {2022})},\
  \Eprint {https://arxiv.org/abs/2106.03890} {arXiv:2106.03890 [astro-ph.HE]}
  \BibitemShut {NoStop}%
\bibitem [{\citenamefont {Oechslin}\ \emph {et~al.}(2004)\citenamefont
  {Oechslin}, \citenamefont {Uryu}, \citenamefont {Poghosyan},\ and\
  \citenamefont {Thielemann}}]{Oechslin:2004yj}%
  \BibitemOpen
  \bibfield  {author} {\bibinfo {author} {\bibfnamefont {R.}~\bibnamefont
  {Oechslin}}, \bibinfo {author} {\bibfnamefont {K.}~\bibnamefont {Uryu}},
  \bibinfo {author} {\bibfnamefont {G.~S.}\ \bibnamefont {Poghosyan}},\ and\
  \bibinfo {author} {\bibfnamefont {F.~K.}\ \bibnamefont {Thielemann}},\
  }\bibfield  {title} {\bibinfo {title} {{The Influence of quark matter at high
  densities on binary neutron star mergers}},\ }\href
  {https://doi.org/10.1111/j.1365-2966.2004.07621.x} {\bibfield  {journal}
  {\bibinfo  {journal} {Mon. Not. Roy. Astron. Soc.}\ }\textbf {\bibinfo
  {volume} {349}},\ \bibinfo {pages} {1469} (\bibinfo {year} {2004})},\ \Eprint
  {https://arxiv.org/abs/astro-ph/0401083} {arXiv:astro-ph/0401083}
  \BibitemShut {NoStop}%
\bibitem [{\citenamefont {Most}\ \emph {et~al.}(2019)\citenamefont {Most},
  \citenamefont {Papenfort}, \citenamefont {Dexheimer}, \citenamefont
  {Hanauske}, \citenamefont {Schramm}, \citenamefont {St\"ocker},\ and\
  \citenamefont {Rezzolla}}]{Most:2018eaw}%
  \BibitemOpen
  \bibfield  {author} {\bibinfo {author} {\bibfnamefont {E.~R.}\ \bibnamefont
  {Most}}, \bibinfo {author} {\bibfnamefont {L.~J.}\ \bibnamefont {Papenfort}},
  \bibinfo {author} {\bibfnamefont {V.}~\bibnamefont {Dexheimer}}, \bibinfo
  {author} {\bibfnamefont {M.}~\bibnamefont {Hanauske}}, \bibinfo {author}
  {\bibfnamefont {S.}~\bibnamefont {Schramm}}, \bibinfo {author} {\bibfnamefont
  {H.}~\bibnamefont {St\"ocker}},\ and\ \bibinfo {author} {\bibfnamefont
  {L.}~\bibnamefont {Rezzolla}},\ }\bibfield  {title} {\bibinfo {title}
  {{Signatures of quark-hadron phase transitions in general-relativistic
  neutron-star mergers}},\ }\href
  {https://doi.org/10.1103/PhysRevLett.122.061101} {\bibfield  {journal}
  {\bibinfo  {journal} {Phys. Rev. Lett.}\ }\textbf {\bibinfo {volume} {122}},\
  \bibinfo {pages} {061101} (\bibinfo {year} {2019})},\ \Eprint
  {https://arxiv.org/abs/1807.03684} {arXiv:1807.03684 [astro-ph.HE]}
  \BibitemShut {NoStop}%
\bibitem [{\citenamefont {Bauswein}\ \emph {et~al.}(2019)\citenamefont
  {Bauswein}, \citenamefont {Bastian}, \citenamefont {Blaschke}, \citenamefont
  {Chatziioannou}, \citenamefont {Clark}, \citenamefont {Fischer},\ and\
  \citenamefont {Oertel}}]{Bauswein:2018bma}%
  \BibitemOpen
  \bibfield  {author} {\bibinfo {author} {\bibfnamefont {A.}~\bibnamefont
  {Bauswein}}, \bibinfo {author} {\bibfnamefont {N.-U.~F.}\ \bibnamefont
  {Bastian}}, \bibinfo {author} {\bibfnamefont {D.~B.}\ \bibnamefont
  {Blaschke}}, \bibinfo {author} {\bibfnamefont {K.}~\bibnamefont
  {Chatziioannou}}, \bibinfo {author} {\bibfnamefont {J.~A.}\ \bibnamefont
  {Clark}}, \bibinfo {author} {\bibfnamefont {T.}~\bibnamefont {Fischer}},\
  and\ \bibinfo {author} {\bibfnamefont {M.}~\bibnamefont {Oertel}},\
  }\bibfield  {title} {\bibinfo {title} {{Identifying a first-order phase
  transition in neutron star mergers through gravitational waves}},\ }\href
  {https://doi.org/10.1103/PhysRevLett.122.061102} {\bibfield  {journal}
  {\bibinfo  {journal} {Phys. Rev. Lett.}\ }\textbf {\bibinfo {volume} {122}},\
  \bibinfo {pages} {061102} (\bibinfo {year} {2019})},\ \Eprint
  {https://arxiv.org/abs/1809.01116} {arXiv:1809.01116 [astro-ph.HE]}
  \BibitemShut {NoStop}%
\bibitem [{\citenamefont {Most}\ \emph {et~al.}(2020)\citenamefont {Most},
  \citenamefont {Jens~Papenfort}, \citenamefont {Dexheimer}, \citenamefont
  {Hanauske}, \citenamefont {Stoecker},\ and\ \citenamefont
  {Rezzolla}}]{Most:2019onn}%
  \BibitemOpen
  \bibfield  {author} {\bibinfo {author} {\bibfnamefont {E.~R.}\ \bibnamefont
  {Most}}, \bibinfo {author} {\bibfnamefont {L.}~\bibnamefont
  {Jens~Papenfort}}, \bibinfo {author} {\bibfnamefont {V.}~\bibnamefont
  {Dexheimer}}, \bibinfo {author} {\bibfnamefont {M.}~\bibnamefont {Hanauske}},
  \bibinfo {author} {\bibfnamefont {H.}~\bibnamefont {Stoecker}},\ and\
  \bibinfo {author} {\bibfnamefont {L.}~\bibnamefont {Rezzolla}},\ }\bibfield
  {title} {\bibinfo {title} {{On the deconfinement phase transition in
  neutron-star mergers}},\ }\href
  {https://doi.org/10.1140/epja/s10050-020-00073-4} {\bibfield  {journal}
  {\bibinfo  {journal} {Eur. Phys. J. A}\ }\textbf {\bibinfo {volume} {56}},\
  \bibinfo {pages} {59} (\bibinfo {year} {2020})},\ \Eprint
  {https://arxiv.org/abs/1910.13893} {arXiv:1910.13893 [astro-ph.HE]}
  \BibitemShut {NoStop}%
\bibitem [{\citenamefont {Weih}\ \emph {et~al.}(2020)\citenamefont {Weih},
  \citenamefont {Hanauske},\ and\ \citenamefont {Rezzolla}}]{Weih:2019xvw}%
  \BibitemOpen
  \bibfield  {author} {\bibinfo {author} {\bibfnamefont {L.~R.}\ \bibnamefont
  {Weih}}, \bibinfo {author} {\bibfnamefont {M.}~\bibnamefont {Hanauske}},\
  and\ \bibinfo {author} {\bibfnamefont {L.}~\bibnamefont {Rezzolla}},\
  }\bibfield  {title} {\bibinfo {title} {{Postmerger Gravitational-Wave
  Signatures of Phase Transitions in Binary Mergers}},\ }\href
  {https://doi.org/10.1103/PhysRevLett.124.171103} {\bibfield  {journal}
  {\bibinfo  {journal} {Phys. Rev. Lett.}\ }\textbf {\bibinfo {volume} {124}},\
  \bibinfo {pages} {171103} (\bibinfo {year} {2020})},\ \Eprint
  {https://arxiv.org/abs/1912.09340} {arXiv:1912.09340 [gr-qc]} \BibitemShut
  {NoStop}%
\bibitem [{\citenamefont {Chatziioannou}\ and\ \citenamefont
  {Han}(2020)}]{Chatziioannou:2019yko}%
  \BibitemOpen
  \bibfield  {author} {\bibinfo {author} {\bibfnamefont {K.}~\bibnamefont
  {Chatziioannou}}\ and\ \bibinfo {author} {\bibfnamefont {S.}~\bibnamefont
  {Han}},\ }\bibfield  {title} {\bibinfo {title} {{Studying strong phase
  transitions in neutron stars with gravitational waves}},\ }\href
  {https://doi.org/10.1103/PhysRevD.101.044019} {\bibfield  {journal} {\bibinfo
   {journal} {Phys. Rev. D}\ }\textbf {\bibinfo {volume} {101}},\ \bibinfo
  {pages} {044019} (\bibinfo {year} {2020})},\ \Eprint
  {https://arxiv.org/abs/1911.07091} {arXiv:1911.07091 [gr-qc]} \BibitemShut
  {NoStop}%
\bibitem [{\citenamefont {Prakash}\ \emph {et~al.}(2021)\citenamefont
  {Prakash}, \citenamefont {Radice}, \citenamefont {Logoteta}, \citenamefont
  {Perego}, \citenamefont {Nedora}, \citenamefont {Bombaci}, \citenamefont
  {Kashyap}, \citenamefont {Bernuzzi},\ and\ \citenamefont
  {Endrizzi}}]{Prakash:2021wpz}%
  \BibitemOpen
  \bibfield  {author} {\bibinfo {author} {\bibfnamefont {A.}~\bibnamefont
  {Prakash}}, \bibinfo {author} {\bibfnamefont {D.}~\bibnamefont {Radice}},
  \bibinfo {author} {\bibfnamefont {D.}~\bibnamefont {Logoteta}}, \bibinfo
  {author} {\bibfnamefont {A.}~\bibnamefont {Perego}}, \bibinfo {author}
  {\bibfnamefont {V.}~\bibnamefont {Nedora}}, \bibinfo {author} {\bibfnamefont
  {I.}~\bibnamefont {Bombaci}}, \bibinfo {author} {\bibfnamefont
  {R.}~\bibnamefont {Kashyap}}, \bibinfo {author} {\bibfnamefont
  {S.}~\bibnamefont {Bernuzzi}},\ and\ \bibinfo {author} {\bibfnamefont
  {A.}~\bibnamefont {Endrizzi}},\ }\bibfield  {title} {\bibinfo {title}
  {{Signatures of deconfined quark phases in binary neutron star mergers}},\
  }\href {https://doi.org/10.1103/PhysRevD.104.083029} {\bibfield  {journal}
  {\bibinfo  {journal} {Phys. Rev. D}\ }\textbf {\bibinfo {volume} {104}},\
  \bibinfo {pages} {083029} (\bibinfo {year} {2021})},\ \Eprint
  {https://arxiv.org/abs/2106.07885} {arXiv:2106.07885 [astro-ph.HE]}
  \BibitemShut {NoStop}%
\bibitem [{\citenamefont {Tootle}\ \emph {et~al.}(2022)\citenamefont {Tootle},
  \citenamefont {Ecker}, \citenamefont {Topolski}, \citenamefont {Demircik},
  \citenamefont {J\"arvinen},\ and\ \citenamefont {Rezzolla}}]{Tootle:2022pvd}%
  \BibitemOpen
  \bibfield  {author} {\bibinfo {author} {\bibfnamefont {S.}~\bibnamefont
  {Tootle}}, \bibinfo {author} {\bibfnamefont {C.}~\bibnamefont {Ecker}},
  \bibinfo {author} {\bibfnamefont {K.}~\bibnamefont {Topolski}}, \bibinfo
  {author} {\bibfnamefont {T.}~\bibnamefont {Demircik}}, \bibinfo {author}
  {\bibfnamefont {M.}~\bibnamefont {J\"arvinen}},\ and\ \bibinfo {author}
  {\bibfnamefont {L.}~\bibnamefont {Rezzolla}},\ }\bibfield  {title} {\bibinfo
  {title} {{Quark formation and phenomenology in binary neutron-star mergers
  using V-QCD}},\ }\href {https://doi.org/10.21468/SciPostPhys.13.5.109}
  {\bibfield  {journal} {\bibinfo  {journal} {SciPost Phys.}\ }\textbf
  {\bibinfo {volume} {13}},\ \bibinfo {pages} {109} (\bibinfo {year} {2022})},\
  \Eprint {https://arxiv.org/abs/2205.05691} {arXiv:2205.05691 [astro-ph.HE]}
  \BibitemShut {NoStop}%
\bibitem [{\citenamefont {Kumar}\ \emph {et~al.}(2023)\citenamefont {Kumar}
  \emph {et~al.}}]{MUSES:2023hyz}%
  \BibitemOpen
  \bibfield  {author} {\bibinfo {author} {\bibfnamefont {R.}~\bibnamefont
  {Kumar}} \emph {et~al.} (\bibinfo {collaboration} {MUSES}),\ }\bibfield
  {title} {\bibinfo {title} {{Theoretical and Experimental Constraints for the
  Equation of State of Dense and Hot Matter}},\ }\href@noop {} {\  (\bibinfo
  {year} {2023})},\ \Eprint {https://arxiv.org/abs/2303.17021}
  {arXiv:2303.17021 [nucl-th]} \BibitemShut {NoStop}%
\bibitem [{\citenamefont {Piekarewicz}\ and\ \citenamefont
  {Centelles}(2009)}]{Piekarewicz:2008nh}%
  \BibitemOpen
  \bibfield  {author} {\bibinfo {author} {\bibfnamefont {J.}~\bibnamefont
  {Piekarewicz}}\ and\ \bibinfo {author} {\bibfnamefont {M.}~\bibnamefont
  {Centelles}},\ }\bibfield  {title} {\bibinfo {title} {{Incompressibility of
  neutron-rich matter}},\ }\href {https://doi.org/10.1103/PhysRevC.79.054311}
  {\bibfield  {journal} {\bibinfo  {journal} {Phys. Rev. C}\ }\textbf {\bibinfo
  {volume} {79}},\ \bibinfo {pages} {054311} (\bibinfo {year} {2009})},\
  \Eprint {https://arxiv.org/abs/0812.4499} {arXiv:0812.4499 [nucl-th]}
  \BibitemShut {NoStop}%
\bibitem [{\citenamefont {Gross-Boelting}\ \emph {et~al.}(1999)\citenamefont
  {Gross-Boelting}, \citenamefont {Fuchs},\ and\ \citenamefont
  {Faessler}}]{Gross-Boelting:1998qhi}%
  \BibitemOpen
  \bibfield  {author} {\bibinfo {author} {\bibfnamefont {T.}~\bibnamefont
  {Gross-Boelting}}, \bibinfo {author} {\bibfnamefont {C.}~\bibnamefont
  {Fuchs}},\ and\ \bibinfo {author} {\bibfnamefont {A.}~\bibnamefont
  {Faessler}},\ }\bibfield  {title} {\bibinfo {title} {{Dirac structure of the
  nucleus-nucleus potential in heavy ion collisions}},\ }\href
  {https://doi.org/10.1016/S0146-6410(99)00061-7} {\bibfield  {journal}
  {\bibinfo  {journal} {Prog. Part. Nucl. Phys.}\ }\textbf {\bibinfo {volume}
  {42}},\ \bibinfo {pages} {65} (\bibinfo {year} {1999})},\ \Eprint
  {https://arxiv.org/abs/nucl-th/9812002} {arXiv:nucl-th/9812002} \BibitemShut
  {NoStop}%
\bibitem [{\citenamefont {Adhikari}\ \emph {et~al.}(2021)\citenamefont
  {Adhikari} \emph {et~al.}}]{PREX:2021umo}%
  \BibitemOpen
  \bibfield  {author} {\bibinfo {author} {\bibfnamefont {D.}~\bibnamefont
  {Adhikari}} \emph {et~al.} (\bibinfo {collaboration} {PREX}),\ }\bibfield
  {title} {\bibinfo {title} {{Accurate Determination of the Neutron Skin
  Thickness of $^{208}$Pb through Parity-Violation in Electron Scattering}},\
  }\href {https://doi.org/10.1103/PhysRevLett.126.172502} {\bibfield  {journal}
  {\bibinfo  {journal} {Phys. Rev. Lett.}\ }\textbf {\bibinfo {volume} {126}},\
  \bibinfo {pages} {172502} (\bibinfo {year} {2021})},\ \Eprint
  {https://arxiv.org/abs/2102.10767} {arXiv:2102.10767 [nucl-ex]} \BibitemShut
  {NoStop}%
\bibitem [{\citenamefont {Lim}\ and\ \citenamefont
  {Schwenk}(2023)}]{Lim:2023dbk}%
  \BibitemOpen
  \bibfield  {author} {\bibinfo {author} {\bibfnamefont {Y.}~\bibnamefont
  {Lim}}\ and\ \bibinfo {author} {\bibfnamefont {A.}~\bibnamefont {Schwenk}},\
  }\bibfield  {title} {\bibinfo {title} {{Symmetry energy and neutron star
  properties constrained by chiral effective field theory calculations}},\
  }\href@noop {} {\  (\bibinfo {year} {2023})},\ \Eprint
  {https://arxiv.org/abs/2307.04063} {arXiv:2307.04063 [nucl-th]} \BibitemShut
  {NoStop}%
\bibitem [{\citenamefont {Lattimer}\ and\ \citenamefont
  {Prakash}(2001)}]{Lattimer:2000nx}%
  \BibitemOpen
  \bibfield  {author} {\bibinfo {author} {\bibfnamefont {J.~M.}\ \bibnamefont
  {Lattimer}}\ and\ \bibinfo {author} {\bibfnamefont {M.}~\bibnamefont
  {Prakash}},\ }\bibfield  {title} {\bibinfo {title} {{Neutron star structure
  and the equation of state}},\ }\href {https://doi.org/10.1086/319702}
  {\bibfield  {journal} {\bibinfo  {journal} {Astrophys. J.}\ }\textbf
  {\bibinfo {volume} {550}},\ \bibinfo {pages} {426} (\bibinfo {year}
  {2001})},\ \Eprint {https://arxiv.org/abs/astro-ph/0002232}
  {arXiv:astro-ph/0002232} \BibitemShut {NoStop}%
\bibitem [{\citenamefont {Yagi}\ and\ \citenamefont
  {Yunes}(2013)}]{Yagi:2013awa}%
  \BibitemOpen
  \bibfield  {author} {\bibinfo {author} {\bibfnamefont {K.}~\bibnamefont
  {Yagi}}\ and\ \bibinfo {author} {\bibfnamefont {N.}~\bibnamefont {Yunes}},\
  }\bibfield  {title} {\bibinfo {title} {{I-Love-Q Relations in Neutron Stars
  and their Applications to Astrophysics, Gravitational Waves and Fundamental
  Physics}},\ }\href {https://doi.org/10.1103/PhysRevD.88.023009} {\bibfield
  {journal} {\bibinfo  {journal} {Phys. Rev. D}\ }\textbf {\bibinfo {volume}
  {88}},\ \bibinfo {pages} {023009} (\bibinfo {year} {2013})},\ \Eprint
  {https://arxiv.org/abs/1303.1528} {arXiv:1303.1528 [gr-qc]} \BibitemShut
  {NoStop}%
\bibitem [{\citenamefont {Yagi}\ and\ \citenamefont
  {Yunes}(2016)}]{Yagi:2015pkc}%
  \BibitemOpen
  \bibfield  {author} {\bibinfo {author} {\bibfnamefont {K.}~\bibnamefont
  {Yagi}}\ and\ \bibinfo {author} {\bibfnamefont {N.}~\bibnamefont {Yunes}},\
  }\bibfield  {title} {\bibinfo {title} {{Binary Love Relations}},\ }\href
  {https://doi.org/10.1088/0264-9381/33/13/13LT01} {\bibfield  {journal}
  {\bibinfo  {journal} {Class. Quant. Grav.}\ }\textbf {\bibinfo {volume}
  {33}},\ \bibinfo {pages} {13LT01} (\bibinfo {year} {2016})},\ \Eprint
  {https://arxiv.org/abs/1512.02639} {arXiv:1512.02639 [gr-qc]} \BibitemShut
  {NoStop}%
\bibitem [{\citenamefont {Most}\ and\ \citenamefont
  {Raithel}(2021)}]{Most:2021ktk}%
  \BibitemOpen
  \bibfield  {author} {\bibinfo {author} {\bibfnamefont {E.~R.}\ \bibnamefont
  {Most}}\ and\ \bibinfo {author} {\bibfnamefont {C.~A.}\ \bibnamefont
  {Raithel}},\ }\bibfield  {title} {\bibinfo {title} {{Impact of the nuclear
  symmetry energy on the post-merger phase of a binary neutron star
  coalescence}},\ }\href {https://doi.org/10.1103/PhysRevD.104.124012}
  {\bibfield  {journal} {\bibinfo  {journal} {Phys. Rev. D}\ }\textbf {\bibinfo
  {volume} {104}},\ \bibinfo {pages} {124012} (\bibinfo {year} {2021})},\
  \Eprint {https://arxiv.org/abs/2107.06804} {arXiv:2107.06804 [astro-ph.HE]}
  \BibitemShut {NoStop}%
\bibitem [{\citenamefont {Alford}\ \emph {et~al.}(2018)\citenamefont {Alford},
  \citenamefont {Bovard}, \citenamefont {Hanauske}, \citenamefont {Rezzolla},\
  and\ \citenamefont {Schwenzer}}]{Alford:2017rxf}%
  \BibitemOpen
  \bibfield  {author} {\bibinfo {author} {\bibfnamefont {M.~G.}\ \bibnamefont
  {Alford}}, \bibinfo {author} {\bibfnamefont {L.}~\bibnamefont {Bovard}},
  \bibinfo {author} {\bibfnamefont {M.}~\bibnamefont {Hanauske}}, \bibinfo
  {author} {\bibfnamefont {L.}~\bibnamefont {Rezzolla}},\ and\ \bibinfo
  {author} {\bibfnamefont {K.}~\bibnamefont {Schwenzer}},\ }\bibfield  {title}
  {\bibinfo {title} {{Viscous Dissipation and Heat Conduction in Binary
  Neutron-Star Mergers}},\ }\href
  {https://doi.org/10.1103/PhysRevLett.120.041101} {\bibfield  {journal}
  {\bibinfo  {journal} {Phys. Rev. Lett.}\ }\textbf {\bibinfo {volume} {120}},\
  \bibinfo {pages} {041101} (\bibinfo {year} {2018})},\ \Eprint
  {https://arxiv.org/abs/1707.09475} {arXiv:1707.09475 [gr-qc]} \BibitemShut
  {NoStop}%
\bibitem [{\citenamefont {Hammond}\ \emph {et~al.}(2021)\citenamefont
  {Hammond}, \citenamefont {Hawke},\ and\ \citenamefont
  {Andersson}}]{Hammond:2021vtv}%
  \BibitemOpen
  \bibfield  {author} {\bibinfo {author} {\bibfnamefont {P.}~\bibnamefont
  {Hammond}}, \bibinfo {author} {\bibfnamefont {I.}~\bibnamefont {Hawke}},\
  and\ \bibinfo {author} {\bibfnamefont {N.}~\bibnamefont {Andersson}},\
  }\bibfield  {title} {\bibinfo {title} {{Thermal aspects of neutron star
  mergers}},\ }\href {https://doi.org/10.1103/PhysRevD.104.103006} {\bibfield
  {journal} {\bibinfo  {journal} {Phys. Rev. D}\ }\textbf {\bibinfo {volume}
  {104}},\ \bibinfo {pages} {103006} (\bibinfo {year} {2021})},\ \Eprint
  {https://arxiv.org/abs/2108.08649} {arXiv:2108.08649 [astro-ph.HE]}
  \BibitemShut {NoStop}%
\bibitem [{\citenamefont {Alford}\ and\ \citenamefont
  {Harris}(2018)}]{Alford:2018lhf}%
  \BibitemOpen
  \bibfield  {author} {\bibinfo {author} {\bibfnamefont {M.~G.}\ \bibnamefont
  {Alford}}\ and\ \bibinfo {author} {\bibfnamefont {S.~P.}\ \bibnamefont
  {Harris}},\ }\bibfield  {title} {\bibinfo {title} {{Beta equilibrium in
  neutron star mergers}},\ }\href {https://doi.org/10.1103/PhysRevC.98.065806}
  {\bibfield  {journal} {\bibinfo  {journal} {Phys. Rev. C}\ }\textbf {\bibinfo
  {volume} {98}},\ \bibinfo {pages} {065806} (\bibinfo {year} {2018})},\
  \Eprint {https://arxiv.org/abs/1803.00662} {arXiv:1803.00662 [nucl-th]}
  \BibitemShut {NoStop}%
\bibitem [{\citenamefont {Alford}\ \emph {et~al.}(2021)\citenamefont {Alford},
  \citenamefont {Haber}, \citenamefont {Harris},\ and\ \citenamefont
  {Zhang}}]{Alford:2021ogv}%
  \BibitemOpen
  \bibfield  {author} {\bibinfo {author} {\bibfnamefont {M.~G.}\ \bibnamefont
  {Alford}}, \bibinfo {author} {\bibfnamefont {A.}~\bibnamefont {Haber}},
  \bibinfo {author} {\bibfnamefont {S.~P.}\ \bibnamefont {Harris}},\ and\
  \bibinfo {author} {\bibfnamefont {Z.}~\bibnamefont {Zhang}},\ }\bibfield
  {title} {\bibinfo {title} {{Beta Equilibrium Under Neutron Star Merger
  Conditions}},\ }\href {https://doi.org/10.3390/universe7110399} {\bibfield
  {journal} {\bibinfo  {journal} {Universe}\ }\textbf {\bibinfo {volume} {7}},\
  \bibinfo {pages} {399} (\bibinfo {year} {2021})},\ \Eprint
  {https://arxiv.org/abs/2108.03324} {arXiv:2108.03324 [nucl-th]} \BibitemShut
  {NoStop}%
\bibitem [{\citenamefont {Alford}\ \emph {et~al.}(2023)\citenamefont {Alford},
  \citenamefont {Haber},\ and\ \citenamefont {Zhang}}]{Alford:2023gxq}%
  \BibitemOpen
  \bibfield  {author} {\bibinfo {author} {\bibfnamefont {M.~G.}\ \bibnamefont
  {Alford}}, \bibinfo {author} {\bibfnamefont {A.}~\bibnamefont {Haber}},\ and\
  \bibinfo {author} {\bibfnamefont {Z.}~\bibnamefont {Zhang}},\ }\bibfield
  {title} {\bibinfo {title} {{Isospin Equilibration in Neutron Star Mergers}},\
  }\href@noop {} {\  (\bibinfo {year} {2023})},\ \Eprint
  {https://arxiv.org/abs/2306.06180} {arXiv:2306.06180 [nucl-th]} \BibitemShut
  {NoStop}%
\bibitem [{\citenamefont {Sawyer}(1989)}]{Sawyer:1989dp}%
  \BibitemOpen
  \bibfield  {author} {\bibinfo {author} {\bibfnamefont {R.~F.}\ \bibnamefont
  {Sawyer}},\ }\bibfield  {title} {\bibinfo {title} {{Bulk viscosity of hot
  neutron-star matter and the maximum rotation rates of neutron stars}},\
  }\href {https://doi.org/10.1103/PhysRevD.39.3804} {\bibfield  {journal}
  {\bibinfo  {journal} {Phys. Rev. D}\ }\textbf {\bibinfo {volume} {39}},\
  \bibinfo {pages} {3804} (\bibinfo {year} {1989})}\BibitemShut {NoStop}%
\bibitem [{\citenamefont {Haensel}\ and\ \citenamefont
  {Schaeffer}(1992)}]{Haensel:1992zz}%
  \BibitemOpen
  \bibfield  {author} {\bibinfo {author} {\bibfnamefont {P.}~\bibnamefont
  {Haensel}}\ and\ \bibinfo {author} {\bibfnamefont {R.}~\bibnamefont
  {Schaeffer}},\ }\bibfield  {title} {\bibinfo {title} {{Bulk viscosity of
  hot-neutron-star matter from direct URCA processes}},\ }\href
  {https://doi.org/10.1103/PhysRevD.45.4708} {\bibfield  {journal} {\bibinfo
  {journal} {Phys. Rev. D}\ }\textbf {\bibinfo {volume} {45}},\ \bibinfo
  {pages} {4708} (\bibinfo {year} {1992})}\BibitemShut {NoStop}%
\bibitem [{\citenamefont {Alford}\ and\ \citenamefont
  {Harris}(2019)}]{Alford:2019qtm}%
  \BibitemOpen
  \bibfield  {author} {\bibinfo {author} {\bibfnamefont {M.~G.}\ \bibnamefont
  {Alford}}\ and\ \bibinfo {author} {\bibfnamefont {S.~P.}\ \bibnamefont
  {Harris}},\ }\bibfield  {title} {\bibinfo {title} {{Damping of density
  oscillations in neutrino-transparent nuclear matter}},\ }\href
  {https://doi.org/10.1103/PhysRevC.100.035803} {\bibfield  {journal} {\bibinfo
   {journal} {Phys. Rev. C}\ }\textbf {\bibinfo {volume} {100}},\ \bibinfo
  {pages} {035803} (\bibinfo {year} {2019})},\ \Eprint
  {https://arxiv.org/abs/1907.03795} {arXiv:1907.03795 [nucl-th]} \BibitemShut
  {NoStop}%
\bibitem [{\citenamefont {Alford}\ \emph {et~al.}(2019)\citenamefont {Alford},
  \citenamefont {Harutyunyan},\ and\ \citenamefont
  {Sedrakian}}]{Alford:2019kdw}%
  \BibitemOpen
  \bibfield  {author} {\bibinfo {author} {\bibfnamefont {M.}~\bibnamefont
  {Alford}}, \bibinfo {author} {\bibfnamefont {A.}~\bibnamefont
  {Harutyunyan}},\ and\ \bibinfo {author} {\bibfnamefont {A.}~\bibnamefont
  {Sedrakian}},\ }\bibfield  {title} {\bibinfo {title} {{Bulk viscosity of
  baryonic matter with trapped neutrinos}},\ }\href
  {https://doi.org/10.1103/PhysRevD.100.103021} {\bibfield  {journal} {\bibinfo
   {journal} {Phys. Rev. D}\ }\textbf {\bibinfo {volume} {100}},\ \bibinfo
  {pages} {103021} (\bibinfo {year} {2019})},\ \Eprint
  {https://arxiv.org/abs/1907.04192} {arXiv:1907.04192 [astro-ph.HE]}
  \BibitemShut {NoStop}%
\bibitem [{\citenamefont {Most}\ \emph {et~al.}(2021)\citenamefont {Most},
  \citenamefont {Harris}, \citenamefont {Plumberg}, \citenamefont {Alford},
  \citenamefont {Noronha}, \citenamefont {Noronha-Hostler}, \citenamefont
  {Pretorius}, \citenamefont {Witek},\ and\ \citenamefont
  {Yunes}}]{Most:2021zvc}%
  \BibitemOpen
  \bibfield  {author} {\bibinfo {author} {\bibfnamefont {E.~R.}\ \bibnamefont
  {Most}}, \bibinfo {author} {\bibfnamefont {S.~P.}\ \bibnamefont {Harris}},
  \bibinfo {author} {\bibfnamefont {C.}~\bibnamefont {Plumberg}}, \bibinfo
  {author} {\bibfnamefont {M.~G.}\ \bibnamefont {Alford}}, \bibinfo {author}
  {\bibfnamefont {J.}~\bibnamefont {Noronha}}, \bibinfo {author} {\bibfnamefont
  {J.}~\bibnamefont {Noronha-Hostler}}, \bibinfo {author} {\bibfnamefont
  {F.}~\bibnamefont {Pretorius}}, \bibinfo {author} {\bibfnamefont
  {H.}~\bibnamefont {Witek}},\ and\ \bibinfo {author} {\bibfnamefont
  {N.}~\bibnamefont {Yunes}},\ }\bibfield  {title} {\bibinfo {title}
  {{Projecting the likely importance of weak-interaction-driven bulk viscosity
  in neutron star mergers}},\ }\href {https://doi.org/10.1093/mnras/stab2793}
  {\bibfield  {journal} {\bibinfo  {journal} {Mon. Not. Roy. Astron. Soc.}\
  }\textbf {\bibinfo {volume} {509}},\ \bibinfo {pages} {1096} (\bibinfo {year}
  {2021})},\ \Eprint {https://arxiv.org/abs/2107.05094} {arXiv:2107.05094
  [astro-ph.HE]} \BibitemShut {NoStop}%
\bibitem [{\citenamefont {Most}\ \emph {et~al.}(2022)\citenamefont {Most},
  \citenamefont {Haber}, \citenamefont {Harris}, \citenamefont {Zhang},
  \citenamefont {Alford},\ and\ \citenamefont {Noronha}}]{Most:2022yhe}%
  \BibitemOpen
  \bibfield  {author} {\bibinfo {author} {\bibfnamefont {E.~R.}\ \bibnamefont
  {Most}}, \bibinfo {author} {\bibfnamefont {A.}~\bibnamefont {Haber}},
  \bibinfo {author} {\bibfnamefont {S.~P.}\ \bibnamefont {Harris}}, \bibinfo
  {author} {\bibfnamefont {Z.}~\bibnamefont {Zhang}}, \bibinfo {author}
  {\bibfnamefont {M.~G.}\ \bibnamefont {Alford}},\ and\ \bibinfo {author}
  {\bibfnamefont {J.}~\bibnamefont {Noronha}},\ }\bibfield  {title} {\bibinfo
  {title} {{Emergence of microphysical viscosity in binary neutron star
  post-merger dynamics}},\ }\href@noop {} {\  (\bibinfo {year} {2022})},\
  \Eprint {https://arxiv.org/abs/2207.00442} {arXiv:2207.00442 [astro-ph.HE]}
  \BibitemShut {NoStop}%
\bibitem [{\citenamefont {Hammond}\ \emph {et~al.}(2023)\citenamefont
  {Hammond}, \citenamefont {Hawke},\ and\ \citenamefont
  {Andersson}}]{Hammond:2022uua}%
  \BibitemOpen
  \bibfield  {author} {\bibinfo {author} {\bibfnamefont {P.}~\bibnamefont
  {Hammond}}, \bibinfo {author} {\bibfnamefont {I.}~\bibnamefont {Hawke}},\
  and\ \bibinfo {author} {\bibfnamefont {N.}~\bibnamefont {Andersson}},\
  }\bibfield  {title} {\bibinfo {title} {{Impact of nuclear reactions on
  gravitational waves from neutron star mergers}},\ }\href
  {https://doi.org/10.1103/PhysRevD.107.043023} {\bibfield  {journal} {\bibinfo
   {journal} {Phys. Rev. D}\ }\textbf {\bibinfo {volume} {107}},\ \bibinfo
  {pages} {043023} (\bibinfo {year} {2023})},\ \Eprint
  {https://arxiv.org/abs/2205.11377} {arXiv:2205.11377 [astro-ph.HE]}
  \BibitemShut {NoStop}%
\bibitem [{\citenamefont {Chabanov}\ and\ \citenamefont
  {Rezzolla}(2023)}]{Chabanov:2023blf}%
  \BibitemOpen
  \bibfield  {author} {\bibinfo {author} {\bibfnamefont {M.}~\bibnamefont
  {Chabanov}}\ and\ \bibinfo {author} {\bibfnamefont {L.}~\bibnamefont
  {Rezzolla}},\ }\bibfield  {title} {\bibinfo {title} {{Impact of bulk
  viscosity on the post-merger gravitational-wave signal from merging neutron
  stars}},\ }\href@noop {} {\  (\bibinfo {year} {2023})},\ \Eprint
  {https://arxiv.org/abs/2307.10464} {arXiv:2307.10464 [gr-qc]} \BibitemShut
  {NoStop}%
\bibitem [{\citenamefont {Madsen}(1992)}]{Madsen:1992sx}%
  \BibitemOpen
  \bibfield  {author} {\bibinfo {author} {\bibfnamefont {J.}~\bibnamefont
  {Madsen}},\ }\bibfield  {title} {\bibinfo {title} {{Bulk viscosity of strange
  quark matter, damping of quark star vibration, and the maximum rotation rate
  of pulsars}},\ }\href {https://doi.org/10.1103/PhysRevD.46.3290} {\bibfield
  {journal} {\bibinfo  {journal} {Phys. Rev. D}\ }\textbf {\bibinfo {volume}
  {46}},\ \bibinfo {pages} {3290} (\bibinfo {year} {1992})}\BibitemShut
  {NoStop}%
\bibitem [{\citenamefont {Reisenegger}\ and\ \citenamefont
  {Bonacic}(2003)}]{Reisenegger:2003pd}%
  \BibitemOpen
  \bibfield  {author} {\bibinfo {author} {\bibfnamefont {A.}~\bibnamefont
  {Reisenegger}}\ and\ \bibinfo {author} {\bibfnamefont {A.~A.}\ \bibnamefont
  {Bonacic}},\ }\bibfield  {title} {\bibinfo {title} {{Bulk viscosity, r-modes,
  and the early evolution of neutron stars}},\ }in\ \href@noop {} {\emph
  {\bibinfo {booktitle} {{International Workshop on Pulsars, AXPs and SGRs
  Observed with BeppoSAX and other Observatories}}}}\ (\bibinfo {year} {2003})\
  \Eprint {https://arxiv.org/abs/astro-ph/0303454} {arXiv:astro-ph/0303454}
  \BibitemShut {NoStop}%
\bibitem [{\citenamefont {Alford}\ \emph {et~al.}(2010)\citenamefont {Alford},
  \citenamefont {Mahmoodifar},\ and\ \citenamefont
  {Schwenzer}}]{Alford:2010gw}%
  \BibitemOpen
  \bibfield  {author} {\bibinfo {author} {\bibfnamefont {M.~G.}\ \bibnamefont
  {Alford}}, \bibinfo {author} {\bibfnamefont {S.}~\bibnamefont
  {Mahmoodifar}},\ and\ \bibinfo {author} {\bibfnamefont {K.}~\bibnamefont
  {Schwenzer}},\ }\bibfield  {title} {\bibinfo {title} {{Large amplitude
  behavior of the bulk viscosity of dense matter}},\ }\href
  {https://doi.org/10.1088/0954-3899/37/12/125202} {\bibfield  {journal}
  {\bibinfo  {journal} {J. Phys. G}\ }\textbf {\bibinfo {volume} {37}},\
  \bibinfo {pages} {125202} (\bibinfo {year} {2010})},\ \Eprint
  {https://arxiv.org/abs/1005.3769} {arXiv:1005.3769 [nucl-th]} \BibitemShut
  {NoStop}%
\bibitem [{\citenamefont {Gavassino}\ \emph {et~al.}(2021)\citenamefont
  {Gavassino}, \citenamefont {Antonelli},\ and\ \citenamefont
  {Haskell}}]{Gavassino:2020kwo}%
  \BibitemOpen
  \bibfield  {author} {\bibinfo {author} {\bibfnamefont {L.}~\bibnamefont
  {Gavassino}}, \bibinfo {author} {\bibfnamefont {M.}~\bibnamefont
  {Antonelli}},\ and\ \bibinfo {author} {\bibfnamefont {B.}~\bibnamefont
  {Haskell}},\ }\bibfield  {title} {\bibinfo {title} {{Bulk viscosity in
  relativistic fluids: from thermodynamics to hydrodynamics}},\ }\href
  {https://doi.org/10.1088/1361-6382/abe588} {\bibfield  {journal} {\bibinfo
  {journal} {Class. Quant. Grav.}\ }\textbf {\bibinfo {volume} {38}},\ \bibinfo
  {pages} {075001} (\bibinfo {year} {2021})},\ \Eprint
  {https://arxiv.org/abs/2003.04609} {arXiv:2003.04609 [gr-qc]} \BibitemShut
  {NoStop}%
\bibitem [{\citenamefont {Celora}\ \emph {et~al.}(2022)\citenamefont {Celora},
  \citenamefont {Hawke}, \citenamefont {Hammond}, \citenamefont {Andersson},\
  and\ \citenamefont {Comer}}]{Celora:2022nbp}%
  \BibitemOpen
  \bibfield  {author} {\bibinfo {author} {\bibfnamefont {T.}~\bibnamefont
  {Celora}}, \bibinfo {author} {\bibfnamefont {I.}~\bibnamefont {Hawke}},
  \bibinfo {author} {\bibfnamefont {P.~C.}\ \bibnamefont {Hammond}}, \bibinfo
  {author} {\bibfnamefont {N.}~\bibnamefont {Andersson}},\ and\ \bibinfo
  {author} {\bibfnamefont {G.~L.}\ \bibnamefont {Comer}},\ }\bibfield  {title}
  {\bibinfo {title} {{Formulating bulk viscosity for neutron star
  simulations}},\ }\href {https://doi.org/10.1103/PhysRevD.105.103016}
  {\bibfield  {journal} {\bibinfo  {journal} {Phys. Rev. D}\ }\textbf {\bibinfo
  {volume} {105}},\ \bibinfo {pages} {103016} (\bibinfo {year} {2022})},\
  \Eprint {https://arxiv.org/abs/2202.01576} {arXiv:2202.01576 [astro-ph.HE]}
  \BibitemShut {NoStop}%
\bibitem [{\citenamefont {Camelio}\ \emph
  {et~al.}(2023{\natexlab{a}})\citenamefont {Camelio}, \citenamefont
  {Gavassino}, \citenamefont {Antonelli}, \citenamefont {Bernuzzi},\ and\
  \citenamefont {Haskell}}]{Camelio:2022ljs}%
  \BibitemOpen
  \bibfield  {author} {\bibinfo {author} {\bibfnamefont {G.}~\bibnamefont
  {Camelio}}, \bibinfo {author} {\bibfnamefont {L.}~\bibnamefont {Gavassino}},
  \bibinfo {author} {\bibfnamefont {M.}~\bibnamefont {Antonelli}}, \bibinfo
  {author} {\bibfnamefont {S.}~\bibnamefont {Bernuzzi}},\ and\ \bibinfo
  {author} {\bibfnamefont {B.}~\bibnamefont {Haskell}},\ }\bibfield  {title}
  {\bibinfo {title} {{Simulating bulk viscosity in neutron stars. I.
  Formalism}},\ }\href {https://doi.org/10.1103/PhysRevD.107.103031} {\bibfield
   {journal} {\bibinfo  {journal} {Phys. Rev. D}\ }\textbf {\bibinfo {volume}
  {107}},\ \bibinfo {pages} {103031} (\bibinfo {year} {2023}{\natexlab{a}})},\
  \Eprint {https://arxiv.org/abs/2204.11809} {arXiv:2204.11809 [gr-qc]}
  \BibitemShut {NoStop}%
\bibitem [{\citenamefont {Camelio}\ \emph
  {et~al.}(2023{\natexlab{b}})\citenamefont {Camelio}, \citenamefont
  {Gavassino}, \citenamefont {Antonelli}, \citenamefont {Bernuzzi},\ and\
  \citenamefont {Haskell}}]{Camelio:2022fds}%
  \BibitemOpen
  \bibfield  {author} {\bibinfo {author} {\bibfnamefont {G.}~\bibnamefont
  {Camelio}}, \bibinfo {author} {\bibfnamefont {L.}~\bibnamefont {Gavassino}},
  \bibinfo {author} {\bibfnamefont {M.}~\bibnamefont {Antonelli}}, \bibinfo
  {author} {\bibfnamefont {S.}~\bibnamefont {Bernuzzi}},\ and\ \bibinfo
  {author} {\bibfnamefont {B.}~\bibnamefont {Haskell}},\ }\bibfield  {title}
  {\bibinfo {title} {{Simulating bulk viscosity in neutron stars. II. Evolution
  in spherical symmetry}},\ }\href
  {https://doi.org/10.1103/PhysRevD.107.103032} {\bibfield  {journal} {\bibinfo
   {journal} {Phys. Rev. D}\ }\textbf {\bibinfo {volume} {107}},\ \bibinfo
  {pages} {103032} (\bibinfo {year} {2023}{\natexlab{b}})},\ \Eprint
  {https://arxiv.org/abs/2204.11810} {arXiv:2204.11810 [gr-qc]} \BibitemShut
  {NoStop}%
\bibitem [{\citenamefont {Israel}\ and\ \citenamefont
  {Stewart}(1979)}]{Israel:1979wp}%
  \BibitemOpen
  \bibfield  {author} {\bibinfo {author} {\bibfnamefont {W.}~\bibnamefont
  {Israel}}\ and\ \bibinfo {author} {\bibfnamefont {J.~M.}\ \bibnamefont
  {Stewart}},\ }\bibfield  {title} {\bibinfo {title} {{Transient relativistic
  thermodynamics and kinetic theory}},\ }\href
  {https://doi.org/10.1016/0003-4916(79)90130-1} {\bibfield  {journal}
  {\bibinfo  {journal} {Annals Phys.}\ }\textbf {\bibinfo {volume} {118}},\
  \bibinfo {pages} {341} (\bibinfo {year} {1979})}\BibitemShut {NoStop}%
\bibitem [{\citenamefont {Gavassino}\ and\ \citenamefont
  {Noronha}(2023)}]{Gavassino:2023xkt}%
  \BibitemOpen
  \bibfield  {author} {\bibinfo {author} {\bibfnamefont {L.}~\bibnamefont
  {Gavassino}}\ and\ \bibinfo {author} {\bibfnamefont {J.}~\bibnamefont
  {Noronha}},\ }\bibfield  {title} {\bibinfo {title} {{Relativistic
  bulk-viscous dynamics far from equilibrium}},\ }\href@noop {} {\  (\bibinfo
  {year} {2023})},\ \Eprint {https://arxiv.org/abs/2305.04119}
  {arXiv:2305.04119 [gr-qc]} \BibitemShut {NoStop}%
\bibitem [{\citenamefont {Chen}\ and\ \citenamefont
  {Piekarewicz}(2014)}]{Chen:2014sca}%
  \BibitemOpen
  \bibfield  {author} {\bibinfo {author} {\bibfnamefont {W.-C.}\ \bibnamefont
  {Chen}}\ and\ \bibinfo {author} {\bibfnamefont {J.}~\bibnamefont
  {Piekarewicz}},\ }\bibfield  {title} {\bibinfo {title} {{Building
  relativistic mean field models for finite nuclei and neutron stars}},\ }\href
  {https://doi.org/10.1103/PhysRevC.90.044305} {\bibfield  {journal} {\bibinfo
  {journal} {Phys. Rev. C}\ }\textbf {\bibinfo {volume} {90}},\ \bibinfo
  {pages} {044305} (\bibinfo {year} {2014})},\ \Eprint
  {https://arxiv.org/abs/1408.4159} {arXiv:1408.4159 [nucl-th]} \BibitemShut
  {NoStop}%
\bibitem [{\citenamefont {Glendenning}(1997)}]{norman1997compact}%
  \BibitemOpen
  \bibfield  {author} {\bibinfo {author} {\bibfnamefont {N.~K.}\ \bibnamefont
  {Glendenning}},\ }\href {https://books.google.com.br/books?id=57XvAAAAMAAJ}
  {\emph {\bibinfo {title} {Compact Stars: Nuclear Physics, Particle Physics
  and General Relativity}}},\ Astronomy and Astrophysics Library\ (\bibinfo
  {publisher} {Springer New York},\ \bibinfo {year} {1997})\BibitemShut
  {NoStop}%
\bibitem [{\citenamefont {Abbott}\ \emph {et~al.}(2019)\citenamefont {Abbott}
  \emph {et~al.}}]{LIGOScientific:2018hze}%
  \BibitemOpen
  \bibfield  {author} {\bibinfo {author} {\bibfnamefont {B.~P.}\ \bibnamefont
  {Abbott}} \emph {et~al.} (\bibinfo {collaboration} {LIGO Scientific,
  Virgo}),\ }\bibfield  {title} {\bibinfo {title} {{Properties of the binary
  neutron star merger GW170817}},\ }\href
  {https://doi.org/10.1103/PhysRevX.9.011001} {\bibfield  {journal} {\bibinfo
  {journal} {Phys. Rev. X}\ }\textbf {\bibinfo {volume} {9}},\ \bibinfo {pages}
  {011001} (\bibinfo {year} {2019})},\ \Eprint
  {https://arxiv.org/abs/1805.11579} {arXiv:1805.11579 [gr-qc]} \BibitemShut
  {NoStop}%
\bibitem [{\citenamefont {Miller}\ \emph {et~al.}(2019)\citenamefont {Miller}
  \emph {et~al.}}]{Miller:2019cac}%
  \BibitemOpen
  \bibfield  {author} {\bibinfo {author} {\bibfnamefont {M.~C.}\ \bibnamefont
  {Miller}} \emph {et~al.},\ }\bibfield  {title} {\bibinfo {title} {{PSR
  J0030+0451 Mass and Radius from $NICER$ Data and Implications for the
  Properties of Neutron Star Matter}},\ }\href
  {https://doi.org/10.3847/2041-8213/ab50c5} {\bibfield  {journal} {\bibinfo
  {journal} {Astrophys. J. Lett.}\ }\textbf {\bibinfo {volume} {887}},\
  \bibinfo {pages} {L24} (\bibinfo {year} {2019})},\ \Eprint
  {https://arxiv.org/abs/1912.05705} {arXiv:1912.05705 [astro-ph.HE]}
  \BibitemShut {NoStop}%
\bibitem [{\citenamefont {Riley}\ \emph {et~al.}(2019)\citenamefont {Riley}
  \emph {et~al.}}]{Riley:2019yda}%
  \BibitemOpen
  \bibfield  {author} {\bibinfo {author} {\bibfnamefont {T.~E.}\ \bibnamefont
  {Riley}} \emph {et~al.},\ }\bibfield  {title} {\bibinfo {title} {{A $NICER$
  View of PSR J0030+0451: Millisecond Pulsar Parameter Estimation}},\ }\href
  {https://doi.org/10.3847/2041-8213/ab481c} {\bibfield  {journal} {\bibinfo
  {journal} {Astrophys. J. Lett.}\ }\textbf {\bibinfo {volume} {887}},\
  \bibinfo {pages} {L21} (\bibinfo {year} {2019})},\ \Eprint
  {https://arxiv.org/abs/1912.05702} {arXiv:1912.05702 [astro-ph.HE]}
  \BibitemShut {NoStop}%
\bibitem [{\citenamefont {Miller}\ \emph {et~al.}(2021)\citenamefont {Miller}
  \emph {et~al.}}]{Miller:2021qha}%
  \BibitemOpen
  \bibfield  {author} {\bibinfo {author} {\bibfnamefont {M.~C.}\ \bibnamefont
  {Miller}} \emph {et~al.},\ }\bibfield  {title} {\bibinfo {title} {{The Radius
  of PSR J0740+6620 from NICER and XMM-Newton Data}},\ }\href
  {https://doi.org/10.3847/2041-8213/ac089b} {\bibfield  {journal} {\bibinfo
  {journal} {Astrophys. J. Lett.}\ }\textbf {\bibinfo {volume} {918}},\
  \bibinfo {pages} {L28} (\bibinfo {year} {2021})},\ \Eprint
  {https://arxiv.org/abs/2105.06979} {arXiv:2105.06979 [astro-ph.HE]}
  \BibitemShut {NoStop}%
\bibitem [{\citenamefont {Riley}\ \emph {et~al.}(2021)\citenamefont {Riley}
  \emph {et~al.}}]{Riley:2021pdl}%
  \BibitemOpen
  \bibfield  {author} {\bibinfo {author} {\bibfnamefont {T.~E.}\ \bibnamefont
  {Riley}} \emph {et~al.},\ }\bibfield  {title} {\bibinfo {title} {{A NICER
  View of the Massive Pulsar PSR J0740+6620 Informed by Radio Timing and
  XMM-Newton Spectroscopy}},\ }\href {https://doi.org/10.3847/2041-8213/ac0a81}
  {\bibfield  {journal} {\bibinfo  {journal} {Astrophys. J. Lett.}\ }\textbf
  {\bibinfo {volume} {918}},\ \bibinfo {pages} {L27} (\bibinfo {year}
  {2021})},\ \Eprint {https://arxiv.org/abs/2105.06980} {arXiv:2105.06980
  [astro-ph.HE]} \BibitemShut {NoStop}%
\bibitem [{\citenamefont {Gavassino}(2023)}]{Gavassino:2023eoz}%
  \BibitemOpen
  \bibfield  {author} {\bibinfo {author} {\bibfnamefont {L.}~\bibnamefont
  {Gavassino}},\ }\bibfield  {title} {\bibinfo {title} {{Relativistic bulk
  viscous fluids of Burgers type and their presence in neutron stars}},\ }\href
  {https://doi.org/10.1088/1361-6382/ace587} {\bibfield  {journal} {\bibinfo
  {journal} {Class. Quant. Grav.}\ }\textbf {\bibinfo {volume} {40}},\ \bibinfo
  {pages} {165008} (\bibinfo {year} {2023})},\ \Eprint
  {https://arxiv.org/abs/2304.05455} {arXiv:2304.05455 [nucl-th]} \BibitemShut
  {NoStop}%
\bibitem [{\citenamefont {Bemfica}\ \emph {et~al.}(2019)\citenamefont
  {Bemfica}, \citenamefont {Disconzi},\ and\ \citenamefont
  {Noronha}}]{Bemfica:2019cop}%
  \BibitemOpen
  \bibfield  {author} {\bibinfo {author} {\bibfnamefont {F.~S.}\ \bibnamefont
  {Bemfica}}, \bibinfo {author} {\bibfnamefont {M.~M.}\ \bibnamefont
  {Disconzi}},\ and\ \bibinfo {author} {\bibfnamefont {J.}~\bibnamefont
  {Noronha}},\ }\bibfield  {title} {\bibinfo {title} {{Causality of the
  Einstein-Israel-Stewart Theory with Bulk Viscosity}},\ }\href
  {https://doi.org/10.1103/PhysRevLett.122.221602} {\bibfield  {journal}
  {\bibinfo  {journal} {Phys. Rev. Lett.}\ }\textbf {\bibinfo {volume} {122}},\
  \bibinfo {pages} {221602} (\bibinfo {year} {2019})},\ \Eprint
  {https://arxiv.org/abs/1901.06701} {arXiv:1901.06701 [gr-qc]} \BibitemShut
  {NoStop}%
\bibitem [{\citenamefont {Philipsen}(2013)}]{Philipsen:2012nu}%
  \BibitemOpen
  \bibfield  {author} {\bibinfo {author} {\bibfnamefont {O.}~\bibnamefont
  {Philipsen}},\ }\bibfield  {title} {\bibinfo {title} {{The QCD equation of
  state from the lattice}},\ }\href
  {https://doi.org/10.1016/j.ppnp.2012.09.003} {\bibfield  {journal} {\bibinfo
  {journal} {Prog. Part. Nucl. Phys.}\ }\textbf {\bibinfo {volume} {70}},\
  \bibinfo {pages} {55} (\bibinfo {year} {2013})},\ \Eprint
  {https://arxiv.org/abs/1207.5999} {arXiv:1207.5999 [hep-lat]} \BibitemShut
  {NoStop}%
\bibitem [{\citenamefont {Douchin}\ and\ \citenamefont
  {Haensel}(2001)}]{Douchin:2001sv}%
  \BibitemOpen
  \bibfield  {author} {\bibinfo {author} {\bibfnamefont {F.}~\bibnamefont
  {Douchin}}\ and\ \bibinfo {author} {\bibfnamefont {P.}~\bibnamefont
  {Haensel}},\ }\bibfield  {title} {\bibinfo {title} {{A unified equation of
  state of dense matter and neutron star structure}},\ }\href
  {https://doi.org/10.1051/0004-6361:20011402} {\bibfield  {journal} {\bibinfo
  {journal} {Astron. Astrophys.}\ }\textbf {\bibinfo {volume} {380}},\ \bibinfo
  {pages} {151} (\bibinfo {year} {2001})},\ \Eprint
  {https://arxiv.org/abs/astro-ph/0111092} {arXiv:astro-ph/0111092}
  \BibitemShut {NoStop}%
\bibitem [{\citenamefont {Alford}\ \emph {et~al.}(2020)\citenamefont {Alford},
  \citenamefont {Harutyunyan},\ and\ \citenamefont
  {Sedrakian}}]{Alford:2020lla}%
  \BibitemOpen
  \bibfield  {author} {\bibinfo {author} {\bibfnamefont {M.}~\bibnamefont
  {Alford}}, \bibinfo {author} {\bibfnamefont {A.}~\bibnamefont
  {Harutyunyan}},\ and\ \bibinfo {author} {\bibfnamefont {A.}~\bibnamefont
  {Sedrakian}},\ }\bibfield  {title} {\bibinfo {title} {{Bulk Viscous Damping
  of Density Oscillations in Neutron Star Mergers}},\ }\href
  {https://doi.org/10.3390/particles3020034} {\bibfield  {journal} {\bibinfo
  {journal} {Particles}\ }\textbf {\bibinfo {volume} {3}},\ \bibinfo {pages}
  {500} (\bibinfo {year} {2020})},\ \Eprint {https://arxiv.org/abs/2006.07975}
  {arXiv:2006.07975 [nucl-th]} \BibitemShut {NoStop}%
\bibitem [{\citenamefont {Rezzolla}\ and\ \citenamefont
  {Zanotti}(2013)}]{Rezzolla_Zanotti_book}%
  \BibitemOpen
  \bibfield  {author} {\bibinfo {author} {\bibfnamefont {L.}~\bibnamefont
  {Rezzolla}}\ and\ \bibinfo {author} {\bibfnamefont {O.}~\bibnamefont
  {Zanotti}},\ }\href@noop {} {\emph {\bibinfo {title} {Relativistic
  hydrodynamics}}}\ (\bibinfo  {publisher} {Oxford University Press},\ \bibinfo
  {address} {New York},\ \bibinfo {year} {2013})\BibitemShut {NoStop}%
\bibitem [{\citenamefont {Alford}\ \emph {et~al.}(2022)\citenamefont {Alford},
  \citenamefont {Brodie}, \citenamefont {Haber},\ and\ \citenamefont
  {Tews}}]{Alford:2022bpp}%
  \BibitemOpen
  \bibfield  {author} {\bibinfo {author} {\bibfnamefont {M.~G.}\ \bibnamefont
  {Alford}}, \bibinfo {author} {\bibfnamefont {L.}~\bibnamefont {Brodie}},
  \bibinfo {author} {\bibfnamefont {A.}~\bibnamefont {Haber}},\ and\ \bibinfo
  {author} {\bibfnamefont {I.}~\bibnamefont {Tews}},\ }\bibfield  {title}
  {\bibinfo {title} {{Relativistic mean-field theories for neutron-star physics
  based on chiral effective field theory}},\ }\href
  {https://doi.org/10.1103/PhysRevC.106.055804} {\bibfield  {journal} {\bibinfo
   {journal} {Phys. Rev. C}\ }\textbf {\bibinfo {volume} {106}},\ \bibinfo
  {pages} {055804} (\bibinfo {year} {2022})},\ \Eprint
  {https://arxiv.org/abs/2205.10283} {arXiv:2205.10283 [nucl-th]} \BibitemShut
  {NoStop}%
\bibitem [{\citenamefont {Typel}\ \emph {et~al.}(2015)\citenamefont {Typel},
  \citenamefont {Oertel},\ and\ \citenamefont {Kl\"ahn}}]{Typel:2013rza}%
  \BibitemOpen
  \bibfield  {author} {\bibinfo {author} {\bibfnamefont {S.}~\bibnamefont
  {Typel}}, \bibinfo {author} {\bibfnamefont {M.}~\bibnamefont {Oertel}},\ and\
  \bibinfo {author} {\bibfnamefont {T.}~\bibnamefont {Kl\"ahn}},\ }\bibfield
  {title} {\bibinfo {title} {{CompOSE CompStar online supernova equations of
  state harmonising the concert of nuclear physics and astrophysics
  compose.obspm.fr}},\ }\href {https://doi.org/10.1134/S1063779615040061}
  {\bibfield  {journal} {\bibinfo  {journal} {Phys. Part. Nucl.}\ }\textbf
  {\bibinfo {volume} {46}},\ \bibinfo {pages} {633} (\bibinfo {year} {2015})},\
  \Eprint {https://arxiv.org/abs/1307.5715} {arXiv:1307.5715 [astro-ph.SR]}
  \BibitemShut {NoStop}%
\bibitem [{\citenamefont {Oertel}\ \emph {et~al.}(2017)\citenamefont {Oertel},
  \citenamefont {Hempel}, \citenamefont {Kl\"ahn},\ and\ \citenamefont
  {Typel}}]{Oertel:2016bki}%
  \BibitemOpen
  \bibfield  {author} {\bibinfo {author} {\bibfnamefont {M.}~\bibnamefont
  {Oertel}}, \bibinfo {author} {\bibfnamefont {M.}~\bibnamefont {Hempel}},
  \bibinfo {author} {\bibfnamefont {T.}~\bibnamefont {Kl\"ahn}},\ and\ \bibinfo
  {author} {\bibfnamefont {S.}~\bibnamefont {Typel}},\ }\bibfield  {title}
  {\bibinfo {title} {{Equations of state for supernovae and compact stars}},\
  }\href {https://doi.org/10.1103/RevModPhys.89.015007} {\bibfield  {journal}
  {\bibinfo  {journal} {Rev. Mod. Phys.}\ }\textbf {\bibinfo {volume} {89}},\
  \bibinfo {pages} {015007} (\bibinfo {year} {2017})},\ \Eprint
  {https://arxiv.org/abs/1610.03361} {arXiv:1610.03361 [astro-ph.HE]}
  \BibitemShut {NoStop}%
\bibitem [{\citenamefont {Typel}\ \emph {et~al.}(2022)\citenamefont {Typel}
  \emph {et~al.}}]{CompOSECoreTeam:2022ddl}%
  \BibitemOpen
  \bibfield  {author} {\bibinfo {author} {\bibfnamefont {S.}~\bibnamefont
  {Typel}} \emph {et~al.} (\bibinfo {collaboration} {CompOSE Core Team}),\
  }\bibfield  {title} {\bibinfo {title} {{CompOSE Reference Manual}},\ }\href
  {https://doi.org/10.1140/epja/s10050-022-00847-y} {\bibfield  {journal}
  {\bibinfo  {journal} {Eur. Phys. J. A}\ }\textbf {\bibinfo {volume} {58}},\
  \bibinfo {pages} {221} (\bibinfo {year} {2022})},\ \Eprint
  {https://arxiv.org/abs/2203.03209} {arXiv:2203.03209 [astro-ph.HE]}
  \BibitemShut {NoStop}%
\bibitem [{\citenamefont {Alford}\ \emph {et~al.}(2012)\citenamefont {Alford},
  \citenamefont {Reddy},\ and\ \citenamefont {Schwenzer}}]{Alford:2011df}%
  \BibitemOpen
  \bibfield  {author} {\bibinfo {author} {\bibfnamefont {M.~G.}\ \bibnamefont
  {Alford}}, \bibinfo {author} {\bibfnamefont {S.}~\bibnamefont {Reddy}},\ and\
  \bibinfo {author} {\bibfnamefont {K.}~\bibnamefont {Schwenzer}},\ }\bibfield
  {title} {\bibinfo {title} {{Bridging the Gap by Squeezing Superfluid
  Matter}},\ }\href {https://doi.org/10.1103/PhysRevLett.108.111102} {\bibfield
   {journal} {\bibinfo  {journal} {Phys. Rev. Lett.}\ }\textbf {\bibinfo
  {volume} {108}},\ \bibinfo {pages} {111102} (\bibinfo {year} {2012})},\
  \Eprint {https://arxiv.org/abs/1110.6213} {arXiv:1110.6213 [nucl-th]}
  \BibitemShut {NoStop}%
\bibitem [{\citenamefont {Harris}(2020)}]{Harris:2020rus}%
  \BibitemOpen
  \bibfield  {author} {\bibinfo {author} {\bibfnamefont {S.~P.}\ \bibnamefont
  {Harris}},\ }\emph {\bibinfo {title} {{Transport in Neutron Star Mergers}}},\
  \href {https://doi.org/10.7936/wrmz-1n98} {Ph.D. thesis},\ \bibinfo  {school}
  {Washington U., St. Louis} (\bibinfo {year} {2020}),\ \Eprint
  {https://arxiv.org/abs/2005.09618} {arXiv:2005.09618 [nucl-th]} \BibitemShut
  {NoStop}%
\bibitem [{\citenamefont {Sa'd}\ and\ \citenamefont
  {Schaffner-Bielich}(2009)}]{Sad:2009hba}%
  \BibitemOpen
  \bibfield  {author} {\bibinfo {author} {\bibfnamefont {B.~A.}\ \bibnamefont
  {Sa'd}}\ and\ \bibinfo {author} {\bibfnamefont {J.}~\bibnamefont
  {Schaffner-Bielich}},\ }\bibfield  {title} {\bibinfo {title} {{Dissipation of
  radial oscillations in compact stars}},\ }\href@noop {} {\  (\bibinfo {year}
  {2009})},\ \Eprint {https://arxiv.org/abs/0908.4190} {arXiv:0908.4190
  [astro-ph.SR]} \BibitemShut {NoStop}%
\bibitem [{\citenamefont {Stephanov}\ and\ \citenamefont
  {Yin}(2018)}]{Stephanov:2017ghc}%
  \BibitemOpen
  \bibfield  {author} {\bibinfo {author} {\bibfnamefont {M.}~\bibnamefont
  {Stephanov}}\ and\ \bibinfo {author} {\bibfnamefont {Y.}~\bibnamefont
  {Yin}},\ }\bibfield  {title} {\bibinfo {title} {{Hydrodynamics with
  parametric slowing down and fluctuations near the critical point}},\ }\href
  {https://doi.org/10.1103/PhysRevD.98.036006} {\bibfield  {journal} {\bibinfo
  {journal} {Phys. Rev. D}\ }\textbf {\bibinfo {volume} {98}},\ \bibinfo
  {pages} {036006} (\bibinfo {year} {2018})},\ \Eprint
  {https://arxiv.org/abs/1712.10305} {arXiv:1712.10305 [nucl-th]} \BibitemShut
  {NoStop}%
\bibitem [{\citenamefont {Kapusta}\ and\ \citenamefont
  {Gale}(2011)}]{Kapusta:2006pm}%
  \BibitemOpen
  \bibfield  {author} {\bibinfo {author} {\bibfnamefont {J.~I.}\ \bibnamefont
  {Kapusta}}\ and\ \bibinfo {author} {\bibfnamefont {C.}~\bibnamefont {Gale}},\
  }\href {https://doi.org/10.1017/CBO9780511535130} {\emph {\bibinfo {title}
  {{Finite-temperature field theory: Principles and applications}}}},\
  Cambridge Monographs on Mathematical Physics\ (\bibinfo  {publisher}
  {Cambridge University Press},\ \bibinfo {year} {2011})\BibitemShut {NoStop}%
\end{thebibliography}%

\end{document}